\documentclass[a4paper,11pt]{article}
\pdfoutput=1 

\usepackage{jheppub} 

\usepackage{bm,amsmath,amssymb,slashed,graphicx,%
            enumerate,alltt,xspace,multirow,xcolor,mathrsfs}
\usepackage{fancyvrb}

\RequirePackage{braket}
\RequirePackage{graphicx}

\RequirePackage[numbers,sort&compress]{natbib}

\makeatletter
\g@addto@macro\bfseries{\boldmath}
\makeatother

\definecolor{labelkey}{rgb}{0,0.5,0.0}

\usepackage{listings}
\definecolor{darkgreen}{rgb}{0,0.4,0}
\newcommand\qg{a} 

\newcommand\qlggset{\{q,l,g,\gamma\}}
\newcommand\qlset{\{q,l\}}
\newcommand\qlgset{\{q,l,\gamma\}}
\newcommand\qset{\{q\}}
\newcommand\qgset{\{q,g\}}
\newcommand\mathd{\mathrm{d}} 
\newcommand\eph{\ensuremath{e_{\mathrm{ph}}}}
\newcommand\ephD{\ensuremath{e_{\mathrm{ph, D}}}}
\newcommand\aph{\ensuremath{\alpha_{\mathrm{ph}}}}
\newcommand\aphD{\ensuremath{\alpha_{\mathrm{ph, D}}}}
\newcommand{\GeV}{\;\mathrm{GeV}}
\newcommand{\TeV}{\;\mathrm{TeV}}
\newcommand{\order}[1]{{\cal O}\left(#1\right)}
\newcommand{\as}{\alpha_s}
\newcommand{\MSbar}{\ensuremath{\overline{\text{MS}}}}

\newcommand{\ki}{\chi}
\newcommand\xbj{x_{\scriptscriptstyle\rm bj}}
\newcommand\mpr{m_{\rm p}}
\newcommand\qperp{q_\perp}
\newcommand\xph{x}
\newcommand\zed{z}
\newcommand\Qtwomax{Q^2_{\uparrow}}
\newcommand\Qtwomin{Q^2_{\downarrow}}
\newcommand\Qtwomaxmin{Q^2_{\uparrow / \downarrow}}
\newcommand\Qtwomaxexp{Q^2_{\rm max}}
\newcommand\Qtwominexp{Q^2_{\rm min}}
\newcommand{\nobracket}{}
\newcommand{\tmop}[1]{\ensuremath{\operatorname{#1}}}

\newcommand\nn{ \nonumber \\ }
\newcommand\rd { \mathrm{d} }
\newcommand\mub{ \mu}
\newcommand\muz{\mu_M}
\newcommand\muzfun{{M^2}}
\newcommand\muzfunNoSqr{{M}}
\newcommand\smu{\left(\mathcal{S}\mu\right)}

\newcommand\Piqa{B_{I,\qg}^{(1,0)}}

\newcommand\Ptwoqa{B_{2,\qg}^{(1,0)}}
\newcommand\Ptwog{B_{2,g}^{(1,0)}}
\newcommand\Ptwoqael{B_{2,\qg}^{(0,1)}}
\newcommand\PLqa{B_{L,\qg}^{(1,0)}}

\newcommand\sigmaprobe[1]{{\sigma_{{l#1}}}}
\newcommand\sigmahatprobe[1]{{\widehat{\sigma}_{l#1}}}

\newcommand\bqed{b_{\rm qed}}
\newcommand\shat{{\hat s}}
\newcommand{\cM}{{\cal M}}

\title{The Photon Content of the Proton}
\preprint{CERN-TH/2017-141}

\author[a]{Aneesh V.~Manohar,}
\author[b]{Paolo Nason,}
\author[c,*]{Gavin P.~Salam,\note[*]{On leave from CNRS, UMR 7589, LPTHE, F-75005, Paris, France}}
\author[c,d]{Giulia Zanderighi}

\emailAdd{amanohar@ucsd.edu}
\emailAdd{paolo.nason@mib.infn.it}
\emailAdd{gavin.salam@cern.ch}
\emailAdd{giulia.zanderighi@cern.ch}

\affiliation[a]{Department of Physics,  University of California at San Diego, 9500 Gilman Drive, La Jolla, CA 92093-0319, USA}
\affiliation[b]{Universit\`a di Milano-Bicocca and INFN, Sezione di
  Milano-Bicocca, Piazza della Scienza 3,20126 Milano, Italy}
\affiliation[c]{CERN, Theoretical Physics Department, CH-1211,  Geneva 23, Switzerland}
\affiliation[d]{Rudolf Peierls Centre for Theoretical, Physics, 1 Keble Road, University of Oxford, UK}

\date{Received: date / Accepted: \today}

\abstract{The photon PDF of the proton is needed for precision
  comparisons of LHC cross sections with theoretical predictions. In a
  recent paper, we showed how the photon PDF could be determined in
  terms of the electromagnetic proton structure functions $F_2$ and
  $F_L$ measured in electron--proton scattering experiments, and gave
  an explicit formula for the PDF including all terms up to
  next-to-leading order.
  In this paper we give details of the derivation. We obtain the
  photon PDF using the factorisation theorem and applying it to
  suitable BSM hard scattering processes. We also obtain the same PDF
  in a process-independent manner using the usual definition of PDFs
  in terms of light-cone Fourier transforms of products of operators.
  We show how our method gives an \emph{exact} representation for the
  photon PDF in terms of $F_2$ and $F_L$, valid to all
  orders in QED and QCD, and including all non-perturbative
  corrections. 
  This representation is then used to give an explicit formula for the
  photon PDF to one order higher than our previous result. We also
  generalise our results to obtain formul\ae\ for the polarised photon
  PDF, as well as the photon TMDPDF. Using our formula, we derive the
  $P_{\gamma i}$ subset of DGLAP splitting functions to order
  $\alpha \alpha_s$ and $\alpha^2$, which agree with known results.
  We give a detailed explanation of the approach that we follow to
  determine a photon PDF and its uncertainty within the above
  framework.
}


\begin{document}

\maketitle


\section{Introduction}
\label{sec:intro}
Hard scattering processes involving hadrons can be computed in terms
of parton distribution functions (PDFs) $f_{a/T}(x,\mu^2)$, the
probability to find a parton $a$ with momentum fraction $x$ in a
hadron target $T$. These distributions depend logarithmically on a
renormalisation/factorisation scale $\mu$ due to radiative
corrections.
In $pp$ scattering at CERN's Large Hadron Collider (LHC), the most
important PDFs needed are those of quarks and gluons in a proton
target. The photon PDF is small compared to that of quarks and gluons,
since it is suppressed by a power of the electromagnetic coupling
$\alpha$. However, knowledge of the photon PDF is becoming more
important as the measurements become increasingly precise: the
uncertainty on the photon PDF is becoming a limiting factor in our
ability to predict certain key reactions at the LHC. Some notable
examples are the production of the Higgs boson through $W/Z$
fusion~\cite{Ciccolini:2007ec}, or in association with an outgoing
weak boson~\cite{Denner:2011id}. For $W^\pm H$ production it is the
largest source of uncertainty~\cite{deFlorian:2016spz}. The photon
distribution is also relevant for the production of
lepton-pairs~\cite{Aad:2015auj,Aad:2016zzw,Accomando:2016tah,Alioli:2016fum,Bourilkov:2016qum},
top-quarks~\cite{Pagani:2016caq}, pairs of weak
bosons~\cite{Luszczak:2014mta,Denner:2015fca,Ababekri:2016kkj,Biedermann:2016yvs,Biedermann:2016guo,Yong:2016njr,Kallweit:2017khh}
and generally enters into electroweak corrections for almost any LHC
process.

Previous results on the photon PDF include 
{MRST2004qed}~\cite{Martin:2004dh}, {NNPDF23\_qed}~\cite{Ball:2013hta}
and CT14qed\_inc~\cite{Schmidt:2015zda} (all available within 
LHAPDF~\cite{Buckley:2014ana}) and the 
HKR16 set~\cite{Harland-Lang:2016kog}. 
These results either have
large uncertainties, or rely upon phenomenologically inspired models
for the contribution to the photon PDF from the low-$Q^2$ region. A
more detailed discussion of previous results is given in
Ref.~\cite{Manohar:2016nzj}.
There, we presented a formula for the photon PDF of the proton
$f_\gamma(x,\mu^2)$ as an integral over proton structure functions
$F_2(x,Q^2)$ and $F_L(x,Q^2)$, including all terms of order
$\mathcal{O}(\alpha)$ and with errors of order
$\mathcal{O}(\alpha^2L)$, $\mathcal{O}(\alpha\alpha_s)$, where $L$ is
the logarithm of $\mu$ divided by some typical hadronic
scale.\footnote{The photon PDF $f_\gamma(x,\mu^2)$ is of order $\alpha
  L$. The result in Ref.~\cite{Manohar:2016nzj} included terms of
  order $\alpha^2 L^2 (\as L)^n$ and $\alpha (\alpha_s L)^n$, but not
  of order $\alpha^2 L$.  By assuming $L\approx 1/\alpha_s$, and
  considering that $\alpha_s^2 \approx \alpha $, we see that the first
  subleading terms to include are of order $\alpha$ (i.e. one power of
  $L$ less than the leading term) and of order $\alpha^2 L^2$.}
It was used to obtain the photon PDF with $\lesssim 3\%$ uncertainty
over a wide range of $x$ values, $10^{-5} \le x \le 0.5$. This reduces
the uncertainty by about a factor of forty over previous photon PDF
determinations such as {MRST2004qed}~\cite{Martin:2004dh} and
{NNPDF23\_qed}~\cite{Ball:2013hta}, which rely on fits to LHC data
and/or modelling. The main idea in our derivation is that a photon
initiated process can be computed two different ways: in terms of a
photon PDF, using the factorisation theorem, or in terms of the deep
inelastic scattering (DIS) hadronic tensor. Equating the two
expressions leads to our result.\footnote{ The crucial observation
  that we used in Ref.~\cite{Manohar:2016nzj} was inspired in part by
  Drees and Zeppenfeld's study of supersymmetric particle production
  at $ep$ colliders~\cite{Drees:1988pp}.}  All information needed to
extract $f_{\gamma}$ is thus contained in $ep$ scattering data. This
point of view is implicit also in
Refs.~\cite{Anlauf:1991wr,Mukherjee:2003yh}.

In this paper, we give additional details on the derivation of the
photon PDF formula. We report the explicit calculations performed for
the two different hard probes that we consider: heavy lepton
production by a flavour violating magnetic moment interaction, and
scalar production via $\gamma\gamma$ collisions.  We also give an
independent derivation of the formula by defining the photon PDF as
the light-cone Fourier transform of the two-point function of
electromagnetic field-strength tensors.  Our derivation gives a
representation for the photon PDF which is \emph{exact}, i.e.\ valid
to all orders in the strong and electromagnetic interactions.
We use this exact representation to obtain expressions for the photon
PDF including all terms of order $\alpha \alpha_s$ and $\alpha^2$, one
order higher in $\alpha_s(\mu)$ or $\alpha(\mu)$ than our original
result~\cite{Manohar:2016nzj}.

In the present work we have slightly revised the photon PDF
calculation and error estimate compared with
Ref.~\cite{Manohar:2016nzj}, also taking into account the new results
presented here (see Sec.~\ref{sec:scale-uncert-res}).
For this reason, we refer to the photon PDF computed using the method
in this paper as LUXqed17, and that computed using the previous
procedure as LUXqed. The ratio of the two is shown in
Fig.~\ref{fig:compare}.  As one can see the ensuing changes are very
minor, and confirm the error estimate given in
Ref.~\cite{Manohar:2016nzj}.

Our formula for the photon PDF can be differentiated w.r.t.\ $\mu$ to
give the $P_{\gamma i}$ subset of the DGLAP splitting functions. The
result in Ref.~\cite{Manohar:2016nzj} was used to obtain the two-loop
order $\alpha \alpha_s$ splitting kernels, which agree with a recent
computation~\cite{deFlorian:2015ujt}.  Here, we present this
calculation in detail, including also the computation of the splitting
kernels of order $\alpha^2$, which agree with
Ref.~\cite{deFlorian:2016gvk}.

We also give some natural extensions of our results. We obtain the
photon transverse momentum dependent PDF (TMDPDF), and by considering
spin-dependent scattering, we obtain the polarised photon PDF
$f_{\Delta \gamma}(x,\mu^2)$ in terms of the polarised structure
functions $g_{1,2}(x,Q^2)$.

The outline of the paper is as follows. Section~\ref{sec:em}
introduces our notation and reviews some known results on QED
corrections to the hadronic tensors, which are important for obtaining
our all-orders formula.
Sec.~\ref{sec:probe} gives the derivation of the photon PDF using
heavy-lepton production via a magnetic moment
interaction. Section~\ref{sec:split-fns} derives the order $\alpha
\alpha_s$ and $\alpha^2$ splitting functions from our photon PDF
formula. The extension of our results to polarised PDFs is given in
Sec.~\ref{sec:spin-dep}. Section~\ref{sec:alt-deriv} gives the
derivation of the photon PDF for the polarised and unpolarised cases
in terms of PDF operators. The result in this section is exact, and is
also used to obtain the photon TMDPDF. Section~\ref{sec:higher-order}
uses the exact result in Sec.~\ref{sec:alt-deriv} to obtain the PDF to
order $\alpha, \alpha \alpha_s$ and $\alpha_s^2$ relative to the
leading order term. In Sec.~\ref{sec:input-data} we discuss the
experimental data inputs used for our numerical evaluation of the
photon PDF. In Sec.~\ref{sec:scale} we give our procedure for
estimating errors from missing higher-order corrections. In
Sec.~\ref{sec:practical-implementation} we give all details about the
practical implementation of the photon PDF formula and how the photon
is matched to other partons. Some results on the photon PDF are given
in Sec.~\ref{sec:photon-pdf-results}. We present our conclusions in
Sec.~\ref{sec:conclusions}.
Useful technical results are given in the
Appendices. Appendix~\ref{app:emcurrent} gives technical details on
QED corrections, App.~\ref{sec:kin} discusses the kinematics for our
exact representation, App.~\ref{app:gamgam} derives the photon PDF
using scalar production in $\gamma \gamma$ collisions,
App.~\ref{sec:coll-q-e-to-L} presents the parton model NLO calculation
for the parton level process $q l \to qL$, App.~\ref{sec:low-Q2-F2-FL}
and~\ref{sec:R} discuss the low-energy behaviour for $F_{2/L}$ and
$R$, App.~\ref{subsec:coeff} summarises the DIS coefficient and
splitting functions and App.~\ref{sec:PDF4LHC15-issues} discusses
issues in PDF4LHC15 at low scales.

\section{Definitions and notation}\label{sec:em}

To aid the reader, it is useful to introduce some elements of our
notation.
We will use dimensional regularisation in $D=4-2\epsilon$ dimensions.
According to the \MSbar\ prescription, the dimensional regularisation
scale $\mu$ is replaced by $\mu \to \smu$, with
\begin{equation}
  {\cal S}^2=\frac{e^{\gamma_{\rm E}}}{4\pi}\,.
  \label{Sdef}
\end{equation}
Subtracting only $1/\epsilon$ poles then gives the renormalised result
in the \MSbar\ scheme.

The notation $i \in \qlggset $, means a sum over quarks \emph{and}
antiquarks, leptons \emph{and} antileptons, photons, and gluons.
Similarly for $i \in \qset$, $i \in \qgset$, $i \in \qlgset$, etc.
Colour multiplicities, when needed, will be denoted by $n_i$ and
written explicitly, $n_i = 1$ for leptons and $n_i = 3$ for quarks.

The perturbative expansion of an object $\cal O$, such as a partonic
cross section $\widehat \sigma$ or a splitting kernel $P$ will be
written as
\begin{align}
\mathcal{O} &= \sum_{r,s} \left[ \frac{\alpha_s(\mu^2)}{2\pi}
  \right]^r \left[ \frac{\alpha(\mu^2)}{2\pi} \right]^s
\mathcal{O}^{(r,s)}\,.
\label{expansion}
\end{align}

One of the results of this paper is a representation for the photon
PDF which is \emph{exact}, including all radiative and
non-perturbative QED and QCD corrections. This requires a careful
treatment of electromagnetic radiative corrections.\footnote{We do not
  consider electroweak corrections in this paper.
  These introduce a number of complexities and will be studied
 elsewhere.
} 
A brief summary of known results on electromagnetic radiative
corrections and the renormalisation of the electromagnetic
current is given in App.~\ref{app:emcurrent}. We
will use the results in the appendix to motivate the definitions in
this section.

We define the physical coupling $\eph$ in terms of the vacuum
polarisation function $\Pi(q^2,\mub^2)$,
\begin{align}
\eph^2(q^2) &= \frac{e^2(\mub^2)}{1 - \Pi(q^2,\mub^2)}\,,
\label{3.15}
\end{align}
which depends on $q^2$, but is independent of $\mub^2$ (see
Eq.~(\ref{A3.17})). $\aph(0) = \eph^2(0)/4\pi$ is the usual fine structure constant
$\simeq 1/137.036$ measured in atomic physics. Eq.~(\ref{3.15}) is used in the spacelike region $q^2<0$, where $\Pi(q^2,\mub^2)$ is real.

The proton hadronic tensor, which enters the cross section formula for
deep-inelastic scattering, is defined as
\begin{align}
W_{\mu\nu}(p,q,s) = \frac{1}{4\pi}\int \mathd^4 z \ e^{i q \cdot z}
\langle p,s| \left[ j_\mu(z), j_\nu(0)\right] | p,s\rangle \,,
\label{eq:hadtens}
\end{align}
where $p$ is the incoming proton momentum, $q$ is the photon momentum
(transferred to the proton) and $s$ denotes the proton spin.
$W_{\mu\nu}$ is the discontinuity of the time-ordered product 
\begin{align}
T_{\mu\nu}(p,q,s) = i\int \mathd^4 z \ e^{i q \cdot z} \langle p,s| T
\left\{ j_\mu(z), j_\nu(0)\right\} | p,s\rangle \,.
\label{eq:hadtensTmunu}
\end{align}
The discontinuity for $q_0 > 0$ gives the proton hadronic tensor,
corresponding to the $j_\mu(z) j_\nu(0)$ term in
Eq.~(\ref{eq:hadtens}).
The discontinuity for $q_0 < 0$ is related by crossing to the
anti-proton hadronic tensor and corresponds to the second term of the
commutator, which will not be needed here.

The conventional decomposition of $W_{\mu \nu}$ into structure
functions, after requiring Lorentz invariance, time reversal, parity,
and current conservation is
\begin{align}
W_{\mu\nu}(p,q,s) &= F_1 \left(-g_{\mu\nu} + {q_\mu
q_\nu\over q^2}\right) + {F_2\over p \cdot q} \left(p_\mu - {p\cdot q \ q_\mu\over
q^2}\right)
\left(p_\nu - {p\cdot q\ q_\nu\over q^2}\right)\cr
\noalign{\smallskip}&+ {ig_1\over p\cdot q}\ \epsilon_{\mu\nu\lambda\sigma}
q^\lambda s^\sigma +
{ig_2\over (p\cdot q)^2}\ \epsilon_{\mu\nu\lambda\sigma}
q^\lambda \left( p\cdot q
\, s^\sigma - s\cdot q\,p^\sigma\right)\ .
\label{3.21}
\end{align}
The proton spin four-vector $s$ is normalised so that $p \cdot s=0$,
$s \cdot s = - \mpr^2$, where $\mpr$ is the proton mass. The structure
functions $F_1$, $F_2$, $g_1$ and $g_2$ are functions of
\begin{align}
\xbj &= \frac{Q^2}{2 p \cdot q}
\label{3.22}
\end{align}
and $Q^2$. We also introduce the longitudinal structure function
\begin{align}
  F_L(\xbj,Q^2) &\equiv \left(1+\frac{4\xbj^2\mpr^2}{Q^2}\right)F_2(\xbj,Q^2)  - 2\xbj F_1(\xbj,Q^2), 
 \label{3.28}
\end{align}
and write our results using $F_2$ and $F_L$ instead of $F_2$ and
$F_1$.

The usual textbook analysis of deep inelastic scattering is at lowest
order in the electromagnetic coupling. In this paper, we are also
interested in higher-order electromagnetic corrections, so we should
be careful about the definition of $W_{\mu\nu}$. We define the
hadronic tensor in Eq.~(\ref{eq:hadtens}) to be given by
one-photon-irreducible graphs in the current channels, and
\emph{including all electromagnetic corrections to the hadronic matrix
  element}. With this definition, $W_{\mu \nu}$ is $\mu$-independent
(see App.~\ref{app:emcurrent}), so the structure functions in the
decomposition Eq.~(\ref{3.21}) are also $\mu$-independent, and depend
only on $\xbj$ and $Q^2$.\footnote{If one had instead used the
  $\MSbar$ current in Eq.~(\ref{eq:hadtens}), the structure functions
  would depend on $\mu$ as well as $\xbj$ and $Q^2$.}
The structure functions and $W_{\mu \nu}$ defined in terms of
one-photon irreducible graphs will be denoted by the label $1\gamma I$
in App.~\ref{app:emcurrent}. In the main text of this paper, we drop
this label for simplicity since we are always referring to the
one-photon irreducible hadronic tensor.

\begin{figure}
\centering
\begin{align*}
\renewcommand{\arraycolsep}{0.5cm}
\begin{array}{ccc}
\includegraphics[scale=0.3]{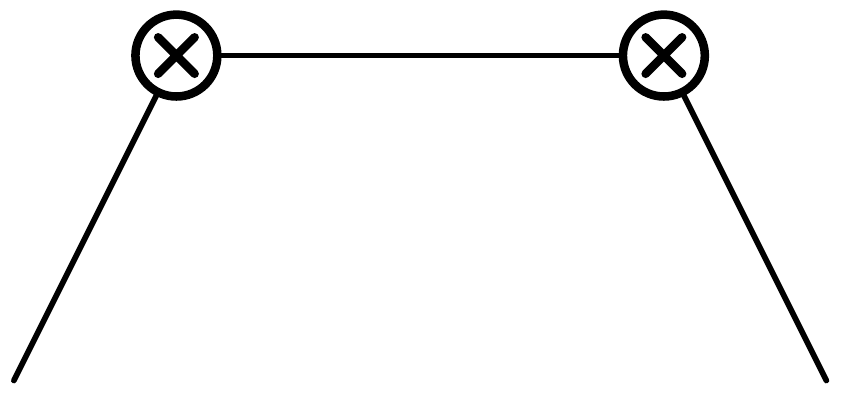} &
\includegraphics[scale=0.3]{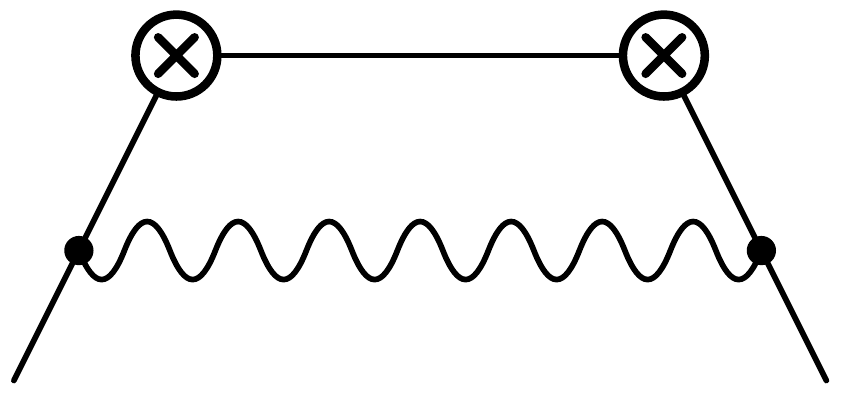} \\
(a) & (b)
\end{array}
\end{align*}
\begin{align*}
\renewcommand{\arraycolsep}{0.5cm}
\begin{array}{ccc}
\includegraphics[scale=0.3]{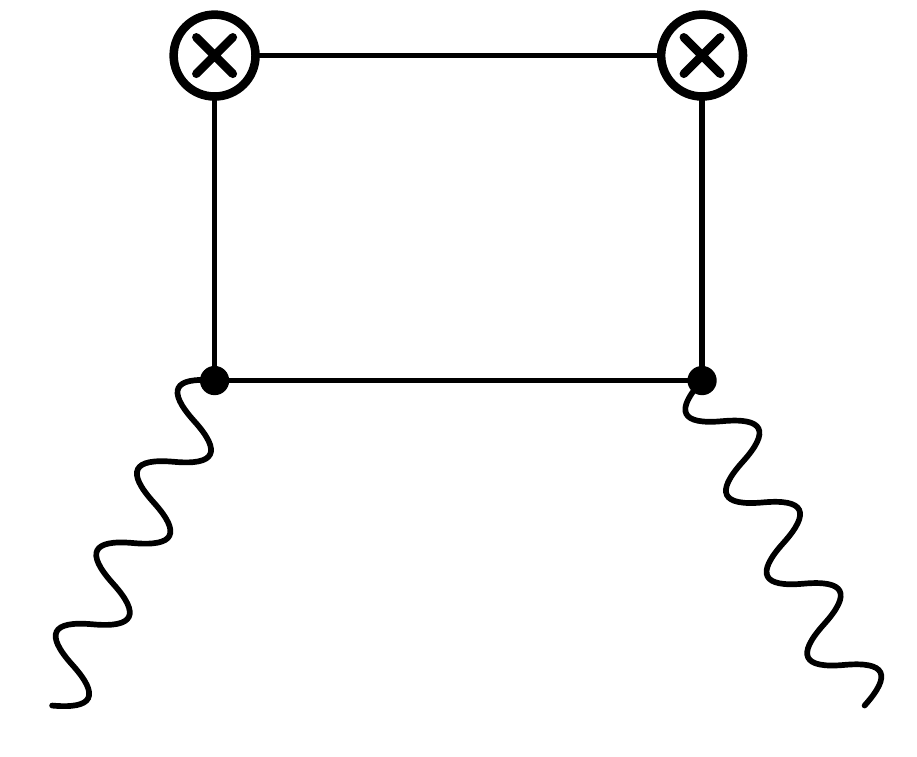} & \includegraphics[scale=0.3]{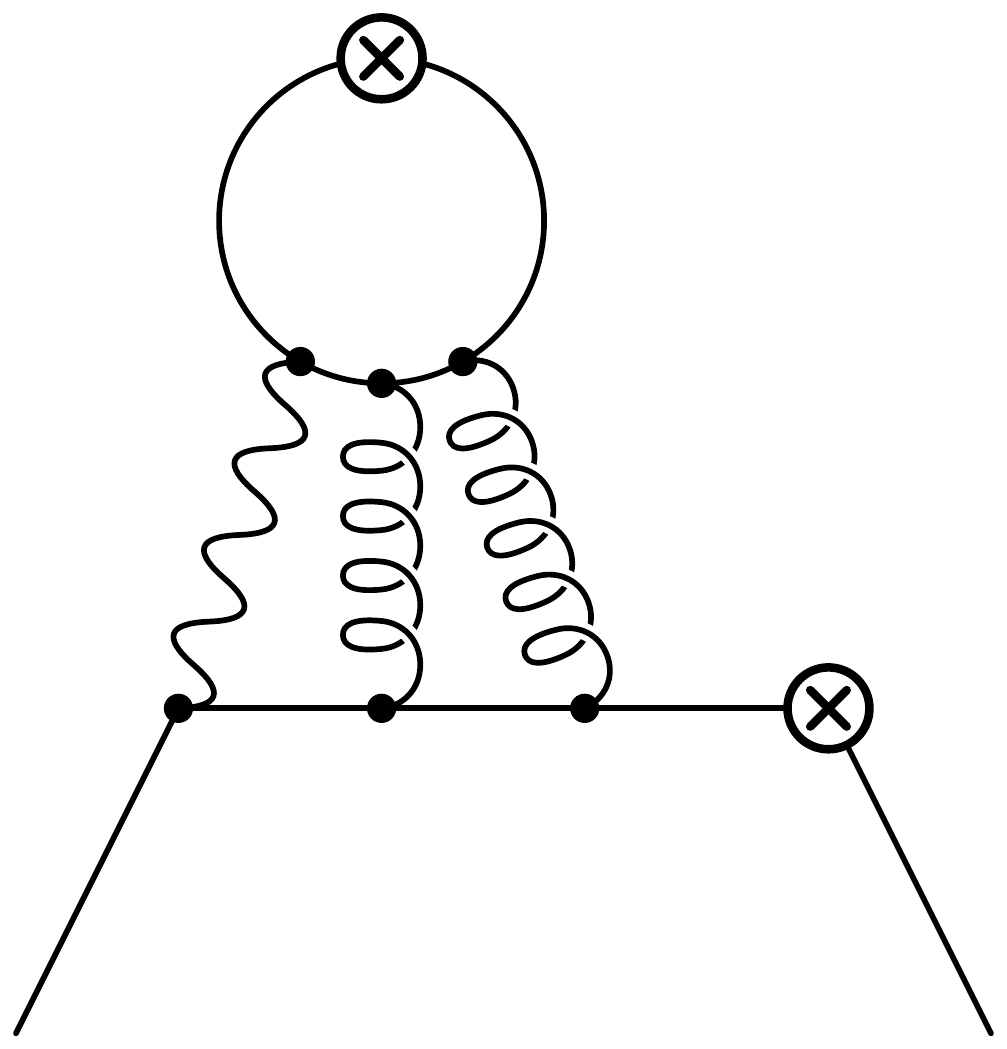} & \includegraphics[scale=0.3]{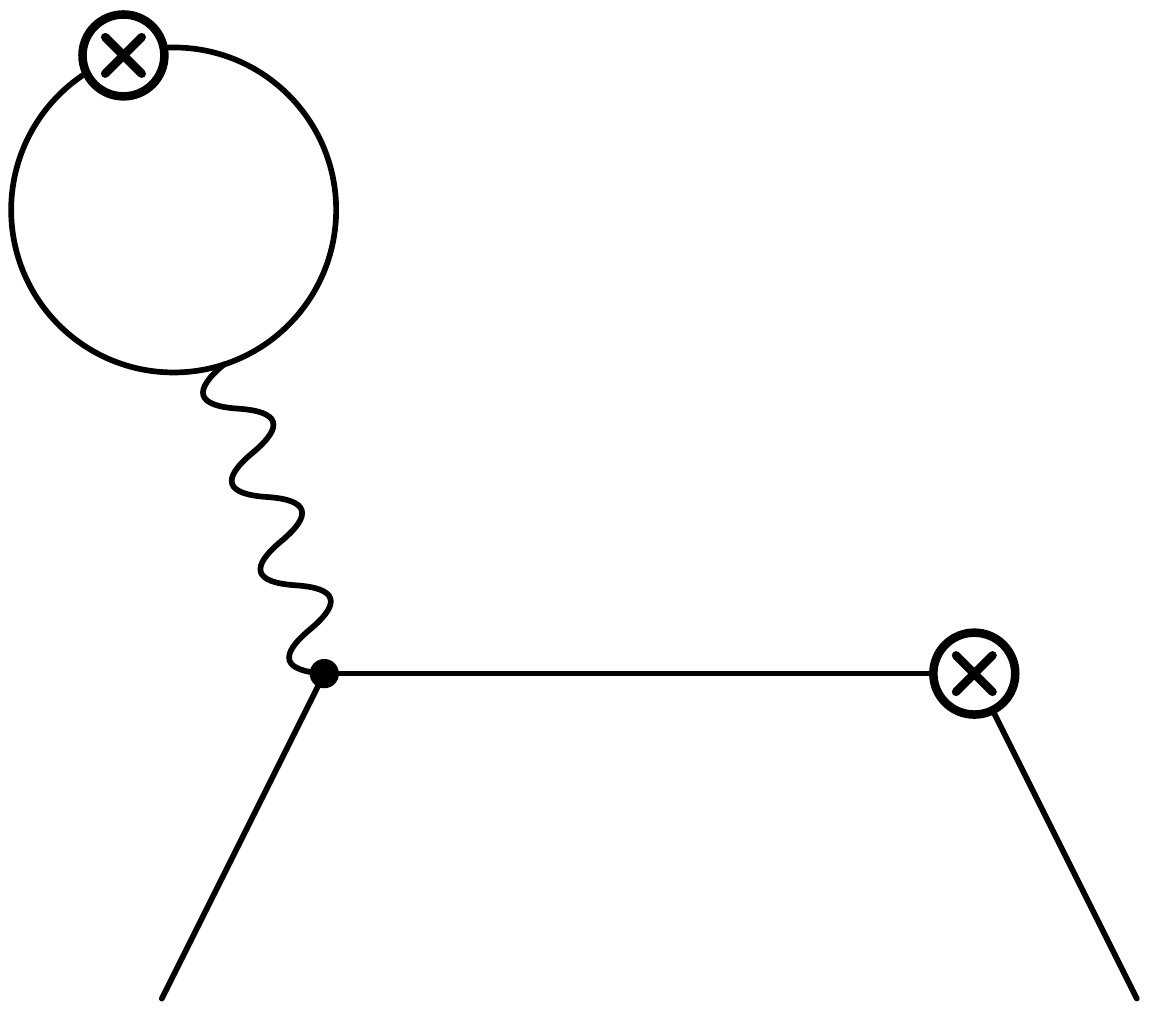} \\
(c) & (d) & (e) \\
\end{array}
\end{align*}
\caption{\label{fig:03:3} Graphs contributing to the OPE coefficients,
  where $\otimes$ is the electromagnetic current, the wavy lines are
  photons, and the spiral lines are gluons; (a) is the lowest order
  graph, (b) is an $\mathcal{O}(\alpha)$ electromagnetic correction to
  the quark coefficient, (c) is an $\mathcal{O}(\alpha)$ contribution
  to the photon coefficient, (d) is an $\mathcal{O}(\alpha
  \alpha_s^2)$ correction to the quark coefficient, and (e) is an
  $\mathcal{O}(\alpha)$ contribution to the quark coefficient, but is
  one-photon-reducible, and with our prescription does not contribute
  to the coefficient functions $C_{2/L}$}.
\end{figure}

The structure functions $F_{2/L}$ can be computed using the QCD
improved parton model formula if $Q^2$ is large,
\begin{equation}
F_{2/L}(x,Q^2) =  \sum_{i\in \qlggset}
\int_{zy,x}  C_{2/L,i}(z, Q^2,\mub)\, f_{i}(y, \mub^2)\,,
\label{3.25}
\end{equation}
where we have introduced the notation
\begin{equation}
\int_{z_1\cdots z_n,x}=\int_0^1 \mathd z_1 \cdots \mathd z_n \,\delta(z_1\cdots z_n-x)\,,
\label{eq:conv}
\end{equation}
as the convolution of short distance coefficient functions
$C_{2/L,i}(x,Q^2,\mub)$ and PDFs $f_{i}(x,\mub^2)$, and the notation
$\in \qlggset $ is defined at the beginning of Sec.~\ref{sec:em}.  A
few sample graphs for $C_{2/L,i}(x,Q^2,\mub)$ are shown in
Fig.~\ref{fig:03:3} (a-d).
%
%
The graph in Fig.~\ref{fig:03:3}(e) with our prescription does not
contribute to the coefficient functions $C_{2/L,i}$, as it is
one-photon reducible.

\section{Photon distribution via a photon-probing process}\label{sec:probe}

In this section we will assume that we have at our disposal a BSM
(beyond standard model) photon probe that couples to SM particles only
through photon exchange.  We neglect weak interactions, and thus do
not worry about making this interaction ${\mathrm SU}(2)\times{\mathrm
  U}(1)$ invariant. The BSM photon probe combined with factorisation
will allow us to determine the photon PDF, which must be independent
of the probe we have chosen.  We verify this by performing the same
computation with a different probe in App.~\ref{app:gamgam}.

The BSM process we consider here involves two neutral spin-1/2
leptons, an incoming lepton $l$ and an outgoing lepton $L$ with masses
zero and $M \gg \mpr$, and momenta $k$ and $k^\prime=k-q$,
respectively, with a transition (i.e.\ flavour violating) magnetic
moment interaction
\begin{align}
\mathcal{L} &= c \frac{e(\mub) }{\Lambda} \overline L\, \sigma^{\mu \nu}  l \, F_{\mu \nu}+ \text{h.c.} \,.
\label{4.1}
\end{align}
where $\sigma^{\mu\nu} = \frac{i}{2}[\gamma^\mu, \gamma^\nu]$.
We formally assume the limit $\Lambda \to \infty$, i.e.\ we work to
lowest order in $1/\Lambda$. The QED Ward identity Eq.~(\ref{zeza})
implies that $c$ is independent of $\mu$ to all orders in
$\alpha$. With the interaction Eq.~(\ref{4.1}), the interaction vertex
is (not including the overall $i$)
\begin{align}
  \frac{ c\, e(\mub) }{\Lambda} \overline
  u(k-q)\left[\slashed{\epsilon},\slashed{q}\right] u(k)\,.
\label{4.5}
\end{align}

We define the spin-averaged leptonic tensor for the $l(k) \to
L(k^\prime)$ transition as
\begin{align}
L^{\mu \nu}(q,k) &= \frac12 \frac{c^2}{\Lambda^2} \text{Tr}[\slashed{k}\, \sigma^{\mu \alpha}q_\alpha \left(\slashed{k}^\prime+M\right)  \sigma^{\nu \beta}q_\beta]\,.
\label{eq:leptens}
\end{align}
Evaluating the traces gives
\begin{align}
L^{\mu \nu}(q,k) &= \frac{4c^2}{\Lambda^2} \biggl[ \left(M^2+Q^2\right) \left( q^\mu q^\nu  - M^2 g^{\mu \nu} \right) +4 Q^2 k^\mu k^\nu -2(M^2+Q^2) \left(k^\mu q^\nu+k^\nu q^\mu\right) \biggr]\,,
\label{4.6}
\end{align}
which satisfies the conditions $q_\mu L^{\mu \nu}=q_\nu L^{\mu
  \nu}=0$.

\begin{figure}
\begin{center}
\includegraphics[width=6cm]{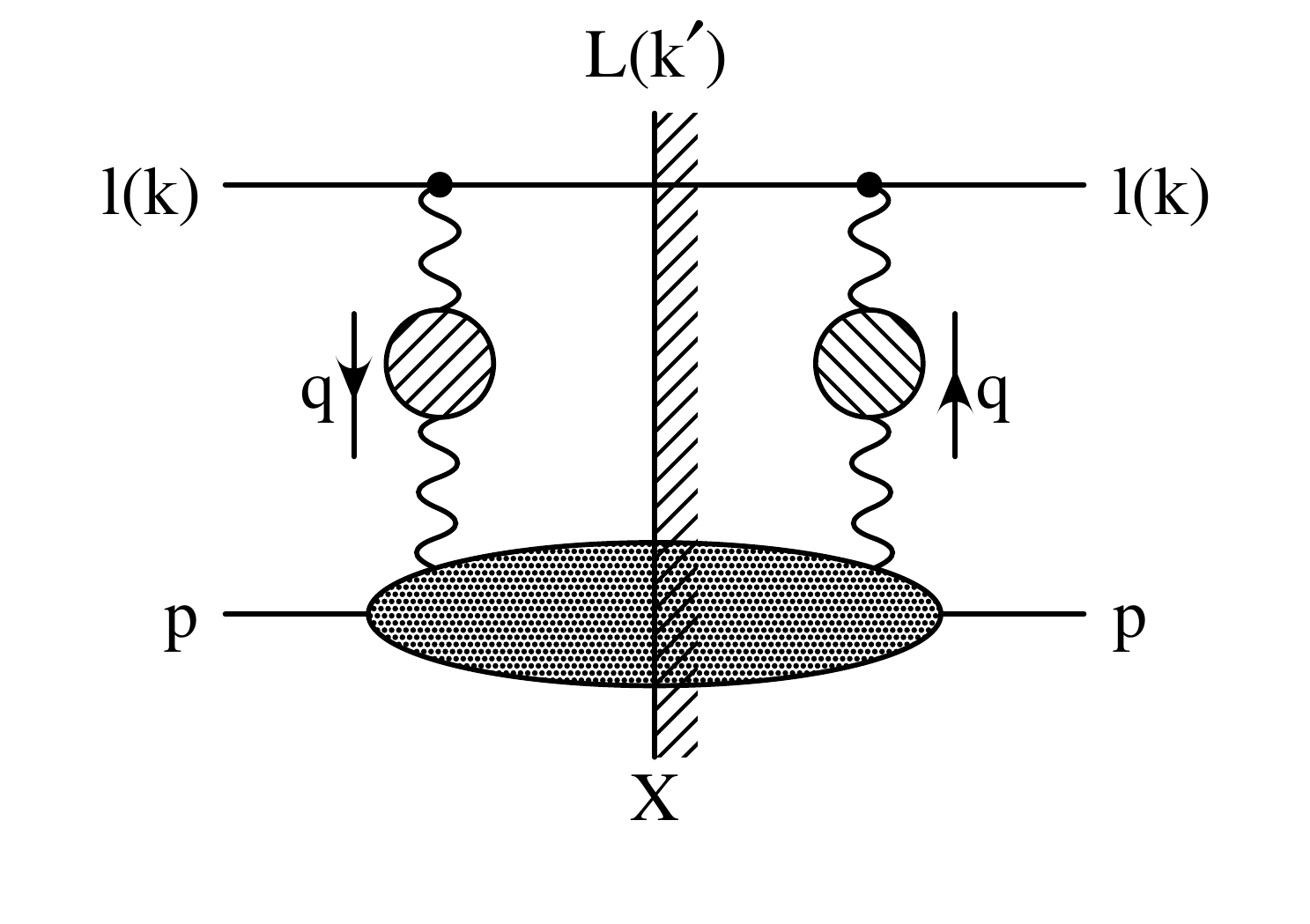}
\end{center}
\caption{\label{fig:04:2} The process $l + p \to L + X$ to leading
  order in $1/\Lambda$, but including QED corrections to all
  orders. The blob in the photon propagators represent the full vacuum
  polarisation correction. The lower hadronic blob also includes QED
  corrections.  }
\end{figure}

We now show how to express a photon initiated process in terms of
hadron structure functions. 
The cross section for the process $l + p \to L + X$, depicted in
Fig.~\ref{fig:04:2},
including all QED corrections, is
\begin{align}
& \sigmaprobe{p} = \frac{1}{4 p\cdot k} \int \frac{\mathd^4
    q}{(2\pi)^4} \frac{\eph^4(q^2)}{q^4} \ (4\pi) W^{\mu\nu}(q,p) \;
  L_{\mu\nu}(q,k) \ 2\pi \delta((k-q)^2-M^2)\theta(k^0-q^0)\nn
  &\hspace{4cm} \times \theta((p+q)^2-\mpr^2)\ \theta(p^0+q^0)\,.
\label{4.4}
\end{align}
This expression is exact.  The vertices on the lepton line in the
figure give the leptonic tensor $L_{\mu \nu}$ of Eq.~(\ref{4.6}), and
on the lower line give the hadronic tensor $4\pi W^{\mu \nu}$, defined
in Eq.~(\ref{eq:hadtens}).  There are vacuum polarisation corrections,
represented as dashed blobs, to each photon propagator connecting the
lepton and hadron sides of the graph, which give a factor of
$1/[1-\Pi(q^2,\mub)]^2$. In addition, there can be arbitrary QED
corrections included in the hadron part of the diagram (except for the
one-particle reducible ones) which are included in our definition of
$W^{\mu \nu}$. QED corrections on the leptonic line are suppressed by
powers of $\Lambda$, and disappear in our limit, as do corrections
associated with the exchange of more than one photon.

Eq.~(\ref{3.15}) has been used to convert the powers of $e^4(\mub)$
together with the factor $1/[1-\Pi(q^2,\mub)]^2$ into $\eph^4(q^2)$.
The $\delta$ and $\theta$ functions for the second line in
Eq.~\eqref{4.4} ensure that $L$ is on-shell, and has positive
energy. The $\theta$ functions on the third line of the same equation
ensure that $X$ has positive energy, and $m_X^2 \ge \mpr^2$, since the
proton is the lightest baryon.  The phase space is given by
\begin{align}
 & \int \frac{\mathd^4 q}{(2\pi)^4} 2\pi
  \delta((k-q)^2-M^2)\theta(k^0-q^0) \theta((p+q)^2-\mpr^2)
  \theta(p^0+q^0) \nn 
  =&\frac{1}{16 \pi^2 M^2}
  \int_\xph^{1-\frac{2 \xph\, \mpr}{M}} \mathd \zed
  \int_{\Qtwomin}^{\Qtwomax} Q^2 \mathd Q^2 \int_{-\pi}^{\pi}
  \frac{\mathd \phi}{2\pi}\,.
  \label{eq:phsp}
 \end{align}
Here $z$ and $x$ are defined as 
\begin{equation}
x=\frac{M^2}{2p \cdot k}\,,\qquad 
z = \frac{x}{\xbj}\,,
\end{equation}
where $\xbj= Q^2/(2 p\cdot q) $ is the usual Bjorken variable. 
The limits on the $Q^2$ integration are given by 
\begin{eqnarray}
  \Qtwomax &=&
  M^2\left(\frac{1-\zed}{\zed}\right)\frac{1-\frac{2\xph^2
      \mpr^2}{(1-\zed)M^2} +\sqrt{1-\frac{4 \xph^2 \mpr^2}{(1-\zed)^2
        M^2}}}{2\left(1+\frac{\xph^2 \mpr^2}{\zed \,M^2}\right)}\,,
  \nn \Qtwomin &=& \frac{\xph^2
    \mpr^2}{1-\zed}\frac{2}{1-\frac{2\xph^2 \mpr^2}{(1-\zed)M^2}
    +\sqrt{1-\frac{4 \xph^2 \mpr^2}{(1-\zed)^2 M^2}}}\,.
    \label{eq:QminQmax}
\end{eqnarray}
Expanding the limits in $\mpr^2/M^2$ gives
\begin{align}
  \Qtwomax &\to \Qtwomaxexp = \frac{ M^2(1-\zed) }{ \zed } \,, &
  \Qtwomin &\to \Qtwominexp = \frac{ \mpr^2 \xph^2 }{1-\zed }\,. 
\label{4.24}
\end{align}
Details of the phase space and kinematics computation are given in
App.~\ref{sec:kin}.

A straightforward calculation, combining the definition of the
hadronic and leptonic tensors Eqs.~(\ref{3.21}) and (\ref{4.6}),
yields
\begin{multline}
  \label{4.6a}
L^{\mu\nu}W_{\mu\nu} =\frac{4 M^4 c^2}{\zed \xph \Lambda^2}  \Bigg[\left(
    -\zed ^2
    -\frac{\zed ^2 Q^2}{2 M^2}
    +\frac{\zed ^2 Q^4}{2 M^4}
  \right)F_L(\xph/\zed, Q^2)
  \\
  +\biggl(
    2 
    -2 \zed 
    + \zed ^2
    + \frac{2 \xph ^2 \mpr^2}{Q^2}
    +\frac{\zed ^2 Q^2}{M^2}
    -\frac{2 \zed  Q^2}{M^2}
    -\frac{2 \xph ^2 Q^2 \mpr^2}{M^4}
  \biggr) F_2(\xph/\zed,Q^2)
  \Bigg]\,,
\end{multline}
that, together with Eq.~(\ref{4.4}) and the phase space Eq.~(\ref{eq:phsp}), gives
\begin{multline}
  \sigmaprobe{p} = 
  \frac{\sigma_0}{2 \pi \alpha(\mub)}   \int_{\xph}^{1-\frac{2 \xph\, \mpr}{M}} 
  \frac{d\zed}{\zed}
  \int^{\Qtwomax}_{\Qtwomin} 
  \frac{dQ^2}{Q^2} \aph^2(-Q^2)
\Bigg[\left(
    -\zed ^2
    -\frac{\zed ^2 Q^2}{2 M^2}
    +\frac{\zed ^2 Q^4}{2 M^4}
  \right)F_L(\xph/\zed,Q^2)
  \\
  +\biggl(
    2 
    -2 \zed 
    + \zed ^2
    + \frac{2 \xph ^2 \mpr^2}{Q^2}
    +\frac{\zed ^2 Q^2}{M^2}
    -\frac{2 \zed  Q^2}{M^2}
    -\frac{2 \xph ^2 Q^2 \mpr^2}{M^4}
  \biggr) F_2(\xph/\zed,Q^2)
  \Bigg]\,,
  \label{4.7}
\end{multline}
where
\begin{equation}
  \sigma_0 \equiv \frac{4 \pi c^2 e^2(\mub)}{\Lambda^2}\,.
 \label{eq:sigma0}
\end{equation}
The final cross section depends on $\xph$, which is fixed by the
external kinematic variables $M$ (the mass of $L$), and $2 p \cdot k =
s-\mpr^2$, where $\sqrt{s}$ is the centre-of-mass energy.

We stress again that Eq.~(\ref{4.7}) is exact as long as $\Lambda$
suppressed terms are neglected, $\aph$ is exact, and $F_{2/L}$ are
defined from the full one-particle irreducible hadronic tensor
\emph{including} QED corrections.

\begin{figure}
\begin{center}
\includegraphics[width=0.8\textwidth]{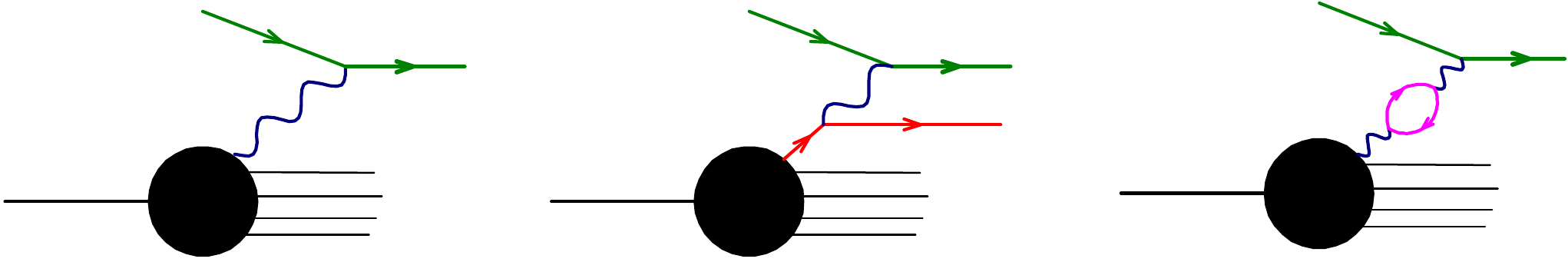}
\end{center}
\caption{\label{fig:04:0} Leading and next-to-leading graphs for the
  process $l + \gamma \to L $ in the QCD improved parton model.
}
\end{figure}

We now turn to the calculation of the cross section using the QCD
improved parton model formula
\begin{equation}
\sigmaprobe{p}(p) = \sum_{j \in \qlggset} \int \mathd y\,
\sigmahatprobe{j}(y p)\, f_{j}(y,\mub^2) \,,
\label{4.8}
\end{equation}
where $f_{j}$ are the PDFs of the proton.
The PDFs are in the $\MSbar$ scheme if the partonic hard scattering
cross sections $\widehat \sigma$ are computed in the $\MSbar$
scheme. Power corrections of the form $\mpr^2/M^2$ or $\mpr^2/(2 p
\cdot k)$ are neglected. Expanding this formula in the coupling
constants yields
\begin{equation}
  \sigmaprobe{p}(p) = \int \mathd y \, \sigmahatprobe{\gamma}^{(0,0)}
  (y p)\, f_{\gamma}(y, \mub^2) + \frac{\alpha(\mu^2)}{2\pi}\sum_{j
    \in \qlset} \int \mathd y\; \sigmahatprobe{j}^{(0,1)}(yp) \,
  f_{j}(y,\mub^2) +\cdots
\label{4.9}
\end{equation}
The diagrams corresponding to $\sigmahatprobe{\gamma}^{(0,0)}$ and
$\sigmahatprobe{j}^{(0,1)}$ are shown in Fig.~\ref{fig:04:0}.
The first diagram is the lowest order term, involving directly the
proton photon-density.  The second diagram is the next-to-leading
correction of relative order $\alpha$, involving the proton
charged-fermions densities.  The contribution corresponding to the
rightmost graph yields zero in the \MSbar{} scheme since we consider
massless fermions, since at this order in collinear factorisation the
photon is implicitly on shell and so the vacuum polarisation is
evaluated at zero virtuality. Thus the index $j$ in the sum is
restricted in practice to the charged fermions only.

At this point a comment is in order. We can systematically compute the
cross section assuming that $\alpha$ and $\alpha_s$ are of the same
size, and that the parton densities themselves are formally all of the
same order. We dub this counting of the order ``democratic", and adopt
it here in what follows, since it is more transparent.  In the
democratic order-counting, the index $i$ appearing in Eq.~(\ref{4.9})
should also run over leptons. Furthermore, neglected terms are of
second order in both $\alpha$ and $\alpha_s$, i.e. of order $\alpha^2$
and $\alpha\alpha_s$ (the $\alpha_s^2$ term being absent), relative to
the Born term.

For phenomenological applications, however, we will take into account
the fact that $\alpha$ is smaller than $\alpha_s$, using as a
guideline the relation $\alpha \approx \alpha_s^2$. We dub this
counting ``phenomenological''.  According to it, the photon density of
the proton is of order $\alpha L$ with respect to a quark density, $L$
being a log of $\mu^2$ over some typical hadronic scale. We can assume
$L\approx 1/\alpha_s$. In this framework the contributions
corresponding to the first and second diagram in Fig.~\ref{4.9} are
respectively of order $\alpha^2 L$, $\alpha^2$, while the last graph
is formally of order $\alpha^3 L\approx\alpha^2 \alpha_s $ (but is
zero in the \MSbar{} scheme).
The next-to-leading correction is of relative order $1/L \sim
\alpha_s$, rather than of order $\alpha$ (as in the democratic
counting), with respect to the Born term.
In the middle diagram of Fig.~\ref{fig:04:0} light leptons can be
excluded, since their PDF is of order $L^2 \alpha^2$, and their
contribution is of order $\alpha^4 L^2$.\footnote{Unless one considers
  the photon content of partially stripped
  ions~\cite{Krasny:2015ffb}.}

The cross section for the process $\sigma (l + q \to L + q)$,
illustrated in the middle graph of Fig.~\ref{fig:04:0}, is easily
computed with standard methods. Details of the calculation are given
in App.~\ref{sec:coll-q-e-to-L}. We get
\begin{eqnarray}
  \sigmahatprobe{\gamma}^{(0,0)}(yp) &=& \sigma_0
  M^2\delta(\shat-M^2)\,, \nn \\ \sigmahatprobe{i}^{(0,1)}(yp) &=&
  e_i^2\, \sigma_0 \frac{\alpha(\mu^2)}{2 \pi}\left[ -2+3z+zp_{\gamma
      q}(z)\left(\log\frac{M^2}{\mu^2}+\log\frac{(1-z)^2}{z}\right)\right]\,,
\label{eq:sigmahatprobe01}
\end{eqnarray}
where $\sigma_0$ is given in Eq.~\eqref{eq:sigma0}, $\shat=ys$,
$z=M^2/\shat=x/y$ and
\begin{align}
p_{\gamma q}(z)&\equiv \frac{1 + (1-z)^2}{z}\,.
\label{4.13}
\end{align}
This yields
\begin{align}
  \sigmaprobe{p}(p) = \sigma_0 \Biggl\{\, x f_{\gamma/p}(x,\mub^2) +
  \frac{\alpha(\mub)}{2 \pi} \int_{\xph}^1 \frac{d\zed}{\zed} \biggl[
    & \zed p_{\gamma q}(\zed) \left( \log \frac{M^2(1-\zed)^2 }{\zed
      \mub^2} \right)+3 \zed-2 \biggr] \times \nn
&   \sum_{i \in \qlset} e_i^2 \frac{\xph}{\zed} f_{i}\left(\frac{\xph}{\zed}, \mub^2\right)
   +{\cal O}(\alpha\as,\alpha^2) \Biggr\} 
   \,,
   \label{eq:sigmaprobeNLO}
\end{align}
where the index $i$ runs over fermions and antifermions (both quarks
and leptons, although the lepton contribution is beyond our accuracy).

We now go back to our exact expression Eq.~(\ref{4.7}). Neglecting
power suppressed $\mpr/M$ terms in the integration limits, it can be
written as
\begin{multline}
  \sigmaprobe{p}(p) = 
  \frac{\sigma_0 }{2 \pi \alpha(\mub) }  \int_{\xph}^1
  \frac{d\zed}{\zed}
  \int^{\Qtwomaxexp}_{\Qtwominexp}
  \frac{dQ^2}{Q^2} \aph^2(-Q^2)   \\
  \Bigg[
    -\zed ^2\, F_L(\xph/\zed,Q^2)
  +\biggl( \zed p_{\gamma q}(\zed) 
    + \frac{2 \xph ^2 \mpr^2}{Q^2}
  \biggr) F_2(\xph/\zed,Q^2)
  \Bigg]
\\
  +\frac{\sigma_0 }{2 \pi \alpha(\mub) }  \int_{\xph}^1
  \frac{d\zed}{\zed}
  \int^{\Qtwomaxexp}_{\Qtwominexp} 
  \frac{dQ^2}{Q^2} \aph^2(-Q^2)
\Bigg[\left(
    -\frac{\zed ^2 Q^2}{2 M^2}
    +\frac{\zed ^2 Q^4}{2 M^4}
  \right)F_L(\xph/\zed,Q^2) \\
  +\biggl(
    \frac{\zed ^2 Q^2}{M^2}
    -\frac{2 \zed  Q^2}{M^2}
    -\frac{2 \xph ^2 Q^2 \mpr^2}{M^4}
  \biggr) F_2(\xph/\zed,Q^2)
  \Bigg]\,, 
  \label{4.18}
\end{multline}
where we have separated the part of the integral which is sensitive to
the full accessible range of $Q^2$ values, including low $Q^2$, in the
first two lines, from the part that is dominated by high $Q^2$ in the
last two lines.  We now simplify this expression in order to match the
accuracy of the corresponding parton model formula
Eq.~(\ref{eq:sigmaprobeNLO}).

Notice that the $F_L$ term in the third line can be dropped, since
$F_L$ is of order $\alpha_s$ or $\alpha$, and would yield to a
contribution beyond our accuracy. This same argument cannot be applied
to the second line of Eq.~(\ref{4.18}), since in this case a
logarithmic integral in $Q^2$ compensates for the extra power of the
strong coupling.  In the last line, within the same accuracy, we can
replace $F_2(x/z,Q^2)$ and $\aph^2(-Q^2)$ with $F_2(x/z,\mub^2)$ and
$\alpha^2(\mub)$ where $\mu$ is a scale of order $M$.  Performing the
$Q^2$ integral and dropping terms suppressed by $m_p^2/M^2$, we obtain
\begin{multline}
  \sigmaprobe{p}(p) = \sigma_0 \Biggl\{
  \frac{1}{2 \pi \alpha(\mub) } \int_{\xph}^1
  \frac{d\zed}{\zed}
  \int^{\Qtwomaxexp}_{\Qtwominexp}
  \frac{dQ^2}{Q^2} \aph^2(-Q^2) \Bigg[
    -\zed ^2\, F_L(\xph/\zed,Q^2) \\
  + \biggl( \zed p_{\gamma q}(\zed)
    + \frac{2 \xph ^2 \mpr^2}{Q^2}
  \biggr) F_2(\xph/\zed,Q^2)
  \Bigg]
  +\frac{\alpha(\mub)}{2 \pi }   
  \int_{\xph}^1
  \frac{d\zed}{\zed}
  (z-2)(1-z)
  F_2(\xph/\zed,\mub^2) +{\cal O}(\alpha\as,\alpha^2)\Biggr\} \,.
  \label{4.18bis}
\end{multline}
We now define a ``physical'' photon PDF
\begin{multline}
  \xph f^{\text{PF}}_{\gamma}(\xph,\mub^2) \equiv \frac{1}{2 \pi
    \alpha(\mub)} \int_\xph^1 \frac{d\zed}{\zed}
  \int^{\frac{\mub^2}{1-\zed}}_{\Qtwominexp} \frac{dQ^2}{Q^2}
  \aph^2(-Q^2) \\ \times \Bigg[ -\zed ^2\, F_L(\xph/\zed,Q^2) +\biggl(
    \zed p_{\gamma q}(\zed) + \frac{2 \xph ^2 \mpr^2}{Q^2} \biggr)
    F_2(\xph/\zed,Q^2) \Bigg]\,.
  \label{4.16}
\end{multline}
The reason for the upper limit on the integration will become
clear later.  
The terminology ``physical'' is associated with the fact that when we
consider the transverse-momentum dependent photon PDF in
Sec.~\ref{sec:tmdpdf}, the latter's integral up to $k_\perp^2 \le
\mu^2$ will coincide with Eq.~(\ref{4.16}).
As we will see below, Eq.~(\ref{4.16}) includes the $\alpha L$
contribution to the photon PDF, but not the total $\alpha$ piece.

Combining Eq.~(\ref{4.16}) with Eq.~(\ref{4.18bis}), we obtain
\begin{multline}
  \sigmaprobe{p}(p) = \sigma_0 \Biggl\{ \xph
  f^{\text{PF}}_{\gamma}(\xph,\mub^2) + \frac{1}{2 \pi \alpha(\mub) }
  \int_{\xph}^1 \frac{d\zed}{\zed} \int^{\Qtwomaxexp}_{\mu^2/(1-z)}
  \frac{dQ^2}{Q^2} \aph^2(-Q^2) \Bigg[ -\zed ^2\, F_L(\xph/\zed,Q^2)
    \\ +\biggl( \zed p_{\gamma q}(\zed) + \frac{2 \xph ^2 \mpr^2}{Q^2}
    \biggr) F_2(\xph/\zed,Q^2) \Bigg] +\frac{\alpha(\mub)}{2 \pi }
  \int_{\xph}^1 \frac{d\zed}{\zed} (z-2)(1-z) F_2(\xph/\zed,\mub^2)
  +{\cal O}(\alpha\as,\alpha^2)\Biggr\} \,.
  \label{4.18tris}
\end{multline}
Since the remaining $Q^2$ integral is now dominated by large $Q^2$
values, we can evaluate it with the same approximations used earlier,
and get
\begin{multline}
  \sigmaprobe{p}(p) = \sigma_0 \Biggl\{  \xph
  f^{\text{PF}}_{\gamma}(\xph,\mub^2) 
   +  \frac{\alpha(\mub)}{2 \pi }
  \int_{\xph}^1
  \frac{d\zed}{\zed}
  \\
  \left[ (z-2)(1-z) + \zed p_{\gamma q}(\zed) \log\left(\frac{M^2(1-z)^2}{\mub^2 z}\right)
\right]
  F_2(\xph/\zed,\mub^2) +{\cal O}(\alpha\as,\alpha^2) \Biggr\} \,.
  \label{4.18x}
\end{multline}

Now consider the parton model formula for the same cross section,
Eq.~(\ref{eq:sigmaprobeNLO}).  Using the lowest order expression in
$\alpha_s(\mub),\alpha(\mub)$
\begin{align}
F_2(\xbj,\mub^2) &= \sum_i e_i^2 \, \xbj\, f_{i}(\xbj,\mub^2)\,,
\label{4.20}
\end{align} 
and comparing Eq.~(\ref{4.18x}) with Eq.~(\ref{eq:sigmaprobeNLO}), we
get
\begin{align}
 \xph f_{\gamma}(\xph,\mub^2) &=  \xph f^{\text{PF}}_{\gamma}(\xph,\mub) + \frac{\alpha(\mub)}{2 \pi}  \int_{\xph}^1
  \frac{d\zed}{\zed}
 \left( -\zed^2 \right) F_2(\xph/\zed,\mub^2) +{\cal O}(\alpha\as,\alpha^2)\,. 
\label{4.22}
\end{align}
The l.h.s.\ is the desired result, the photon PDF in the $\MSbar$
scheme. The r.h.s.\ expresses it as an integral over lepton-proton
scattering structure functions. The first term is given in
Eq.~(\ref{4.16}), and we refer to the second term as the
``$\MSbar$-conversion term''. The upper integration limit in
Eq.~(\ref{4.16}) was chosen to be $\mub^2/(1-\zed)$ so that the
logarithms cancelled between Eq.~(\ref{4.18x}) and
Eq.~(\ref{eq:sigmaprobeNLO}). Writing the first term explicitly gives
\begin{multline}
  \xph f_{\gamma}(\xph,\mub^2) = \frac{1}{2 \pi \alpha(\mub)} \int_\xph^1
  \frac{d\zed}{\zed}
  \biggl\{  \int^{\frac{\mub^2}{1-\zed}}_{\Qtwominexp} 
  \frac{dQ^2}{Q^2} \aph^2(-Q^2)
\Bigg[
    -\zed ^2\,
  F_L(\xph/\zed,Q^2) \\
  +\biggl( \zed p_{\gamma q}(\zed)
    + \frac{2 \xph ^2 \mpr^2}{Q^2}
  \biggr) F_2(\xph/\zed,Q^2)
  \Bigg] 
 - \alpha^2(\mub)\, \zed^2  F_2(\xph/\zed,\mub^2)\biggr\}
  + \mathcal{O}(\alpha\alpha_s,\alpha^2)\,.
  \label{eq:master}
\end{multline}
Formula Eq.~\eqref{eq:master} is independent of the particular probe
process that we have chosen. In App.~\ref{app:gamgam} we verify this
by carrying out the same derivation using as a probe the production of
a scalar particle via photon-photon fusion and using PDF operators in
Sec.\ref{sec:alt-deriv}.
A more direct method of obtaining Eq.~\eqref{eq:master} is illustrated
in Sec.~\ref{sec:alt-deriv}, and will be used in
Sec.~\ref{sec:higher-order} to extend its accuracy to higher orders.

Examining Eq.~(\ref{eq:master}) according to our ``phenomenological''
counting, the dominant terms under the $Q^2$ integration are of order
$\alpha L$, and the \MSbar-conversion term is of order $\alpha$. For
the contribution of terms under the $Q^2$ integration, we should be
careful to include terms of relative order $\alpha L$ in both $\aph$
and $F_2$.  These, together with the $L$ coming from the logarithmic
integration, yield corrections of order $\alpha^2 L^2\approx \alpha
\as^2 L^2 \approx \alpha$.  Thus $\aph(-Q^2)$ can be replaced with
$\alpha(Q^2)$, since they differ by terms of relative higher order in
$\alpha$ without powers of $L$ enhancement.  We plan to extract
$F_{2/L}$ at low and moderate $Q^2$ directly from data, while for high
$Q^2$ we will compute it using available PDF parametrisations.  Thus
the discussion of its accuracy is more delicate, and we postpone it to
Sec.~\ref{sec:input-data}.

\section{Connection with splitting functions}
\label{sec:split-fns}

From our formula for $f_\gamma$, Eq.~(\ref{eq:master}), we can derive
formul\ae\ for the splitting functions of the photon. To this end we
will adopt the ``democratic'' counting, taking Eq.~(\ref{eq:master})
at full leading order accuracy in $\alpha$ and $\alpha_s$, and
treating the two couplings as if they were of the same order.
Neglected terms are therefore of order $\alpha^2$ and
$\alpha\alpha_s$, while $\alpha_s^2$ is obviously absent.

In order to simplify our notation we will use the following
conventions.  When we write the PDFs or the \MSbar{} couplings without
a scale argument, we imply that the scale is $\mu^2$.  Furthermore we
will adopt the normalisation convention
\begin{align}
\mub^2 \frac{\mathd }{\mathd \mub^2} f_{a}(x) = \sum_b \int_{zy,x} P_{ab}(z) f_b(y)\,,\label{7.8}
\end{align}
and the convolution is defined in Eq.~(\ref{eq:conv}).

The DIS coefficient functions have the expansion
\begin{align}
  C_{2/L,i} &= \sum_{r,s} \left( \frac{\alpha_s}{2\pi}\right)^r \left(
  \frac{\alpha}{2\pi}\right)^s C^{(r,s)}_{2/L,i} \,.
\label{7.8b}
\end{align}
where
\begin{equation}
  F_{2/L}(x) = \sum_{i\in\qlggset} x \int_{yz,x} 
  C_{2/L,i}(z) f_{i}(y)\,.
\end{equation}
We also define
\begin{align}
  \aph (- Q^2) &= \alpha (\mu^2) \sum_{r,s} \left(
  \frac{\alpha_s}{2\pi}\right)^r \left( \frac{\alpha}{2\pi}\right)^s
  c_{\tmop{ph}}^{(r,s)}(Q^2,\mu^2)\,,
\end{align}
with
\begin{align}
    c_{\tmop{ph}}^{(0,0)}(Q^2,\mu^2)&= 1\,, \nn
    c_{\tmop{ph}}^{(1,0)}(Q^2,\mu^2)&= 0\,, \nn
    c_{\tmop{ph}}^{(0,1)}(Q^2,\mu^2)
    &=\left(\frac{1}{2}\sum_{i\in\qset} e_i^2\right)
    \left(-\frac{10}{9}+\frac{2}{3} \log\frac{Q^2}{\mu^2}\right).
\end{align}

The factor of $1/2$ inside the parenthesis is due to the fact that we
sum over all charged fermions \emph{and} antifermions.  In the
following we will also use the notation
\begin{equation}
  c_{\tmop{ph}}^{(0,1)}\equiv c_{\tmop{ph}}^{(0,1)}(\mu^2,\mu^2) \,.
\end{equation}
The RG equation for the QED coupling is
\begin{align}
\mub^2 \frac{\mathd \alpha}{\mathd \mub^2} &= - \frac12\alpha
\sum_{r,s} \left( \frac{\alpha_s}{2\pi}\right)^r \left(
\frac{\alpha}{2\pi}\right)^s \bqed^{(r,s)} \,.
\label{7.11}
\end{align}
We have
\begin{align}
b_{\tmop{qed}}^{(0, 1)} &= - \frac{4}{3} \left(\frac{1}{2}
\sum_{i\in\qlset} n_i e_i^2\right), & b_{\tmop{qed}}^{(0, 2)} &= -
2\left(\frac{1}{2} \sum_{i\in\qlset} n_i e_i^4\right), &
b_{\tmop{qed}}^{(1,1)}&=-2C_F\left(\frac{1}{2} \sum_{i\in\qset}
e_i^2\right)\,,
\end{align}
where $n_i$ is 3 for quarks and 1 for leptons.
Furthermore we define
\begin{align}
  p_{qq}(x) &\equiv C_F \left(\frac{1+x^2}{1-x}\right)_+\,,
  & p^{\rm qed}_{qq}(x) &\equiv \left(\frac{1+x^2}{1-x}\right)_+\,,\nn 
  p_{qg}(x) &\equiv T_F \left(x^2+(1-x)^2 \right)\,,
  &p_{q\gamma}(x) &\equiv x^2+(1-x)^2 \,.
  \end{align}
We now begin by re-writing Eq.~(\ref{eq:master}) with a suitable
change in the upper limit of integration and dropping power-suppressed
terms, in order to ease the derivation of the evolution equation:
\begin{eqnarray}
  &  & x f_{\gamma } (x,\mub^2) =\nonumber \\
  &  & \frac{1}{2 \pi \alpha(\mub^2)} \int_{z y, x} \Biggl\{
  \int_{Q^2_{\min}}^{\mu^2} \frac{\mathd Q^2}{Q^2} \aph^2(-Q^2)
  \left[ - z^2 F_L (y, Q^2) + \left( zp_{\gamma q}(z) + \frac{2 x^2 m_p^2}{Q^2}
  \right) F_2 (y, Q^2) \right] \nonumber \\
  &  &  + \int_{\mu^2}^{\frac{\mu^2}{1 - z}} \frac{\mathd Q^2}{Q^2}
  \aph^2(-Q^2)  zp_{\gamma q}(z) F_2 (y, Q^2) - \alpha^2(\mub^2) z^2
  F_2 (y,\mub^2) \Biggr\} + \mathcal{O} (\alpha \as, \alpha^2)\nonumber \\
  & = & \frac{1}{2 \pi \alpha(\mub^2) } \int_{z y, x} \Biggl\{
  \int_{Q^2_{\min}}^{\mu^2} \frac{\mathd Q^2}{Q^2} \aph^2(-Q^2)
  \left[ - z^2 F_L (y, Q^2) + \left(zp_{\gamma q}(z) + \frac{2 x^2 m_p^2}{Q^2}
  \right) F_2 (y, Q^2) \right] \nonumber \\
  &  & \nobracket - \alpha^2(\mub^2)  [z^2 + \log (1 - z) z p_{\gamma q} (z)]
  F_2 (y,\mub^2) \Biggr\} + \mathcal{O} (\alpha \as, \alpha^2)\,.
\end{eqnarray}
To find the evolution equation, we compute
\begin{eqnarray}
  &  & \frac{x}{\alpha} \frac{\mathd \left(\alpha  f_{\gamma} (x)\right)}{\mathd \log \mu^2} =
  \int_{z y, x} \Biggl\{ \frac{\alpha^2_{\tmop{ph}} (- \mu^2)}{2 \pi
  \alpha } [- z^2 F_L (y) + z p_{\gamma q} (z) F_2 (y)]
  \nonumber \\
  &  &  - (z^2 + \log (1 - z) z p_{\gamma q} (z)) \left[ \frac{1}{\pi} 
  \frac{\mathd \alpha }{\mathd \log \mu^2} F_2 (y) +
  \frac{\alpha}{2 \pi} \frac{\mathd F_2 (y)}{\mathd \log \mu^2}
  \right] \Biggr\}\nonumber  + \mathcal{O} ( \alpha \alpha_s^2,\alpha^2 \alpha_s,\alpha^3)\,,
\end{eqnarray}
where we have neglected terms suppressed by powers of $m_p^2/\mu^2$
and, for ease of notation, we have omitted the $\mub$-dependence in
the coupling and parton densities.  In the first line we must use for
$\aph(-\mu^2)$ and $F_{2/L}$ expressions that are accurate at order
$\alpha_s$ and $\alpha$:
\begin{eqnarray}
  \aph(-\mu^2) &=& \alpha  \left(1+\frac{\alpha}{2\pi}
  c_{\rm ph}^{(0,1)} \right)\,, \\
  F_L (y) & = & y \int_{v w, y} \Biggl\{
  \frac{\alpha_s}{2 \pi} \sum_{i\in\qset}C_{L, i}^{(1, 0)} (v) f_{i}(w) + \frac{\alpha}{2 \pi}
  \sum_{i\in \qlset}C^{(0, 1)}_{L, i} (v) f_{i}(w)  \nonumber \\
  &+& \frac{\alpha_s}{2 \pi} C_{L,g}^{(1, 0)} (v) f_{g} (w)
  + \frac{\alpha}{2 \pi} C_{L, \gamma}^{(0,1)} (v) f_{\gamma} (w) \Biggr\} , \\
  F_2 (y) & = & y  \int_{v w, y}  \Biggl\{ \sum_{i\in\qlset} e_i^2 f_{i} (w) \delta (1 -
  v) 
  +  \frac{\alpha_s}{2 \pi}\sum_{i\in\qset} C_{2, i}^{(1, 0)} (v) f_{i}(w)  \nonumber \\
  &+&  \frac{\alpha}{2 \pi} \sum_{i\in\qlset} C^{(0, 1)}_{2, i} (v)  f_{i} (w) 
  +   \frac{\alpha_s}{2 \pi} C_{2, g}^{(1, 0)} (v) f_{g} (w) + \frac{\alpha
    }{2 \pi} C_{2, \gamma}^{(0, 1)} (v) f_{\gamma} (w) \Biggr\} \,, 
\end{eqnarray}
where the coefficient functions are given in App.~\ref{subsec:coeff}.
In the second line, the derivatives of $F_2$ and $\alpha$ are only
needed to leading order.  We have
\begin{eqnarray}
  \frac{\mathd F_2 (y)}{\mathd \log \mu^2} & = & y\, \int_{v
  w, y}  \Bigl\{ \sum_{i\in\qset} e_i^2  \frac{\alpha_s}{2 \pi} [p_{q q} (v) f_i (w) + p_{q
  g} (v) f_g (w)] \nonumber \\
  &  & +  \sum_{i\in\qlset} e_i^4 \frac{\alpha}{2 \pi} [p_{q q}^{\tmop{qed}} (v) f_i (w)
    + n_i p_{q \gamma} (v) f_{\gamma} (w)] \Bigr\}, \\
  \frac{\mathd \alpha}{\mathd \log \mu^2}&=&-\frac{1}{2} b_{\rm qed}^{(0,1)} \alpha
  \frac{\alpha}{2\pi}\,.
\end{eqnarray}
Inserting these expressions we get
\begin{eqnarray}
  &  & \frac{\mathd f_{\gamma} (x)}{\mathd \log \mu^2} = - \frac{1}{2}
  \left[ - b_{\tmop{qed}}^{(0, 1)}  \frac{\alpha }{2 \pi} -
  b_{\tmop{qed}}^{(1, 1)} \frac{\alpha  \alpha_s }{(2 \pi)^2} -
  b_{\tmop{qed}}^{(0, 2)} \left(  \frac{\alpha }{2 \pi} \right)^2 \right]
  f_{\gamma} (x) + \nonumber \\
  & + & \int_{z v w, x} \frac{\alpha}{2 \pi}
  \Biggl\{ - z \bigg[  \frac{\alpha_s}{2 \pi}
  \sum_{i\in\qset}C_{L, i}^{(1, 0)} (v) f_i(w) + \frac{\alpha}{2 \pi}\sum_{i\in\qlset}  C^{(0, 1)}_{L, i} (v)
  f_i (w) + \frac{\alpha_s}{2 \pi} C_{L, g}^{(1, 0)} (v) f_g (w)
   \nonumber \\
  & + &  \frac{\alpha}{2 \pi}  C_{L, \gamma}^{(0, 1)} (v) 
  f_\gamma(w) \bigg] 
   +  p_{\gamma q} (z) \bigg[ \left(1+2\frac{\alpha}{2\pi}c_{\rm ph}^{(0,1)}\right)
    \sum_{i\in\qlset} e_i^2 f_i (w) \delta (1 - v) \nonumber \\
    &+& 
  \frac{\alpha_s}{2 \pi} \sum_{i\in\qset}C_{2, i}^{(1, 0)} (v) f_i(w) + \frac{\alpha}{2 \pi}
  \sum_{i\in\qlset} C^{(0, 1)}_{2, i} (v) f_i (w) + \frac{\alpha_s}{2 \pi} C_{2,
  g}^{(1, 0)} (v) f_g (w) \nonumber \\
  &  &  + \frac{\alpha}{2 \pi} C_{2, \gamma}^{(0, 1)} (v) f_\gamma (w) \bigg]
   -  (z + \log (1 - z)
  p_{\gamma q} (z)) \times \biggl[ - \frac{\alpha }{2 \pi}
  b_{\tmop{qed}}^{(0, 1)} \sum_{i\in\qlset} e_i^2 f_i (w) \delta (1 - v)  \nonumber\\
  & + &  \frac{\alpha_s }{2 \pi} \sum_{i\in\qset} e_i^2  (p_{q q} (v)
    f_i (w) + p_{q g} (v) f_g (w)) \nonumber \\
    &+&\frac{\alpha }{2 \pi} \sum_{i\in\qlset} e_i^4
  (p_{q q}^{\tmop{qed}} (v) f_i (w) + n_i p_{q \gamma} (v)
  f_{\gamma} (w)) \bigg) \biggr] \Biggr\} 
        \;+\; \order{\as\alpha^2, \as^2\alpha, \alpha^3}
        \,. \label{eq:evolutionFgamma}
\end{eqnarray}
We can now read off from Eq.~(\ref{eq:evolutionFgamma}) the splitting function
coefficients, 
\begin{align}
&P^{(0,1)}_{\gamma \gamma}(z)  =   \frac12 \bqed^{(0,1)} \delta(1-\zed) \,, \label{eq:pgaga01}\\  
&P^{(0,1)}_{\gamma i}(z)  =   e_i^2 p_{\gamma q}(\zed) \,,  \label{eq:pgai01} \\
&P^{(1,1)}_{\gamma \gamma}(z)  =   \frac12 \bqed^{(1,1)}   \delta(1-\zed) \,, \label{eq:pgaga11} \\
&P^{(0,2)}_{\gamma \gamma}(z) =\frac{1}{2} b^{(0,2)}_{\tmop{qed}}\delta(1-z) 
 + \int_{wy,z} \biggl[ -w C_{L,\gamma}^{(0,1)}(y) + p_{\gamma q}(w) C_{2,\gamma}^{(0,1)}(y) \nn &
\quad\quad\quad  -\bigl(w+\log(1-w)p_{\gamma q}(w)\bigr)\sum_i n_i e_i^4  p_{q\gamma}(y) \biggr] \,, \label{eq:pgaga02} \\
&P^{(1,1)}_{\gamma q}(z)  =\!\!  \int_{wy,z}\biggl[ p_{\gamma q}(w) C_{2,q}^{(1,0)}(y)   - (w+ \log(1-w) p_{\gamma q}(w)) e_q^2  p_{qq}(y) 
-w C_{L,q}^{(1,0)}(y) \biggr] \,, \label{eq:pgaq01} \\
&P^{(0,2)}_{\gamma i}(z)  =   e_i^2 p_{\gamma q}(z) 2c_{\rm ph}^{(0,1)}+ (z+\log(1-z)\, p_{\gamma q}(z)) e_i^2 b_{\rm qed}^{(0,1)}
\nonumber \\
&+  \int_{wy,z}\biggl[ p_{\gamma q}(w) C_{2,i}^{(0,1)}(y)   - (w+\log(1-w) p_{\gamma q}(w))  e_i^4  p_{qq}^{\rm qed}(y) 
-w C_{L,i}^{(0,1)}(y) \biggr] \,, \quad i \in \qlset \label{eq:pgai02} \\
&P^{(1,1)}_{\gamma g}(z)  = \int_{wy,z} \biggl[ p_{\gamma q}(w) C_{2,g}^{(1,0)}(y)   - (w+\log(1-w) p_{\gamma q}(w))  \sum_{i\in\qset} e_i^2 p_{qg}(y) 
-w C_{L,g}^{(1,0)}(y) \biggr] \,. \label{eq:pgag11}
\end{align}
The order $\alpha$ kernels $P^{(0,1)}$ agree with the known
expressions given in Ref.~\cite{Altarelli:1977zs}. Notice that
$P^{(0,1)}_{\gamma \gamma}$ has the correct sign, because our photon
PDF Eq.~(\ref{eq:master}) is proportional to $1/\alpha(\mub)$, not
$\alpha(\mub)$.  The order $\alpha_s$ and $\alpha$ coefficient
functions for $F_{2/L}$ are summarised in App.~\ref{subsec:coeff}.
Evaluating the convolutions for Eqs.~(\ref{eq:pgaq01})
and~(\ref{eq:pgag11}), we get full agreement with the recent
calculation in Ref.~\cite{deFlorian:2015ujt}, Eqs.~(35, 36) and (30).
Similarly, evaluating the integrals of the second order QED splitting
functions, Eqs.~(\ref{eq:pgaga02}) and (\ref{eq:pgai02}), we find full
agreement with formulae (3.9) and (3.21) of
Ref.~\cite{deFlorian:2016gvk}.
It is remarkable that our expression for the photon PDF gives the
DGLAP evolution kernel to one higher order in the couplings than was
present in the input coefficient functions.
Specifically we obtained the (two-loop) order $\alpha \alpha_s$ and
$\alpha^2$ $P_{\gamma i}$ splitting kernels using coefficient
functions at (one-loop) order $\alpha_s$ and $\alpha$.

\section{Spin dependent case}\label{sec:spin-dep}

There is an extensive experimental and theoretical effort to
understand the polarised gluon distribution in the proton, and $\Delta
g$, the gluon contribution to the proton
spin~\cite{Carlitz:1988ab,Altarelli:1988nr,Jaffe:1989jz,Aidala:2012mv}. $\Delta
\gamma$ is the photon analogue of $\Delta g$. Since photons and gluons
both couple to quarks via a $\gamma^\mu$ interaction, $\Delta \gamma$
could shed some light on $\Delta g$.

The results of Sec.~\ref{sec:probe} can be readily generalised to the
spin-dependent case, to obtain the polarised photon PDF of a proton of
helicity $H$, the difference between the probabilities to find photons
with spin parallel and anti-parallel to the proton spin in a
longitudinally polarised proton target.

The derivation here of the polarised photon PDF follows that for the
unpolarised PDF, using the polarisation asymmetry in the cross
sections instead of the spin-averaged cross sections.  We start with
the same interaction Lagrangian as before, Eq.~(\ref{5.1}). The
difference between the cross sections for right-handed and left-handed
leptons to scatter off a longitudinally polarised proton with helicity
$H$ to order $\alpha \sigma_0$ is, using Eq.~(\ref{3.21}) for $W_{\mu
  \nu}$,
\begin{multline}
 \frac12 \left(\sigma_{l_R p_H} - \sigma_{l_L p_H} \right) 
 =\frac{1}{2\pi \alpha(\mub^2) }  \sigma_0
   \int 
  \frac{\mathd\zed}{\zed}
  \int^{Q_\text{max}^2}_{Q_\text{min}^2} 
  \frac{\mathd Q^2}{Q^2}  \aph^2(-Q^2) \\
\biggl\{  H\biggl(4-2 \zed -\frac{4 m_p^2 \xph^2}{Q^2} - \frac{4 m_p^2 \xph^2 Q^2}{M^4} - \frac{8 m_p^2 \xph^2}{M^2} - \frac{2 \zed   Q^2}{M^2} \biggr) \xph g_1(\xph/\zed,Q^2) \\
  -H \left( \frac{8 m_p^2 \xph^2}{\zed M^2} + \frac{8 m_p^2 \xph^2}{\zed Q^2} \right) \xph  g_2(\xph/\zed,Q^2)  \biggr\}\,.
\label{5.1}
\end{multline}

To obtain the photon polarised parton density $f_{\Delta \gamma}$, we
use the factorisation formula
\begin{align}
\sigma_{l_h p_H}(p)
&= \sum_{i_s} \int \mathd x\; \widehat \sigma_{l_h i_s}(xp)  f_{i_s/p_H}(x,\mu^2)\,, 
\label{eq:fact_form_pol}
\end{align}
for the scattering of a lepton $l$ with helicity $h$ off a target $p$
with helicity $H$. The sum on $i_s$ is over all partons and their
helicities. We use the notation $f_{\Delta q}=f_{q_R/p_R}-f_{q_L/p_R}$
for the polarised parton distributions. Note that
$f_{q_R/p_R}=f_{q_L/p_L}$ and $\sigma_{l_R p_R}=\sigma_{l_L p_L}$,
etc.\ by parity invariance.
Since $f_{\Delta \gamma}$ is of order $\alpha$, we need the photon
hard-scattering cross section to lowest order,
\begin{align}
& \frac12 \left[ \widehat \sigma_{l_R \gamma_R}- \widehat \sigma_{l_L
      \gamma_R}\right] = \sigma_0\, M^2 \delta(s-M^2)\,,
\label{eq:sigma_pol_LO}
\end{align}
where $\sigma_0$ is given in Eq.~(\ref{eq:sigma0}).  Observe that the
total (spin averaged) cross section is given by
\begin{align}
& \frac12 \left[ \widehat \sigma_{l_R \gamma_R}+ \widehat \sigma_{l_L
      \gamma_R}\right] = \sigma_0\, M^2 \delta(s-M^2)\,,
\end{align}
and this implies, together with Eq.~(\ref{eq:sigma_pol_LO}), that
$\widehat \sigma_{l_L \gamma_R}=0$.  This result is easily
understood. We can look at the cross section in the centre-of-mass
frame where we collide a right moving left-handed lepton onto a left
moving right-handed photon, forming a heavy lepton at rest. By angular
momentum conservation, the heavy lepton must have $J_z=3/2$ along the
collision axis, which is not possible for a spin-1/2 particle.  This
simple argument confirms the correctness of the sign in
Eq.~(\ref{eq:sigma_pol_LO}).

The polarisation asymmetry to order $\alpha \sigma_0$ is
\begin{align}
& \frac12 \left[ \sigma_{l_R q_R}- \sigma_{l_L q_R}\right] =
  \frac{\alpha(\mub) \sigma_0}{2\pi}\left\{ \xph
  (2-\xph)\left[-\frac{1}{\epsilon}+\ln
    \frac{M^2(1-\xph)^2}{\xph\mub^2}\right]-3\xph(1-\xph) \right\}\,,
\label{5.3}
\end{align}
where the $1/\epsilon$ term is an infrared divergence, and we have
used the 't~Hooft-Veltman scheme for $\gamma_5$.
Combining with Eqs.~\eqref{eq:fact_form_pol} and
\eqref{eq:sigma_pol_LO} and performing the $\MSbar$ subtraction we get
\begin{eqnarray}
  \frac12 \left[  \sigma_{l_R p_H}- \sigma_{l_L p_H}\right] &=& \sigma_0 H
  \Biggl\{\, x f_{\Delta\gamma}(x,\mub^2) +
   \frac{\alpha(\mub)}{2 \pi}  \int_{\xph}^1
   \frac{d\zed}{\zed}
   \biggl[ 
     z(2-z) \left( \log \frac{M^2(1-\zed)^2 }{\zed \mub^2} \right)\nonumber \\
    && \phantom{aaa} - 3 \zed(1-\zed) \biggr] \times
   \sum_i e_i^2 \frac{\xph}{\zed} f_{\Delta i}\left(\frac{\xph}{\zed}, \mub^2\right)
   +{\cal O}(\alpha\as,\alpha^2) \Biggr\} 
   \,. \label{eq:sigmapolprobeNLO}
\end{eqnarray}

We now have all the ingredients necessary to determine the polarised
photon PDF. We follow the derivation for the unpolarised case given in
Sec.~\ref{sec:probe}. Define, as before, the polarised PDF in a
``physical factorisation'' scheme
\begin{multline}
f^{\text{PF}}_{\Delta \gamma}(\xph,\mub^2)
 =\frac{1}{2\pi \alpha(\mub^2) } 
   \int 
  \frac{\mathd \zed}{\zed}
  \int^{\frac{\mub^2}{1-\zed}}_{Q_\text{min}^2} 
  \frac{\mathd Q^2}{Q^2}  \aph^2(-Q^2) \\
 \biggl\{ \left(4-2 \zed - \frac{4 m_p^2 \xph^2}{Q^2} \right)  g_1(\xph/\zed,Q^2) 
- \left( \frac{8 m_p^2 \xph^2}{\zed Q^2} \right)   g_2(\xph/\zed,Q^2)  \biggr\}\,,
\label{5.10ph}
\end{multline}
by dropping the $1/M^2$ suppressed terms in Eq.~(\ref{5.1}).

As in Sec.~\ref{sec:probe}, the integral in Eq.~(\ref{5.1}) can be
written as the integral for the terms included in Eq.~(\ref{5.10ph}),
divided into the range $Q_\text{min}^2 \to \mub^2/(1-\zed)$ and
$\mub^2/(1-\zed) \to Q_\text{max}^2$, plus the integral over the
remaining terms.  The remaining integrals only get contributions from
$Q^2$ of order $M^2$, since $\mub \sim M$, and so contain no large
logarithms. Since $Q^2$ is large over the integration region, we can
replace $g_1(\xph/\zed,Q^2)$ by $g_1(\xph/\zed,\mub^2)$ and
$\aph^2(-Q^2)$ by $\alpha(\mub^2)$ up to corrections of order
$\alpha_s(\mub^2)$ and $\alpha(\mub^2)$. The integral is now trivial,
and gives
\begin{multline}
 \frac12 \left[ \sigma_{l_R p_H} - \sigma_{l_L p_H} \right] = H
 \sigma_0 \xph f^{\text{PF}}_{\Delta \gamma}(\xph,\mub^2) +
 \frac{\alpha(\mub^2)}{2\pi } H \sigma_0 \int \frac{\mathd\zed}{\zed}
 \biggl\{ \\ +2(2-\zed) \log \frac{M^2(1-z)^2}{\mub^2 z} - 2(1 - \zed)
 \biggr\} \xph g_1(\xph/\zed,\mub^2)\,.
\label{eq:sigmapolhad}
\end{multline}
The $g_1$ structure function is
\begin{align}
  g_1(x,Q^2) &= \frac12\sum_q e_q^2
  \int_{yz,x} \bigl\{ C_{\Delta q}(y) \Delta q(z,Q^2)  
+ C_{\Delta g}(y) \otimes \Delta g (z,Q^2) \bigr\}\,,
\label{5.14}
\end{align}
where $C_{\Delta q}=\delta(1-z)$ to lowest order in $\alpha_s$. Comparing
Eqs.~(\ref{eq:sigmapolprobeNLO}, \ref{eq:sigmapolhad}) we get
\begin{multline}
f_{\Delta \gamma}(\xph,\mub^2) =\frac{1}{2\pi \alpha(\mub^2) } \int
\frac{\mathd \zed}{\zed} \Biggl\{
\int^{\frac{\mub^2}{1-\zed}}_{Q_\text{min}^2} \frac{\mathd Q^2}{Q^2}
\aph^2(-Q^2) \\ \Bigl[ \left(4-2 \zed - \frac{4 m_p^2 \xph^2}{Q^2}
  \right) g_1(\xph/\zed,Q^2) - \left( \frac{8 m_p^2 \xph^2}{\zed Q^2}
  \right) g_2(\xph/\zed,Q^2) \Bigr]\, \\ +\alpha^2(\mu^2)\Bigl[
  \left(2(2-z) \log\frac{M^2(1-z)^2}{\mu^2 z}-2(1-z)\right)
  \\ -\left(2(2-z) \log\frac{M^2(1-z)^2}{\mu^2 z}-6(1-z)\right)
  \Bigr]g_1(\xph/\zed,Q^2) \biggr\} \\ =\frac{1}{2\pi \alpha(\mub^2) }
\int \frac{\mathd \zed}{\zed} \Biggl\{
\int^{\frac{\mub^2}{1-\zed}}_{Q_\text{min}^2} \frac{\mathd Q^2}{Q^2}
\aph^2(-Q^2) \\ \Bigl[ \left(4-2 \zed - \frac{4 m_p^2 \xph^2}{Q^2}
  \right) g_1(\xph/\zed,Q^2) - \left( \frac{8 m_p^2 \xph^2}{\zed Q^2}
  \right) g_2(\xph/\zed,Q^2) \Bigr]\, \\ +\alpha^2(\mu^2)\Bigl[ 4(1-z)
  \Bigr]g_1(\xph/\zed,Q^2) \biggr\}
\label{5.10}\,,
\end{multline} 
for the polarised photon PDF.  Note that the log terms cancel. Taking
the first moment of our result, Eq.~(\ref{5.10}) gives $\Delta
\gamma$, the photon contribution to the proton spin.

As in the unpolarised case, we have checked that the process $\gamma
\gamma \to S$ gives the same result. Furthermore, we can easily derive
from these expressions a number of polarised splitting functions
involving the photon. The expressions are similar to those in
Eqs.~(\ref{eq:pgaga01}--\ref{eq:pgag11}), but we don't give them here.

\section{Alternative derivation using PDF operators}
\label{sec:alt-deriv}

\subsection{Collinear photon PDF}
\label{sec:alt-deriv-collinear}

We have derived the photon PDF by using factorisation applied to two
different processes, $l + p \to L + X$ and $\gamma + p \to S + X$ in
Secs.~\ref{sec:probe}, \ref{sec:spin-dep} and App.~\ref{app:gamgam},
and shown that we get an $\MSbar$ PDF that is process
independent. PDFs can also be defined as the matrix element of a PDF
operator, which is a light-cone Fourier transform of a two-point
function~\cite{Collins:1989gx}, and does not depend on using a probe
process. This has the advantage that it yields much simpler
calculations. In this section, we derive the photon PDF using PDF
operators.  We will exploit the simplicity of this method when going
to higher orders in the next section.

The operator PDF definition is manifestly process independent, and
directly gives the $\MSbar$ PDF.
The photon PDF operator can be obtained by analogy with the gluon PDF
operator in Ref.~\cite{Collins:1989gx} by replacing the gluon
field-strength tensor $G_{\mu\nu}$ by the photon field-strength
$F_{\mu \nu}$, and dropping the Wilson line, since $F_{\mu \nu}$ is
gauge invariant:
\begin{align}
f_\gamma(x,\mub^2) &= -\frac{1}{4 \pi x p^+} \int_{-\infty}^\infty d w
e^{-i x w p^+} \ \braket{p | F^{n \lambda}(w n) F^{n}{}_\lambda(0) +
  F^{n \lambda}(0) F^{n}{}_\lambda(w n) | p}_c \,.
 \label{6.1}
\end{align}
We have introduced a light-like vector $n$, and we have defined
$p^+=p\cdot n$.  We use the notation $F^{n \lambda} \equiv n_\sigma
F^{\sigma \lambda}$, etc.\ in this section.
The subscript $c$ is a reminder that only the connected matrix element
contributes.  The operators are at the same $x^+$ and $x_\perp$
coordinates, and Fourier transformed along the $x^-$ direction.
The $\mu$ dependence arises because the matrix element has
divergences, and is renormalised in the $\MSbar$ scheme (so that the
PDF is in the $\MSbar$ scheme).  All other results in this section are
also renormalised in the $\MSbar$ scheme and, as in Eq.~\eqref{6.1},
we do not explicitly show the counterterms.

The polarised photon PDF can be obtained in the same way from the
polarised gluon PDF~\cite{Manohar:1990kr,Manohar:1990jx},
\begin{align}
f_{\Delta \gamma}(x,\mub^2) &= \frac{i}{4 \pi x p^+}
\int_{-\infty}^\infty d w \, e^{-i x w p^+} \ \braket{p | F^{n
    \lambda}(w n) \widetilde F^{n}{}_\lambda(0) -F^{n \lambda}(0)
  \widetilde F^{n}{}_\lambda(w n) | p}_c \,,
 \label{6.2}
\end{align}
$ \widetilde F_{\alpha \beta} = \frac12 \epsilon_{\alpha \beta \lambda
  \sigma} F^{\lambda \sigma}$, with $\epsilon_{0123}=+1$, and we
conventionally assume that the proton has positive helicity. The
$\epsilon$-tensor is defined to live in the four physical dimensions,
according to the 't Hooft-Veltman scheme.

\begin{figure}
\begin{align*}
{\setlength{\arraycolsep}{0.15cm}
\begin{array}{ccc}
\begin{minipage}{3.5cm} \includegraphics[width=3.5cm]{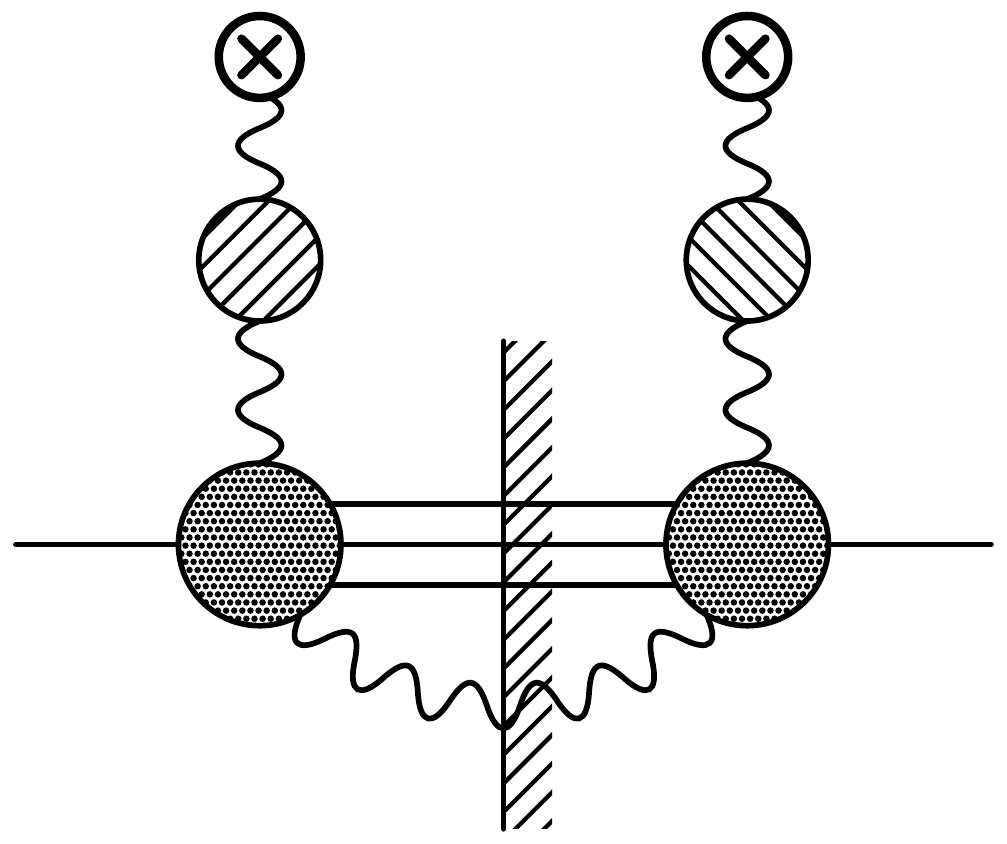} \end{minipage} &  &
\begin{minipage}{3.5cm} \includegraphics[width=3.5cm]{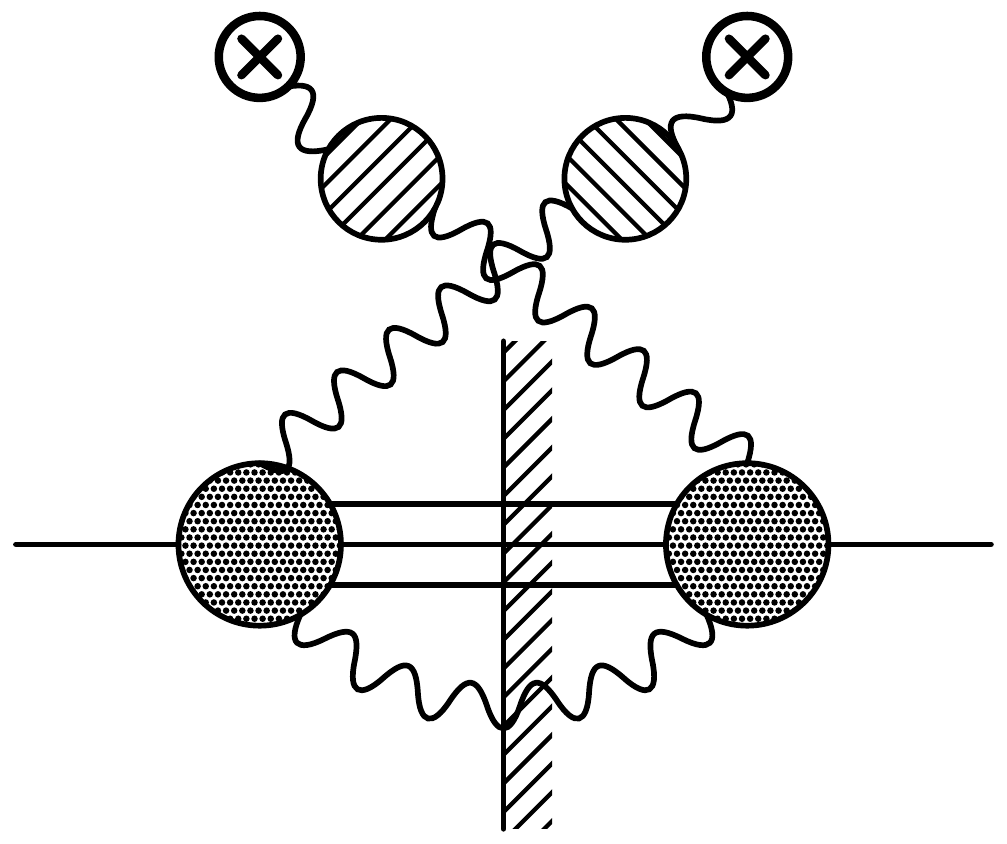} \end{minipage} \\[5pt]
(a) && (b)
\end{array}
}
\end{align*}
\caption{\label{fig:pdfop} Matrix elements of the PDF operator in a
  proton state. The graphs include arbitrary radiative corrections to
  the hadron tensor, represented schematically by the single photon
  loop correction.}
\end{figure}

The PDFs can be computed from Eqs.~(\ref{6.1}, \ref{6.2}). The leading
order graphs are shown in Fig.~\ref{fig:pdfop}.
The key observation is that the lower part of the diagram, the
interaction of the photon with the proton target, is precisely the
definition of the hadronic tensor $W_{\mu \nu}(p,q)$, because the
photon interacts with the proton through the electromagnetic current
$j^\mu$ that appears in Eq.~(\ref{3.21}).%
\footnote{It is this factorisation property which allows us to compute
  the photon PDF.  There are expressions analogous to
  Eqs.~(\ref{6.1},\ref{6.2}) with $F_{\mu \nu}$ replaced by
  $G_{\mu\nu}$ which give the gluon PDF. Since the gluon is a colour
  adjoint, there is also a colour Wilson line between the two field
  strength tensors in the operator product. The gluon analogue of graphs
  Fig.~\ref{fig:pdfop} cannot be written in terms of experimentally
  measured proton structure functions, since the gluon does not couple
  to the proton via gauge invariant colour singlet operators. In
  addition, there are graphs with gluon exchange between the Wilson
  line and the proton.}
Note that Eq.~(\ref{6.1}) involves the ordinary product of operators,
not the time-ordered product. The matrix element in Eq.~(\ref{6.1})
can be evaluated by comparing with the expression for $W_{\mu \nu}$ in
Eq.~(\ref{eq:hadtens}), which corresponds to taking the discontinuity
of the time-ordered product, $T_{\mu\nu}$ in
Eq.~(\ref{eq:hadtensTmunu}), i.e.\ evaluating the cut graphs in
Fig.~\ref{fig:pdfop}.
The diagrams give
\begin{subequations}
\label{6.3}
\begin{align}
f_\gamma(x,\mub^2) & =-\frac{e^2(\mub^2) \smu^{2\epsilon}}{x p^+}  \int \frac{\mathd^D q}{(2\pi)^D}  \left[ 2\pi \delta(q^+ + x p^+) + 2\pi \delta(q^+ - x p^+)  \right]    \times \nn
& \left[(n \cdot q) g^{\lambda \mu}-q^\lambda g^{n \mu} \right]\left[(n \cdot q) g^{\lambda \nu}-q^\lambda g^{n \nu} \right]  \frac{1 }{\left(q^2\left[1 - \Pi_D(q^2,\mub^2)\right]\right)^2} \times \nn & \left[ W^{(D)}_{\mu \nu}(p,q) +W^{(D)}_{\nu \mu}(p,-q)\right]\,, \label{6.3a} \\
f_{\Delta \gamma}(x,\mub^2) & =\frac{ie^2(\mub^2) \smu^{2\epsilon}}{x p^+}  \int \frac{\mathd^D q}{(2\pi)^D}  \left[ 2\pi \delta(q^+ + x p^+)- 2\pi \delta(q^+ - x p^+)  \right]  \times   \nn
& \left[ (n \cdot q) g^{\lambda \mu}-q^\lambda g^{n \mu} \right] \left[ \epsilon_{n\lambda\alpha \beta} q^\alpha g^{\beta \nu} \right]  \frac{1 }{\left(q^2\left[1 - \Pi_D(q^2,\mub^2)\right]\right)^2} \times \nn & \left[ W^{(D)}_{\mu \nu}(p,q) + W^{(D)}_{\nu \mu}(p,-q)\right]\,,
\label{6.3b}
\end{align}
\end{subequations}
where $\cal S$ is defined in Eq.~(\ref{Sdef}), and $\Pi_D(q^2,\mub^2)$
is the $\MSbar$ subtracted polarisation function computed in $D$
dimensions (i.e. the $D=4$ limit is not yet taken).

The two $\delta$-functions arise from the two operator orderings in
Eqs.~(\ref{6.1},\ref{6.2}), the next factor in each equation is the
Feynman rule for the field-strength tensors, and the last factor is
from the photon propagators and the proton matrix element, which, by
definition is the one-photon irreducible hadron tensor $W^{(D)}_{\mu
  \nu}$.  The two contributions are from the direct and crossed
graphs. The formula Eq.~(\ref{6.3}) is exact, and has no QCD or QED
corrections, provided that all QED corrections on the hadronic side
are included in the definition of the structure functions.  We stress
that also the hadronic tensor appearing in Eq.~(\ref{6.3}) is $\MSbar$
subtracted, but is evaluated in $D$ dimensions. The limit $D\to 4$ can
only be taken after the overall $\MSbar$ counterterm is added to
Eq.~(\ref{6.3}).

The vacuum polarisation factors can be written in terms of $\aph$
using Eq.~(\ref{3.15}). The two $W^{(D)}_{\mu \nu}$ terms can be
combined by letting $q \to - q$. The hadronic tensor $W^{(D)}_{\mu
  \nu}$ is non-zero for $0 \le -q^2/(2p \cdot q) \le 1$, so only the
$q^+ <0$ piece contributes, and we find
\begin{subequations}
\begin{eqnarray}
f_\gamma(x,\mub^2) & =&-\frac{2 \smu^{-2\epsilon} }{e^2(\mub^2) x p^+}
\int \frac{\mathd^D q}{(2\pi)^D} \left[ 2\pi \delta(q^+ + x p^+) \right]
\frac{\ephD^4(q^2)}{\left(q^2\right)^2}
\left[ (n\cdot q)^2 W^{(D) \lambda}_{\lambda}
+ q^2 W^{(D)}_{nn} \right] \,, \nn \label{6.4a} \\
f_{\Delta \gamma}(x,\mub^2)
&=&\frac{2i \smu^{-2\epsilon}}{e^2(\mub^2) x p^+} \int \frac{\mathd^D q}{(2\pi)^D}
\left[ 2\pi \delta(q^+ + x p^+) \right]
\frac{\ephD^4(q^2) }{\left(q^2\right)^2} (n \cdot q) \epsilon_{n\mu q \nu}
W^{(D)}_{\mu \nu}(p,q) \,.
\label{6.4b}
\end{eqnarray}
\label{6.4}
\end{subequations}
Here $W^{(D)}_{nn}=W^{(D)}_{\mu \nu} n^\mu n^\nu$ and $\epsilon_{n \mu
  q \nu}= \epsilon_{\alpha \mu \beta \nu} n^\alpha q^\beta$.  The
physical coupling constant $\ephD$, in analogy to Eq.~\eqref{3.15}, is
defined as
\begin{align}
\ephD^2(q^2) &= \frac{e^2(\mub^2)\left(\mu \mathcal{S}\right)^{2 \epsilon}}{1 - \Pi_D(q^2,\mub^2)} \,, &
\alpha_D(q^2) &= \frac{\alpha(\mub^2)\left(\mu \mathcal{S}\right)^{2 \epsilon}}{1 - \Pi_D(q^2,\mub^2)} \,.
\label{eq:ephD}
\end{align}
The $D$ dimensional $\epsilon$ tensor vanishes if its indices are
outside 4 dimensions.  The contractions with $W_{\mu \nu}$ in
$D=4-2\epsilon$ dimensions are
\begin{subequations}
\begin{align}
& (n\cdot q)^2 W^{(D)\lambda}_{\lambda}  + q^2 W^{(D)}_{nn} 
= (n \cdot q)^2 \biggl[ \frac{1}{\xbj} F_{L, D}  + \left(-\frac{1}{\xbj} - \frac{2 (n \cdot p)}{(n \cdot q)} - \frac{2 (n \cdot p)^2 \xbj}{(n \cdot q)^2} - \frac{2 m_p^2 \xbj}{Q^2} \right) F_{2, D} \nn 
& \hspace{10cm} + 2 \epsilon F_{1, D} \biggr] \,, \label{6.5a} \\
&  \epsilon_{n\mu q \nu}   W^{(D)}_{\mu \nu}(p,q)  
= \frac{4 i \xbj\, g_{1, D}}{Q^2} \left[Q^2_4 (n \cdot s) + (n \cdot q)(q \cdot s) \right] \nn
&  \hspace{5cm} - \frac{4 i \xbj\, g_{2, D}}{Q^2} \left[2 \xbj (n \cdot p)(q \cdot s)\frac{Q^2_4}{Q^2}-Q^2_4 (n \cdot s)\right] ,
\label{6.5b}
\end{align}
\label{6.5}
\end{subequations}
where $\xbj=Q^2/(2 p \cdot q)$ and the structure functions
($F_{2/1/L,D}$ and $g_{1/2,D}$) are evaluated at $\xbj, Q^2$ in $D$
dimensions with standard $\MSbar$ subtraction (i.e., as in the case of
the physical electromagnetic coupling, the limit $D=4$ is not yet
taken).
The second expression depends on $Q^2_4$ involving only the component
of $Q$ in $D=4$ dimensions. We have
\begin{align}
Q^2_4 &= Q^2 - Q^2_{-2\epsilon} 
\end{align}
where $Q^2_{-2\epsilon}$ is the fractional dimension part of $Q$.  Let
$q^+ =-xp^+$ and $z= x/\xbj$ then, for a right-handed proton moving
along the $z$ axis, $n=(1,0,0,-1)$, $p=(E,0,0,p)$, $s=(p,0,0,E)$,
\begin{align}
n \cdot s &= n \cdot p\,, &
(q\cdot s) &=(q \cdot p) - m_p^2 \frac{n \cdot q}{n \cdot p}\,,
\label{6.6}
\end{align}
and Eq.~(\ref{6.5}) becomes
\begin{subequations}
\begin{eqnarray}
&& (n\cdot q)^2 W^{(D)\lambda}_{\lambda} + q^2 W^{(D)}_{nn} = -(p^+)^2
  \frac{x}{z} \biggl[-z^2 F_{L, D} + \left(z^2-2z+2 + \frac{2 m_p^2
      x^2}{Q^2} \right) F_{2, D} - 2 \epsilon z x F_{1, D} \biggr]\,,
  \nn \label{6.7a}\\ && (n \cdot q) \epsilon_{n\mu q \nu} W^{(D)}_{\mu
    \nu}(p,q) = -(p^+)^2 \frac{x}{z}(4 i x)
  \biggl[\left(1-\frac{z}{2}-\frac{m_p^2 x^2}{Q^2}
    -\frac{Q^2_{-2\epsilon}}{Q^2} \right)g_{1, D} \nn && \hspace{8cm}
    - \frac{2m_p^2 x^2}{Q^2 z} \left(1-\frac{Q^2_{-2\epsilon}}{Q^2}
    \right) g_{2, D} \biggr]\,,
\label{6.7b}
\end{eqnarray}
\label{6.7}
\end{subequations}
where the structure functions are evaluated at $x/z,Q^2$.

The $q$ integral can be performed by switching to light-cone
coordinates,
\begin{align}
\int \frac{\mathd^D q}{(2\pi)^D} 
&= \frac{1}{(4\pi)^{D/2-1}} \frac{1}{\Gamma(D/2-1)} \frac12 \int_{-\infty}^{\infty}
\frac{\mathd q^+}{2\pi}  \int_{-\infty}^\infty \frac{\mathd q^-}{2\pi}  \int _{-q^+ q^-}^\infty \left(Q^2+q^+ q^-\right)^{D/2-2} \mathd Q^2\,.
\label{6.8}
\end{align}
We also have
\begin{align}
\xbj &= \frac{x}{z}=\frac{Q^2}{2 p \cdot q} = \frac{Q^2}{p^+ q^- - x m_p^2}  
\implies q^- = \frac{1}{p^+}\left( \frac{z}{x} Q^2 + x m_p^2 \right)\,,
\label{6.9}
\end{align}
which is used to replace the integration variable $q^-$ by $z$. The
$q^+$ integral is evaluated using the $\delta$-function, leaving the
integrals over $z$ and $Q^2$.  $Q^2_{-2\epsilon}$ is in the $\perp$
direction, and all $\perp$ directions are equivalent, so we can make
the replacement
\begin{align}
Q^2_{-2\epsilon} &\to \frac{D-4}{D-2} Q^2_\perp =  \frac{D-4}{D-2} \left(Q^2+q^+ q^-\right)
\end{align}
in evaluating the integrals.

The kinematically allowed region in the $z,Q^2$ plane is
\begin{align}
x & \le z &
Q^2 & \ge -q^+ q^- \implies Q^2 \ge \frac{m_p^2 x^2}{1-z}\,.
\label{6.10}
\end{align}
The first inequality follows because $W_{\mu \nu}$ vanishes for $w \ge
1$, and the second from $Q^2_\perp \ge 0$.

The resulting integrals are
\begin{subequations}
\begin{align}
 f_\gamma(x,\mub^2) 
&= \frac{8 \pi }{x \alpha(\mub^2) \smu^{2\epsilon}} \frac{1}{(4\pi)^{D/2}} \frac{1}{\Gamma(D/2-1)} \times \nn
& \int_x^1 \frac{\mathd z}{z} \int_{\frac{m_p^2 x^2}{1-z}}^\infty  \frac{\mathd Q^2}{Q^2} \aphD^2(q^2) \left(Q^2(1-z)-x^2 m_p^2\right)^{D/2-2} \times \nn
& \biggl\{- z^2 F_{L,D}(x/z,Q^2) + \left[ 2-2z +  z^2  + \frac{2 m_p^2 x^2}{Q^2}\right] F_{2, D}(x/z,Q^2)  - 2 \epsilon z x F_{1, D} (x/z,Q^2)\biggr\} \,, \label{6.11a}\\
f_{\Delta \gamma} (x,\mub^2) 
 &=  \frac{8 \pi }{\alpha(\mub^2) \smu^{2\epsilon}} \frac{1}{(4\pi)^{D/2}} \frac{1}{\Gamma(D/2-1)} \times \nn
& \int_x^1 \frac{\mathd z}{z}  \int_{\frac{m_p^2 x^2}{1-z}}^\infty  \frac{\mathd Q^2}{Q^2} \aphD^2(q^2) \left(Q^2(1-z)-x^2 m_p^2\right)^{D/2-2} \times \nn
& \biggl\{ \left(4-2 z-\frac{4m_p^2 x^2}{Q^2} - 4 \frac{D-4}{D-2} \frac{ Q^2(1-z)-x^2 m_p^2 }{ Q^2 }  \right)g_{1, D}(x/z,Q^2) \nn
&- \frac{8m_p^2 x^2}{Q^2 z}
\left(1-\frac{D-4}{D-2}\frac{ Q^2(1-z)-x^2 m_p^2 }{ Q^2 } \right) g_{2, D}(x/z,Q^2)  \biggr\}\,, 
\label{6.11b}
\end{align}
\label{6.11}
\end{subequations}
with $\aphD(q^2) = \ephD^2(q^2)/(4\pi)$.  This expression for the PDF
is \emph{exact}, and includes QED radiative corrections if the
structure functions are the one-photon-irreducible ones.

We split the $Q^2$ integral into $m_p^2 x^2/(1-z) \to \mub^2/(1-z)$
and $ \mub^2/(1-z) \to \infty$,
\begin{subequations}
  \label{eq:fboth-split}
  \begin{align}
    \label{eq:fgamma-split}
    f_\gamma(x,\mub^2) &= f^{\text{PF}}_\gamma(x,\mub^2) +
                         f^{\text{con}}_\gamma(x,\mub^2) \,,
    \\
    \label{eq:fDeltagamma-split}
    f_{\Delta\gamma}(x,\mub^2) &= f^{\text{PF}}_{\Delta\gamma}(x,\mub^2) +
                                 f^{\text{con}}_{\Delta\gamma}(x,\mub^2)\,.
  \end{align}
\end{subequations}
The first part is finite, and gives (by definition of $f_{\gamma,\Delta \gamma}^{\text{PF}}$)
\begin{subequations}
  \label{6.12}
  \begin{align}
      \label{6.12a}      
    f^{\text{PF}}_\gamma(x,\mub^2)
    &= \frac{1 }{2 \pi \alpha(\mub^2)  x} 
      \int_x^1 \frac{\mathd z}{z}  \int_{\frac{m_p^2 x^2}{1-z}}^{\frac{\mub^2}{1-z}}  \frac{\mathd Q^2}{Q^2}  \aph^2(q^2)
      \nn
    & \biggl\{- z^2 F_L(x/z,Q^2) + \left[ 2-2z  +  z^2  + \frac{2
      m_p^2 x^2}{Q^2}\right] F_2(x/z,Q^2)\biggr\}\,, 
    \\
    \label{6.12b}
      f^{\text{PF}}_{\Delta \gamma} (x,\mub^2) 
    & = \frac{1 }{2\pi \alpha(\mub^2) } 
      \int_x^1 \frac{\mathd z}{z}  \int_{\frac{m_p^2 x^2}{1-z}}^{\frac{\mub^2}{1-z}}  \frac{\mathd Q^2}{Q^2}  \aph^2(q^2) \nn
    & \biggl\{ \left(4-2 z - \frac{4m_p^2 x^2}{Q^2} \right)g_1(x/z,Q^2) - \frac{8m_p^2 x^2}{Q^2 z} g_2(x/z,Q^2)  \biggr\}\,,
  \end{align}
\end{subequations}
which is the result obtained previously for what was referred to as
the ``physical factorisation'' PDF, Eqs.~\eqref{4.16} and
\eqref{5.10ph}. We automatically get the correct integrand for the
PDF, without dropping any terms, as had to be done for the BSM probes
in Sec.~\ref{sec:probe}.

The ``$\MSbar$ -conversion'' term is the remaining integral over $
\mub^2/(1-z) \to \infty$ plus the counterterms.  Since $Q^2 \gtrsim
\mub^2$, we can set $m_p^2 \to 0$, up to neglected power corrections,
and use Eq.~(\ref{3.28}) to replace $2xF_1$ by $F_2-F_L$,
\begin{subequations}
\label{6.13}%
\begin{align}
 f_\gamma^{\text{con}}(x,\mub^2)
& =   \frac{8 \pi  \smu^{-2\epsilon}}{x \alpha(\mub^2)} \frac{1}{(4\pi)^{D/2}} \frac{1}{\Gamma(D/2-1)}  \int_x^1 \frac{\mathd z}{z}  (1-z)^{D/2-2}  \int_{\frac{\mub^2}{1-z}}^\infty  \frac{\mathd Q^2}{Q^2}  \nn
 & \left(Q^2\right)^{D/2-2}  \aphD^2(q^2)  \biggl\{ - z^2 (1-\epsilon) F_{L, D}(x/z,Q^2) \nn & + \left[ 2-2z  +  z^2 -\epsilon z^2  \right] F_{2, D}(x/z,Q^2) \biggr\}\,, \label{6.13a}\\
 f_{\Delta \gamma}^{\text{con}}(x,\mub^2)
& = \frac{8 \pi  \smu^{-2\epsilon}}{\alpha(\mub^2) } \frac{1}{(4\pi)^{D/2}} \frac{1}{\Gamma(D/2-1)}  \int_x^1 \frac{\mathd z}{z}  (1-z)^{D/2-2}  \int_{\frac{\mub^2}{1-z}}^\infty   \frac{\mathd Q^2}{Q^2}   \nn
 & \left(Q^2\right)^{D/2-2}  \aphD^2(q^2)  \biggl\{  \left(4 - 2 z  + 4 \frac{\epsilon}{1-\epsilon} (1-z) \right)g_{1, D}(x/z,Q^2)  \biggr\}\,.
\label{6.13b}
\end{align}
\end{subequations}
Eq.~(\ref{6.13}) is still exact. The integral is over large values of
$Q$, so the $q$ dependence of $\aphD$, $F_{2/L, D}$ and $g_{1/2, D}$
can be computed in perturbation theory.
The integral does not involve any large logarithms, since it only
involves the scale $\mub$.\footnote{In fact, the integral can instead
  generate a further $1/\epsilon$ pole, but not logarithms.}

Introducing the dimensionless variable $s$,
\begin{align}
  s & = \frac{Q^2(1-z)}{\mu^2} \,,
\label{6.17}
\end{align}
the integrals become
\begin{subequations}
\label{6.113}
\begin{align}
 f_\gamma^{\text{con}}(x,\mub^2)
& =   \frac{\smu^{-2\epsilon}}{2 \pi x \alpha(\mub^2)  \mub^{2\epsilon}} \frac{e^{\epsilon \gamma_E}}{\Gamma(1-\epsilon)}  \int_x^1 \frac{\mathd z}{z}  \int_{1}^\infty  \frac{\mathd s}{s^{1+\epsilon}}  \aphD^2(-\mub^2 s/(1-z))  \times  \nn
 & \biggl\{ - z^2 (1-\epsilon) F_{L, D}(x/z,\mub^2 s/(1-z)) + \left[ 2-2z  +  z^2 -\epsilon z^2 \right] F_{2, D}(x/z,\mub^2 s/(1-z)) \biggr\}\,,  \label{6.113a}\\
 f_{\Delta \gamma}^{\text{con}}(x,\mub^2)
& =\frac{\smu^{-2\epsilon}}{2 \pi  \alpha(\mub^2)  \mub^{2\epsilon}} \frac{e^{\epsilon \gamma_E}}{\Gamma(1-\epsilon)}  \int_x^1 \frac{\mathd z}{z}    \int_{1}^\infty  \frac{\mathd s}{s^{1+\epsilon}} \aphD^2(-\mub^2 s/(1-z)) \times \nn
 &   \biggl\{  \left(4 - 2 z  + 4 \frac{\epsilon}{1-\epsilon} (1-z) \right)g_{1, D}(x/z,\mub^2 s/(1-z))  \biggr\}\,.
\label{6.113b}
\end{align}
\end{subequations}
The factor of $(1-z)^{D/2-2}$ has cancelled, which is why
$\mub^2/(1-z)$ was chosen as the intermediate $Q^2$ value to split the
integrals.

For the remainder of this section, we restrict to lowest order in
$\alpha$ and $\alpha_s$. At this order, the structure functions do not
depend on $Q^2$, and can be evaluated at $\mub^2$ without incurring
any large logarithms, and $\aphD(q^2) \approx\alpha(\mu^2) (\mu {\cal
  S})^{2\epsilon}$ from Eq.~(\ref{eq:ephD}).
Since $F_L$ is order $\alpha_s$, it can be dropped. Evaluating the $s$
integral, and expanding in $\epsilon$ gives
\begin{subequations}
\label{6.14}
\begin{align}
f_\gamma^{\text{con}}(x,\mub^2)
& = \frac{\alpha(\mub^2)}{2 \pi x}
 \int_x^1 \frac{\mathd z}{z}  \biggl\{   \frac{1}{\epsilon} \left[ 2-2z  +  z^2  \right] F_2(x/z,\mub^2)\biggr\} -   z^2 F_2 (x/z,\mub^2)\biggr\} \,, \label{6.14a}\\
f_{\Delta \gamma}^{\text{con}}(x,\mub^2)
& = \frac{\alpha(\mub^2)}{2\pi}  \int_x^1 \frac{\mathd z}{z} \biggl\{   \frac{1}{\epsilon}   \left(4 - 2 z \right)g_1(x/z,\mub^2)+4 (1-z) g_1(x/z,\mub^2) \biggr\}   \,.
\label{6.14b}
\end{align}
\end{subequations}
The $1/\epsilon$ term is absorbed by the $\MSbar$ counterterm, and one
obtains in the $\MSbar$ scheme
\begin{subequations}
\label{6.15}
\begin{align}
f_\gamma^{\MSbar\, \text{con}}(x,\mub^2)
& = \frac{\alpha(\mub^2)}{2 \pi x}
 \int_x^1 \frac{\mathd z}{z}  (   -  z^2 )   F_2 (x/z,\mub^2)
  +\mathcal{O}(\alpha^2,\alpha\alpha_s) \,, \label{6.15a} \\
  f_{\Delta \gamma}^{\MSbar\, \text{con}}(x,\mub^2)
& = \frac{\alpha(\mub^2)}{2 \pi }
 \int_x^1 \frac{\mathd z}{z}  4(1-z)    g_1 (x/z,\mub^2)
  +\mathcal{O}(\alpha^2,\alpha\alpha_s) \,.
\label{6.15b}
\end{align}
\end{subequations}
The PDFs in the $\MSbar$ scheme are thus  
\begin{subequations}
\label{6.16}
\begin{align}
f_\gamma (x,\mub^2) &= f_\gamma^{\text{PF}}(x,\mub^2) + f_\gamma^{\MSbar\, \text{con}}(x,\mub^2) +\mathcal{O}(\alpha^2,\alpha\alpha_s) 
\,, \label{6.16a} \\
f_{\Delta \gamma} (x,\mub^2) &= f_{\Delta
  \gamma}^{\text{PF}}(x,\mu^2) + f_{\Delta
  \gamma}^{\MSbar\, \text{con}}(x,\mub^2)
+\mathcal{O}(\alpha^2,\alpha\alpha_s) \,, 
\label{6.16b}
\end{align}
\end{subequations}
with $f_{\gamma/\Delta \gamma}^{\text{PF}}(x,\mu^2)$ defined in
Eq.~\eqref{6.12}.  The results agree with our previous results using
the $l \gamma \to L$ and $\gamma \gamma \to S$ processes.

The derivation in this section shows the PDFs are process independent,
since it does not use any physical process. It also shows how to
systematically extend the result to higher orders. The PF term
Eq.~(\ref{6.12}) is exact, and the only corrections are to the
$\MSbar$-conversion terms Eq.~(\ref{6.113}), which does not contain
any large logarithms. The result Eq.~(\ref{6.16}) can be extended to
higher orders using the perturbation expansion for $\aphD$ and for the
structure functions in $D$ dimensions.  The result to one higher order
is given in Sec.~\ref{sec:higher-order}. Unlike the method of the
earlier sections, we do not have to compute any hard scattering cross
sections to higher order when using PDF operators.

\subsection{Derivation using an abstract probe}

Using the operator definition of the photon PDF avoids the
complication of having to compute a specific probe process. On the
other hand, the freedom of choosing an appropriate probe process can
be exploited to achieve a similar simplicity. In particular we may
consider the following photon probe tensor
\begin{eqnarray}
L^{\mu\nu}(q,p,n)&=& \left[-(q^+)^2 g^{\mu\nu}+q^+(q^\mu n^\nu+q^\nu n^\mu)-n^\mu n^\nu q^2\right]
\nonumber \\
 &\times& \delta(q^++xp^+) \theta(-q^2) \theta(\mu^2-|\mathbf{q}_\perp^2|),
 \label{eq:absprobe}
\end{eqnarray}
to be contracted with the hadronic tensor in Eq.~(\ref{4.4}). This
probe does not refer to any specific BSM process, and indeed it is
difficult to imagine a physical process that would give rise to this
tensor. However, this does not matter as long as the tensor has the
appropriate transversality properties, which implies that the
``cross section'' can be computed both in terms of the proton
electromagnetic structure functions and in terms of the parton model
formula.

The calculation in terms of structure functions yields
\begin{eqnarray}
  \sigma_{\rm SF} &=& \frac{1}{2 p^0} \int \frac{\mathd^4 q}{(2 \pi)^4} 
  \frac{\eph^4(q^2)}{e^2(\mu^2)}  \frac{1}{q^4} 4 \pi W^{\mu \nu} (q, p)
  L^{\mu \nu} (q, p,n)
\nonumber \\
   &=&-\frac{2\pi}{p^0} \int \! \frac{\mathd^4 q}{(2 \pi)^4} \delta(q^++xp^+)\theta(\mu^2-| \mathbf{q}_\perp^2|)
   \frac{\eph^4(q^2)}{e^2(\mu^2)}\frac{1}{q^4}
   \left[(n\cdot q)^2 W_\lambda^\lambda+q^2 W_{nn}\right]. \label{eq:sigmaabsprobe}
\end{eqnarray}
The $1/(2p^0)$ is from the incident proton flux, since the proton
state is normalised to $2p^0$.

The parton model calculation yields instead
\begin{eqnarray}
  \sigma_{\rm PM} &=& \sigma_{\rm PM}^{(0)} + \sigma_{\rm PM}^{({\rm HO})}\,,\nonumber \\
  \sigma_{\rm PM}^{(0)} &=&
  \int \mathd y\, \frac{1}{2yp^0} f_\gamma(y,\mu^2) [(yp^+)^2] \delta(-yp^++xp^+) =  \frac{p^+}{2p^0} x f_\gamma(x,\mu^2)\,, \nonumber \\
  \sigma_{\rm PM}^{({\rm HO})} &=&-\frac{2\pi}{p^0} \int \frac{\mathd^D q}{(2 \pi)^D} \delta(q^++xp^+)\theta(\mu^2-|\mathbf{q}_\perp^2|) \nonumber \\
&\times&   \frac{\smu^{-2\epsilon}\ephD^4(q^2)}{e^2(\mu^2)}\frac{1}{q^4}
   \left[(n\cdot q)^2 W_{\lambda}^{(D)\lambda}+q^2 W^{(D)}_{nn}\right]+\mbox{c.t.}\,. \label{eq:sigmaabsprobePM}
\end{eqnarray}
In the $\sigma_{\rm PM}^{(0)}$ term, the partonic cross section was
evaluated by setting $q=-py$ in Eq.~(\ref{eq:absprobe}), contracting
it with a $-g_{\mu\nu}$, dividing by the number of photon
polarisations in $D$ dimensions, and dividing by the incident parton
flux $2 yp^0$. The $\sigma_{\rm PM}^{({\rm HO})}$ term must be
computed in $D$ dimensions, for incoming massless particles. This
leads to the presence of collinear singularities, that are removed
according to the usual factorisation procedure. In order to write the
HO term in the form of Eq.~(\ref{eq:sigmaabsprobePM}) we have
exploited the fact that all higher-order corrections must begin with a
dressed photon attached to the $L_{\mu\nu}$ tensor on one side, and to
the $W_{\mu\nu}$ tensor on the other side. Again we must assume that
the renormalisation and factorisation procedure has been carried out
for the $W_{\mu\nu}$ tensor and the dressed photon, but their
expressions must remain in $D$ dimension until we perform the
\MSbar{} subtraction of the divergence arising from the $q$
integration. Our final result is
\begin{equation}
f_\gamma(x,\mu^2)=\frac{2p^0}{xp^+}\left(\sigma_{\rm SF}-\sigma_{\rm PM}^{\rm (HO)}\right).
\end{equation}
We now notice that the $\sigma_{\rm SF}$ term corresponds exactly to
$f_\gamma^{\rm PF}$, while the $\sigma_{\rm PM}^{\rm (HO)}$ term is
directly related to $f_\gamma^{\rm con}$. The first correspondence
follows from the fact that the $\theta(\mu^2-|\mathbf{q}_\perp^2|)$
condition is equivalent to $\theta(\mu^2-Q^2(1-z))$. In fact, using
Eq.~(\ref{6.9}), we find
\begin{equation}
|\mathbf{q}_\perp^2|=-q^2+q^+q^-=Q^2+(-xp^+)\frac{1}{p^+}\frac{z}{x}Q^2=(1-z)Q^2\,,
\end{equation}
up to corrections of order $m_p^2/\mu^2$.  The second correspondence
is slightly more subtle, since in $f_\gamma^{\rm con}$ the $q$
integration is restricted by $|\mathbf{q}_\perp|^2>\mu^2$, while in
$\sigma_{\rm PM}^{({\rm HO})}$ it is restricted by
$|\mathbf{q}_\perp|^2<\mu^2$.  On the other hand, the unrestricted
integration evaluated in the parton model limit (i.e.  neglecting all
masses) is zero in perturbation theory, since it leads to scaleless
integrals which vanish in dimensional regularisation. Thus
\begin{equation}
f_\gamma^{\rm con}=-\frac{2p^0}{xp^+}\sigma_{\rm PM}^{\rm (HO)}\,,
\end{equation}
demonstrating the equivalence of the two procedures.

\subsection{The photon TMDPDF}\label{sec:tmdpdf}
The analysis of the previous section can readily be generalised to
obtain the photon transverse momentum dependent PDF (TMDPDF)
$f_\gamma(x,\mathbf{k}_\perp,\mub^2)$. The TMDPDF (see
Ref.~\cite{Angeles-Martinez:2015sea} for a review on TMDPDFs) is given
by Eq.~(\ref{6.1}), with the two field-strength tensors separated by
$\mathbf{x}_\perp$ and Fourier transforming in $\mathbf{x}_\perp$,
\begin{align}
  & f_\gamma(x,\mathbf{k}_\perp,\mub^2) = -\frac{1}{4 \pi x p^+} \int \rd^{D-2} \mathbf{x}_\perp \int_{-\infty}^\infty d w  e^{-i x w p^+} \times \nonumber \\ 
  & \qquad \qquad \qquad \qquad e^{-i \mathbf{x}_\perp \cdot \mathbf{k}_\perp} \ \braket{p | F^{n \lambda}(w n+\mathbf{x}_\perp) F^{n}{}_\lambda(0) + F^{n \lambda}(\mathbf{x}_\perp) F^{n}{}_\lambda(w n) | p}_c \,.
 \label{6.22}
\end{align}
In momentum space, it is given by Eq.~(\ref{6.3}) with an insertion of
$(2\pi)^{D-2} \delta^{(D-2)}(\mathbf{k}_\perp-\mathbf{q}_\perp)$.  The
derivation proceeds as in the previous section. The $q$ integral is
written as
\begin{align}
\int \frac{\mathd^D q}{(2\pi)^D} &= \frac12 \int_{-\infty}^{\infty}
\frac{\mathd q^+}{2\pi} \int_{-\infty}^\infty \frac{\mathd
  q^-}{2\pi}\int \frac{\rd^{D-2} \mathbf{q}_\perp}{(2\pi)^{D-2}}
\label{6.18}
\end{align}
and the $\mathbf{q}_\perp$ and $q^+$ integrals are done using the
$\delta$-functions, leaving only the $q^-$ integral. In terms of the
variable $z=x/w$ one has
\begin{align}
q^- &= \frac{ \mathbf{k}_\perp^2 z + x^2 \mpr^2}{p^+ x (1-z)}\,,
\label{6.19}
\end{align}
so that the $q^-$ integral can be replaced by one over $z$. 
Defining 
\begin{align}
{\cal Q}^2 &\equiv \frac{ \mathbf{k}_\perp^2 + x^2 \mpr^2}{1-z}\,,
\label{6.21}
\end{align}
this leads to 
\begin{align}
 f_\gamma(x,\mathbf{k}_\perp,\mub^2)
 & = \frac{2}{\alpha(\mub^2)  x} 
 \int_x^1 \frac{\mathd z}{z} \frac{\aph^2(q^2)}{\mathbf{k}_\perp^2+x^2 \mpr^2}
\nn
& \biggl\{- z^2 F_L(x/z,{\cal Q}^2) + \left[ 2-2z  +  z^2  + \frac{2 m_p^2 x^2}{{\cal Q}^2}\right] F_2(x/z,{\cal Q}^2)\biggr\}\,, \nn
 f_{\Delta \gamma}(x,\mathbf{k}_\perp,\mub^2)
& = \frac{2 }{ \alpha(\mub^2) } 
\int_x^1 \frac{\mathd z}{z}  \frac{\aph^2(q^2)}{\mathbf{k}_\perp^2+x^2 \mpr^2} \nn
& \biggl\{ \left(4-2 z - \frac{4m_p^2 x^2}{{\cal Q}^2} \right)g_1(x/z,{\cal Q}^2) - \frac{8m_p^2 x^2}{{\cal Q}^2 z} g_2(x/z,{\cal Q}^2)  \biggr\}\,.
\label{6.20}
\end{align}
Eq.~(\ref{6.20}) is an exact expression for the TMDPDF if
one-photon-irreducible structure functions are used.

The TMDPDF is connected with our physical factorisation component of
the collinear PDF through the following simple relation:
\begin{equation}
  \label{eq:PF-from-TMD}
  f_\gamma^\text{PF}(x,\mu^2) = \int  \frac{\rd^2
    \mathbf{k}_\perp}{(2\pi)^2}  
  \,\, f_\gamma(x, \mathbf{k}_\perp, \mu^2)\,
  \Theta(\mu^2 - \mathbf{k}_\perp^2)\,.
\end{equation}

\section{The Photon PDF to higher order}
\label{sec:higher-order}

In the previous sections, we obtained the photon PDFs $f_\gamma$ and
$f_{\Delta \gamma}$ up to corrections of order $\alpha(\mub)
\alpha_s(\mub)$ and $\alpha^2(\mub)$, as given in
Eqs.~(\ref{eq:master}) and (\ref{5.10}). In Sec.~\ref{sec:alt-deriv},
we wrote an exact formula for the photon PDFs, Eq.~(\ref{6.11}), and
we expressed it as the sum of two terms: $f_{\gamma,\Delta
  \gamma}^{\text{PF}}$ in Eq.~(\ref{6.12}) and $f_{\gamma,\Delta
  \gamma}^{\text{con}}$ in Eq.~(\ref{6.113}). The physical PDF
$f_{\gamma,\Delta \gamma}^{\text{PF}}$ is a finite integral that can
be evaluated explicitly using measured DIS structure functions, while
$f_{\gamma,\Delta \gamma}^{\text{con}}$ is an infinite integral, that
needs to be renormalised in the $\MSbar$ scheme to obtain the $\MSbar$
PDF. The renormalised $f_{\gamma,\Delta \gamma}^{\MSbar\, \text{con}}$
was computed in perturbation theory in Eq.~(\ref{6.15}) to order
$\alpha(\mub)$ to reproduce the earlier results in
Eq.~(\ref{eq:master}) and Eq.~(\ref{5.10}). In principle, one can
extend the computation of $f_{\gamma,\Delta \gamma}^{\MSbar\,
  \text{con}}$ to arbitrarily high orders in perturbation theory. In
this section, we do this to order $\alpha(\mub) \alpha_s(\mub)$ and
$\alpha^2(\mub)$, and adopt our ``democratic counting'' for the order
in perturbation theory.
For the sake of simplicity, we will limit ourselves
to the unpolarised case.

We start with the expression Eq.~(\ref{6.113a}) for
$f_\gamma^{\text{con}}(x,\mub^2)$, which we can write in compact form
as
\begin{align}
& f_{\gamma}^{\text{con}}(\chi,\mub^2)
 =  \frac{\alpha(\mub)}{2\pi } \frac{e^{\epsilon
     \gamma_E}}{\Gamma(1-\epsilon)} 
 \int_{xyz,\chi} \int_{1}^\infty  \frac{\mathd s}{s^{1+\epsilon}}    \left[\frac{1}{1- \Pi(-\mub^2 s/(1-z),\mub)}\right]^2  \nn
 & \hspace{1cm} \biggl\{ \sum_{I\in \{2,L\}} \sum_{a\in\qlggset} \left[p_{I} (z) + \epsilon r_{I}(z) \right] \times  \mathscr{F}_{I,\qg} (x, \mub^2 s/(1-z)) ,\mub^2,\epsilon)\,f_\qg(y, \mub^2) \biggr\}\,.
\label{9.1}
\end{align}
In the sum, $I$ is over $2,L$, while $\qg$ is over all partons.
The functions $p_{I}$ and $r_{I}$ are
\begin{align}
p_2(z) &= p_{\gamma q}(z), &
p_L(z) &= - z, &
\\
r_2(z) &= - z, & 
r_L(z) &= z\,.
\label{9.2}
\end{align}
We stress that the expression $[p_I+\epsilon r_I]$ is exact, i.e. it
has no higher order corrections in $\epsilon$.  The coefficient
functions $\mathscr{F}$ are defined in such a way that
\begin{align}
\frac{F_{2, D}(x,Q^2)}{x} &= \int_{zy,x} \sum_{a\in \qlggset\}} \mathscr{F}_{2,\qg} (z, Q^2,\mub^2,\epsilon) f_\qg(y,\mub^2) , \nn
\frac{F_{L, D}(x,Q^2)}{x} &= \int_{zy,x} \sum_{a\in\qlggset}  \mathscr{F}_{L,\qg} (z, Q^2,\mub^2,\epsilon)  f_\qg(y,\mub^2),
\label{9.3}
\end{align}
consistently with the notation of Ref.~\cite{Zijlstra:1992qd}.  They
are perturbatively calculable, since $\mub$ is a hard scale.

We now focus upon the ${\cal O}(\alpha \alpha_s)$ term. We only need
$\qg\in\{q,g\}$.  As usual we define series expansions for
$\mathscr{F}_{2/L,\qg}$ and $f_\gamma^{\text{con}}$,
\begin{align}
  \label{eq:F2LExpansion}
  \mathscr{F}_{2/L,\qg} &= \sum_{i,j=0}^\infty 
                          \left(\frac{\alpha_s}{2\pi}\right)^i 
                          \left(\frac{\alpha  }{2\pi}\right)^j
                          \mathscr{F}_{2/L,\qg}^{(i,j)}\,,
                          \\
  \label{eq:FconExpansion}
  f_\gamma^{\text{con}} &= \sum_{i,j=0}^\infty 
                          \left(\frac{\alpha_s}{2\pi}\right)^i 
                          \left(\frac{\alpha  }{2\pi}\right)^j
                          f_\gamma^{(i,j)}\,.
\end{align}
At lowest order the only non-vanishing coefficient function is
\begin{equation}
\mathscr{F}_{2,q}^{(0,0)}(x) = C^{(0,0}_{2,i}(x),\quad\quad
C^{(0,0}_{2,i}(x)\equiv e_i^2 \delta(1-x)\quad \mbox{for}\;i\in\qlset\,.
\label{9.4}
\end{equation}
For the computation of the ${\cal O}(\alpha_s\alpha)$ corrections we
need the coefficient functions to order $\alpha_s$, including terms of
order $\epsilon$,
\begin{align}
&  \mathscr{F}_{I,\qg}^{(1,0)}(x, Q^2,\mub^2,\epsilon) =
  \left(\frac{\mub^2}{Q^2}\right)^\epsilon \Big(-\frac{1}{\epsilon} \Piqa(x) + C_{I,\qg}^{(1,0)}(x) -  \epsilon  a_{I,\qg}^{(1,0)}(x)\Big)
+ \left[ \frac{1}{\epsilon} \Piqa(x) \right]_{\text{c.t.}}\,,
\label{9.5}
\end{align}
where
\begin{equation}
    \Ptwoqa=\left\{\begin{array}{ll} p_{qq} e_a^2\; & \mbox{for}\; a\in\qset \\[5pt]
                                 p_{qg} \sum_{i\in\qset} e_i^2\; & \mbox{for}\; a=g
                                 \end{array}\right. ,\qquad \PLqa=0\;. 
  \label{9.51}
\end{equation}
The ``$\text{c.t.}$'' suffix on the square bracket is there to remind
us that the enclosed expression is an \MSbar{} counterterm.  The
$C_{I,a}$ and $a_{I,a}$ coefficients (taken from
Ref.~\cite{Zijlstra:1992qd}) are given in App.~\ref{subsec:coeff}.

We can substitute the coefficient functions in Eq.~(\ref{9.1}) and
evaluate the integral. Note that
\begin{align}
  \frac{\mub^2}{Q^2} \to 
  \frac{1-z}{s}\,,
\label{9.16}
\end{align}
so that the $\mathd s$ integration yields a $1/(2\epsilon)$ for all
terms of Eq.~(\ref{9.5}) with the exception of the terms in square
brackets (the counterterms), where it yields a $1/\epsilon$.  The
leading term is order $\alpha(\mub)/(2\pi)$:
\begin{align}
  f^{\substack{(0,1) }}_{\gamma}(\chi,\mub^2) &=\int_{xyz,\chi} \sum_{I\in \{2,L\}} \sum_{\qg\in\qlset} \Biggl\{
 \biggl( \frac{1}{\epsilon} p_I(z)+r_I(z) \biggr) C_{I,\qg}^{(0,0)}(x) \nn
 &+ \left[-\frac{1}{\epsilon} p_I(z) C_{I,\qg}^{(0,0)}(x)
   \right]_{\text{c.t.}}  \Biggr\}  f_{\qg}(y, \mub^2)\,.
\label{9.6}
\end{align}
Explicitly, the finite part is 
\begin{align}
f^{\substack{(0,1) }}_\gamma(x,\mub^2) &=\sum_{a\in\qlset} e_a^2 \int_x^1 \frac{\mathd z}{z} 
\left(-z \right)  f_a(x/z,\mub^2)   \,, 
\label{9.6a}
\end{align}
and gives the result obtained earlier in Eq.~(\ref{6.15a}) when $F_2$
there is replaced by its leading order approximation.
The infinite pieces are cancelled by the counterterms.

The $\alpha (\mub) \alpha_s(\mub)/(2\pi)^2$ piece is
\begin{align}
  f^{\substack{(1,1) }}_{\gamma}(\chi,\mub)
  &=\sum_{I\in\{2,L\}}\sum_{a\in\{q,g\}}\int_{xyz,\chi} \Biggl\{  \frac{1}{2\epsilon^2} p_I(z)  \Piqa(x) \nn
  & + \frac{1}{2\epsilon} \left[(r_I(z)- p_I(z) \log(1-z)) \Piqa(x) + p_i(z) C_{I,\qg}^{(1,0)}(x) \right]
  \nn
  & - \biggl[ \frac12 r_I(z) \log(1-z) + p_I(z) \biggl(
    \frac14 \log(1-z)  +\frac{\pi}{24} \biggr) \biggr]  \Piqa(x) \nn
&    -\frac{1}{2} p_I(z) \log(1-z)  C_{I,\qg}^{(1,0)}(x)  + p_I(z)  \frac12 a_{I,\qg}^{(1,0)}(x)
    -\frac12 r_I(z) C_{I,\qg}^{(1,0)}(x) \Biggr\} f_a(y, \mub^2)\,.
    \label{9.6b}
\end{align}
The infinite terms of Eq.~(\ref{9.6b}) are cancelled by the $\MSbar$
counterterms for the photon PDF.  The finite part of Eq.~(\ref{9.6b})
gives instead the $\MSbar$-conversion factor at order $\alpha
\alpha_s$.  In explicit form, we get:
\begin{align}
& f^{\substack{(1,1) }}_\gamma(\chi,\mub^2) = \sum_{\qg\in\qgset} \int_{xyz,\chi} \Bigg\{
 \frac14\Bigg[    2  z  \log(1-z)   - p_{\gamma q}(z) \Bigg(  \log^2(1-z)
  + \frac{\pi^2}{6} \Bigg)
\Bigg]
 \Ptwoqa(x) \nn
 &\phantom{aaa} +\frac12 \left[ p_{\gamma q}(z) \log(1-z)  - z \right]
 C^{(1,0)}_{2,\qg}(x) -\frac12 z \left[  \log(1-z) -1 \right]  C^{(1,0)}_{L,\qg}(x) \nn
&\phantom{aaa} -\frac12 p_{\gamma q}(z)  a^{(1,0)}_{2,\qg}(x)  
 + \frac12 z  a^{(1,0)}_{L,\qg}(x) \Bigg\} f_\qg(y, \mub^2)\,.
 \label{eq:fcon11}
\end{align}
Differentiating the finite parts of the photon PDF at order $\alpha
\alpha_s$ gives the $\alpha \alpha_s^2$ splitting functions, following
the procedure in Sec.~\ref{sec:split-fns}. The expression is lengthy,
involving triple convolutions, and is not given explicitly here.

The $\alpha^2$ corrections are given by the same equations, with the
QCD corrections to the splitting functions and to the coefficient
functions replaced by the QED ones. The only new feature in the QED
correction is the vacuum polarisation contribution from $\aph$,
\begin{align}
\Pi_D(q^2,\mub) &= \frac{\alpha(\mub^2)}{4 \pi} b_{\rm qed}^{(0,1)} \left[ \frac{6\,\Gamma^2(2-\epsilon)\Gamma(\epsilon) e^{\epsilon \gamma_E}}{\Gamma(4-2\epsilon)} \left( \frac{\mub^2}{-q^2} \right)^\epsilon - \left[\frac{1}{\epsilon}\right]_{\text{c.t.}}\right] \,,
\label{9.13}
\end{align}
which gives an additional contribution
\begin{align}
&  f^{\substack{(0,2), \text{vac} }}_{\gamma}(\chi,\mub^2)   = \sum_{\qg\in\qlset}\int_{xyz,\chi}  b_{\rm qed}^{(0,1)}
  \biggl\{ - \frac{p_2(z) }{2\epsilon^2} + \frac{1}{6\epsilon}\left(p_2(z)(5+3\log(1-z))-3r_2(z) \right) \nn
  & + p_2(z) \bigg(\frac{1}{4} \log^2(1-z)  +\frac{14}{9}
  +\frac{5}{6}\log(1-z) \bigg) \nn
  & + r_2(z) \left(\frac{1}{2}\log(1-z)+\frac{5}{6}\right)  \biggl\} C_{2,\qg}^{(0,0)}(x) f_\qg(y, \mub^2)\,.
\end{align}
The final result for the $f_\gamma^{\substack{(0,2)}}$, reinstating the sum over all flavours,
and remembering that also leptons contribute in the democratic counting, is
  \begin{align}
    f^{\substack{(0,2)}}_\gamma(\chi,\mub) =& \sum_{a\in \qlgset} \int_{xyz,\chi} \Biggl\{
    \frac14\Bigg[    2  z  \log(1-z)   - p_{\gamma q}(z) \Bigg(  \log^2(1-z)
      + \frac{\pi^2}{6} \Bigg)
      \Bigg]
    \Ptwoqael(x) \nn
    & +\frac12 \left[ p_{\gamma q}(z) \log(1-z)  - z \right]
    C^{(0,1)}_{2,a}(x) -\frac12 z \left[  \log(1-z) -1 \right]  C^{(0,1)}_{L,a}(x) \nn
    & -\frac12 p_{\gamma a}(z)  a^{(0,1)}_{2,a}(x)  
    + \frac12 z  a^{(0,1)}_{L,a}(x)\Biggr\} f_a(y, \mub^2)\, 
      \nn
&  + \sum_{a\in \qlset} b_{\rm qed}^{(0,1)} e_a^2 \int_{yz,\chi} \biggl[
    \frac14 p_{\gamma q}(z)  \log^2(1-z) + \frac56 p_{\gamma q}(z)  \log(1-z) \nn
    & + \frac{14}{9} p_{\gamma q}(z)  - \frac12 z \log(1-z) - \frac56 z \biggr] f_a(y, \mub^2)\,,
    \label{eq:fcon02}
\end{align}
where
\begin{equation}
  \Ptwoqael=\left\{\begin{array}{ll} p^{\rm qed}_{qq} e_a^4\; & \mbox{for}\; a\in\qlset \\[5pt]
                                 p_{q\gamma} \sum_{i\in\qlset} n_i e_i^4\; & \mbox{for}\; a=\gamma.
                                 \end{array}\right.
  \label{9.52}
\end{equation}

In summary, the final result for the photon PDF is
\begin{subequations}
  \label{9.11}
  \begin{align}
    \label{9.11a}
    f_{\gamma} (x,\mub^2) 
    &= f^{\text{PF}}_{\gamma} (x,\mub^2) + f^{\text{con}}_\gamma(x,\mub^2) \,,
      \\
    \label{9.11b}
    &= f^{\text{PF}}_{\gamma} (x,\mub^2) 
      + \!\! \sum_{r \ge 0,s \ge 1} \left( \frac{\alpha_s}{2\pi}\right)^r 
      \left( \frac{\alpha}{2\pi}\right)^s  f^{\substack{(r,s)  }}_\gamma(x,\mub^2)\,.
  \end{align}
\end{subequations}
The $f^{\text{PF}}_{\gamma}$ contribution was given in
Eq.~(\ref{6.12a}), while the $f_\gamma^{(0,1)}$, $f_\gamma^{(1,1)}$
and $f_\gamma^{(0,2)}$ terms are given in Eqs.~(\ref{9.6a},
\ref{eq:fcon11}, \ref{eq:fcon02}).
For $s\ge 2$, the r.h.s.\ contains
the photon PDF $f_{\gamma}$, and Eq.~(\ref{9.11}) can be solved
iteratively for $f_\gamma$.  The QCD correction can be included by
simply adding the $f^{\substack{(r,s) }}_\gamma$ term to the final
result. Including QED corrections is more difficult. The formalism
presented here uses the one-photon-irreducible structure functions.
To include QED corrections in a systematic way without double-counting
requires experimental knowledge of the elastic form factors and DIS
structure functions in the one-particle irreducible definition. This
requires removing two-photon-exchange (TPE) contributions to the
scattering cross sections. However, electromagnetic corrections only
on the hadronic side should not be removed. This makes the analysis
simpler than the usual one presented in the literature, in which QED
corrections on the hadronic side are also removed (see,
e.g.\ \cite{Bernauer:2013tpr}). Furthermore, $\aph(Q^2)$ should also
be evaluated including corrections of order $\alpha^2$ (in our NLO
result it was enough to include corrections of order $\alpha^2 L$).
This requires the knowledge of the hadronic vacuum polarisation, that,
on the other hand, is extracted from $e^+e^-$ data (see the review on
electroweak physics in Ref.~\cite{Olive:2016xmw}).

In analogy with our analysis in Sec.~\ref{sec:split-fns}, the
$f_\gamma^{(1,1)}$ and $f_\gamma^{(0,2)}$ coefficients, together with
the two-loop QCD and QED coefficient functions could be used to
obtain the $P_{\gamma i}^{(r,s)}$ splitting functions for $r+s=3$ and
$s\ge 1$.
We note that the complementary $P_{q \gamma}^{(2,1)}$ and
$P_{g \gamma}^{(2,1)}$ splitting functions have been given in
Ref.~\cite{Vogt:2005dw}. 

\section{Inputs for the unpolarised photon distribution}
\label{sec:input-data}

To evaluate the photon parton density we require information on the
$F_2$ and $F_L$ structure functions over the full $x,Q^2$ kinematic
range.
Our evaluation will be up to the accuracy outlined in
section~\ref{sec:probe}, i.e.\ using $L \sim \ln \mu^2/m_p^2$, we
include terms $\alpha L (\as L)^n$ at lowest order and $\alpha (\as
L)^n$, $\alpha^2 L^2 (\as L)^n$ corrections at higher order.
As we proceed we will highlight issues that arise if one wishes to go
to higher accuracy.

The $F_2$ and $F_L$ structure functions are most commonly determined
from electron--proton scattering data.
A first comment concerns the treatment of electromagnetic corrections
to the structure functions.
We use the prescription described in Sec.~\ref{sec:em}, whereby all
electromagnetic corrections to the interaction of a proton with a
single photon are included, with the exception of the photon vacuum
polarisation contributions.
The treatments of electromagnetic corrections to data differ in their
details according to the kinematic region considered. However two
features emerge: the experimental analyses always correct for
electromagnetic radiation from the incoming electron and they always
take out the photon vacuum polarisation contributions.
These are the two elements that are required for consistency with our
accuracy.
A more detailed discussion of QED corrections to DIS measurements is
given in section~9 of Ref.~\cite{Blumlein:2012bf}.

We separate the data inputs according to the kinematic region and the
corresponding final state in $ep$ scattering.
The main kinematic variables for the separation will be $Q^2$ and
$W^2$ where
\begin{equation}
  \label{eq:Wsq}
  W^2 = m_p^2 + \frac{1-\xbj}{\xbj} Q^2 \,, 
\end{equation}
is the squared invariant mass of the outgoing system associated with
the hadronic side of the collision.

\subsection{Elastic contribution}

In our definition, the elastic contribution corresponds to the region
of $W < \mpr + m_{\pi^0}$.
In particular it includes configurations where one or more photons are
radiated from the proton.%
\footnote{For the determination of the
  structure functions we find it useful to think of a process in which
  there can be at most one exchanged photon between the probe and the
  proton, as in the process of Sec.~\ref{sec:probe}.
  In actual electron-proton scattering experiments there can be two or
  more exchanged photons, either real of virtual.
  These corrections are beyond our accuracy and cannot be classified
  in terms of the usual electromagnetic structure functions, since
  they correspond to a more complex tensor structure.}
Experimental data on elastic scattering is usually corrected for
radiation from the proton, since the measurements are performed with
the goal of extracting the electric and magnetic Sachs form factors of
the proton, $G_E$ and $G_M$ respectively.
The correspondence between the form factors and the structure
functions (see e.g.\ Eqs.(19) and (20) of Ref.~\cite{Ricco:1998yr}) is
given by
\begin{subequations}
\label{eq:F2L-elastic}
\begin{align}
  F_2^\text{el}(\xbj,Q^2) &= \frac{[G_E(Q^2)]^2 + [G_M(Q^2)]^2 \tau}{1+\tau} \delta(1-\xbj)\,,
  \\
  F_L^\text{el}(\xbj,Q^2) &= \frac{[G_E(Q^2)]^2}{\tau} \delta(1-\xbj)\,,
\end{align}
\end{subequations}
where $\tau = {Q^2}/{(4m_p^2)}$.
The approximation of $F_2$ and $F_L$ as $\delta$-functions neglects
precisely the photon radiation from the proton.
However, most of the radiation is soft and so cancels in inclusive
quantities and all that changes when going beyond the
$\delta$-function approximation is a relative $\order{\alpha}$
correction (free of any logarithms) to the photon distribution, which
is beyond our accuracy. Collinear logarithms are absent since the
proton form factors fall off rapidly with $Q^2$.

Substituting Eq.~(\ref{eq:F2L-elastic}) into Eq.~(\ref{eq:master}), one
obtains  
\begin{multline}
  x f^\text{el}_{\gamma}(x,\mu^2)
  =\frac{1}{2 \pi \alpha (\mu^2)} \int_{\frac{x^2 m_p^2}{1 - x}}^\infty
      \frac{\mathd Q^2}{Q^2} \aph^2 (-Q^2) \Bigg\{ \left(
      1 - \frac{x^2 m_p^2}{Q^2 (1 - x)} \right) \frac{2 (1 - x) G_E^2 (Q^2)}{1 +
      \tau}  \\
  + \left( 2 - 2 x + x^2 + \frac{2 x^2 m_p^2}{Q^2} \right) \frac{G_M^2 (Q^2)
      \tau}{1 + \tau} \Bigg\}\,,
      \label{eq:elastic-component}
\end{multline}
for the elastic component of the photon distribution.

Note that we have set the upper limit of the integration to infinity,
rather than make it $\mu^2$-dependent as in Eq.~(\ref{eq:master}).
For large $\mu^2$ values, because of the $~1/Q^4$ scaling of the form
factors (see below), the resulting difference is a higher-twist
effect.
Using an infinite upper limit ensures the absence of higher-twist
contamination.
For the same reason, the last term in Eq.~(\ref{eq:master}) (the
\MSbar{}-conversion) does not appear in
Eq.~(\ref{eq:elastic-component}).

A widely used approximation for the $G_{E,M}$ form factors is the
dipole form,
\begin{equation}
  \label{eq:dipole-form-factor}
  G_E^\text{dip}(Q^2) = \frac{1}{(1+Q^2/m_\text{dip}^2)^{2}} \,,\qquad
  G_M^\text{dip}(Q^2) = \mu_p G_E(Q^2)\,,
\end{equation}
where $m_\text{dip}^2=0.71\GeV^2$ and $\mu_p\simeq 2.793$ is the
anomalous magnetic moment of the proton.
For $Q^2 = 0$ this form yields the exact results $G_E(0) = 1$ and
$G_M(0)=\mu_p$, while elsewhere it is an approximation.

The dipole form is of interest in part because it provides insight
into the asymptotic behaviours of the elastic component of the photon
distribution.
At small $x$, to leading order in $\alpha$, one finds
\begin{equation}
  \label{eq:elastic-small-x}
  x f^\text{el}_{\gamma}(x,\mu^2) = \frac{2\alpha}{\pi} \left( \ln
    \frac{1}{x} + \order{1}\right)\,,
\end{equation}
dominated by the electric component. 
This behaviour is correct even beyond the dipole model.
At large $x$, the magnetic component dominates and one finds,
\begin{equation}
  \label{eq:elastic-large-x}
  x f^\text{el,dip}_{\gamma}(x,\mu^2) =
  \frac{\alpha^2\!\left(\frac{\mpr^2}{1-x}\right)}{8\pi \alpha(\mu^2)}
  \frac{\mu_p^2 m_\text{dip}^8}{\mpr^8} 
  (1-x)^4
  + \order{(1-x)^5}\,,
\end{equation}
including running coupling effects.
The scaling as $(1-x)^4$ holds insofar as the true high-$Q^2$
behaviour of the magnetic form factor remains $G_M(Q^2) \sim
1/Q^4$. 

\begin{figure}
  \centering
  \includegraphics[width=0.55\textwidth,page=2]{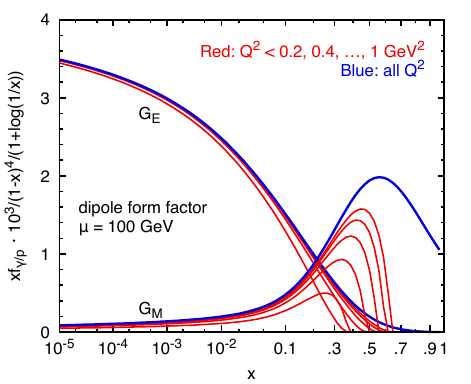}
  \caption{The electric and magnetic elastic components of the photon
    distribution, with the dipole form factor approximation, placing a
    cut on the allowed $Q^2$ integration range, to help illustrate
    which $Q^2$ values contribute.
    The result is normalised so as to divide out the leading $\ln 1/x$
    and $(1-x)^4$ behaviours at small and large-$x$ respectively.
  }
  \label{fig:Q2-decomposition-elastic-dipole}
\end{figure}

The scaling behaviours are illustrated in
Fig.~\ref{fig:Q2-decomposition-elastic-dipole}, which shows the
electric and magnetic elastic contributions to the photon
distribution, normalised to the asymptotic small and large-$x$ trends,
and obtained with the dipole approximation to the form factors.
The red curves show the elastic contribution with an upper $Q^2$
cutoff of $0.2-1.0$\,$\text{GeV}^2$, in steps of
$0.2$\,$\text{GeV}^2$, and the blue curve shows the total contribution
with no $Q^2$ cutoff.
The dominance of the elastic component is clear at small $x$ and the
magnetic component takes over for $x \gtrsim 0.2$.
For phenomenologically interesting $x$ values, $x \lesssim 0.5$, most
of the contribution comes from the region of $Q^2 < 1 \GeV^2$.

\begin{figure*}
  \centering
  \includegraphics[page=1,width=0.5\textwidth]{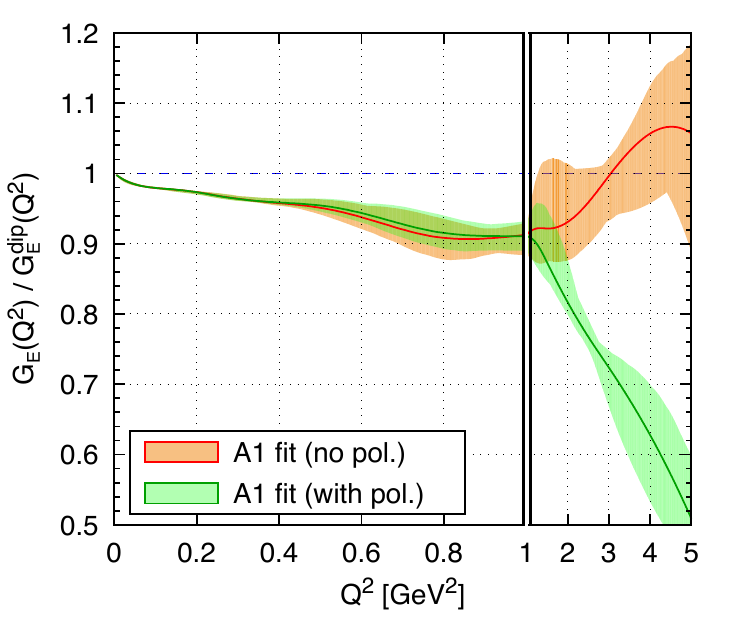}%
  \hfill%
  \includegraphics[width=0.5\textwidth]{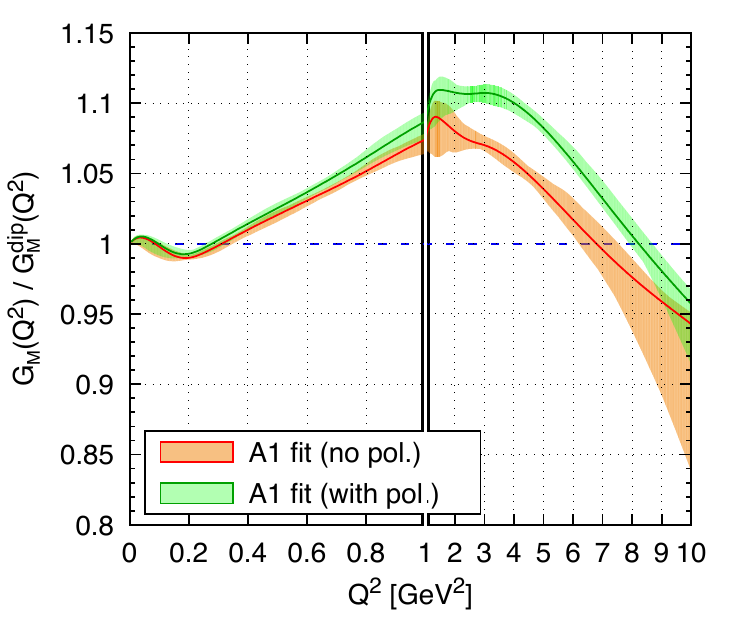}%
  \caption{Elastic form factors (ratio to standard dipole form) for
    the electric (left) and magnetic (right) as fitted by the A1
    collaboration~\cite{Bernauer:2013tpr} with and without polarised
    data. 
    Note the change in scale at $Q^2=1\GeV^2$ along the $x$ axis.
  }
  \label{fig:elastic-FF}
\end{figure*}

For accurate results, the dipole approximation,
Eq.~(\ref{eq:dipole-form-factor}), is not sufficient.
The most recent extensive experimental study of the form factors comes
from the A1 collaboration~\cite{Bernauer:2013tpr}.
The A1 data itself is limited to $Q^2 \lesssim 1\GeV^2$, however the
work includes fits to global data up to $Q^2 \sim 10\GeV^2$.
The electric and magnetic form factor fits are shown in
Fig.~\ref{fig:elastic-FF}, normalised to the dipole form.
The A1 paper includes two classes of fits, one for just unpolarised
data, and one that also includes polarised data.\footnote{In each case
  we use the fit labelled ``SplinesWithVariableKnots'' as distributed
  with the arXiv version of the paper~\cite{Bernauer:2013tpr}.}
Both fits show clear deviations from the dipole form. 
The fit with polarised data is the recommended default because it
provides additional constraints on two-photon-exchange (TPE)
contributions and we take it as our default.
Given the delicacy of the treatment of the TPE contribution, we use the
central value of the unpolarised fit as an error estimate that comes in
addition (in quadrature) to the quoted uncertainty on the global
polarised fit.
Note that the fits extend only up to $Q^2 = 10\GeV^2$ and beyond this
point we use the dipole shape, normalised to the fitted $G_{E/M}(Q^2)$
at $Q^2=10\GeV^2$. 
We treat the fit uncertainties on the elastic and magnetic components
as $100\%$ correlated, which is the most conservative assumption,
because they both enter with the same sign in
Eq.~(\ref{eq:elastic-component}). 

\begin{figure}
  \centering
  \includegraphics[width=0.48\textwidth,page=2]{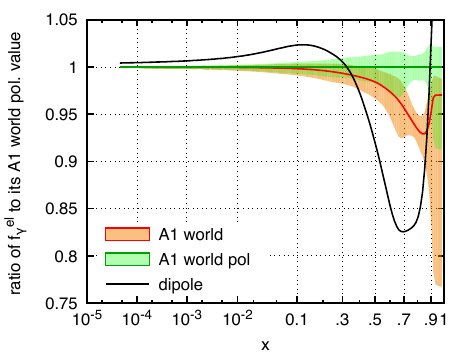}\hfill
  \caption{Elastic contribution to $f_{\gamma}(x,Q^2)$ using various
    fits for the form factors, normalised to the result obtained with
    the A1 world fit including polarised data.
    The ratio for the the A1 world fit freezes above $x=0.9$ because
    the A1 fits extends only up $Q^2 = 10\GeV^2$ and beyond that scale
    we simply extrapolate the results for $G_{E/M}(Q^2)$ using the
    standard-dipole shape, normalised to $G_{E/M}(10\GeV^2)$.
   }
  \label{fig:elastic-component-ratios}
\end{figure}

The impact of the A1 fits on the elastic contribution, relative to the
dipole form, is shown in Fig.~\ref{fig:elastic-component-ratios}. 
The deviation from the dipole form is manifest, with an impact of up
to $10\%$ for $x \lesssim 0.5$. 
In this $x$-range, the difference between the results using fits
with and without polarised data is smaller than the intrinsic
uncertainty bands for the individual fits.
Note that since the data constrains the form factors for
$Q^2 \lesssim 10\GeV^2$, the elastic contribution is effectively known
only for $x \lesssim 0.9$, cf.\ the form of the lower limit in
Eq.~(\ref{eq:elastic-component}).

We have also computed the elastic contribution using a preliminary
unpublished fit to the elastic form factors by Lee, Arrington and
Hill.\footnote{We thank G.~Lee and R.~Hill for providing us with fit
  files for the form factors (see Ref.~\cite{Lee:2015jqa} for a
  discussion of their approach).}
There is some change in the elastic contribution to $f_\gamma$, which
is negligible for small $x$, and becomes comparable to our overall
total error (for the elastic plus inelastic components) for $x\gtrsim
0.5$.

\subsection{Low-$Q^2$ region}

In addition to the elastic component, we need the inelastic part.
This corresponds to the region of $W > \mpr + m_{\pi^0}$.
We split the inelastic part into several sub-regions.
At low $Q^2$, the structure functions cannot be computed
from parton distribution functions and we must
rely on direct measurements and theoretical model-independent
constraints.

Results for low-$Q^2$ scattering tend to be given either as a
differential scattering cross section or in terms of some subset of
$F_2$, $F_L$, $R_{L/T}$, $\sigma_T$ and $\sigma_L$, only two of which
are independent.
They are related through the following equations (using the same
normalisation convention adopted by the HERMES collaboration in
Ref.~\cite{Airapetian:2011nu}),
\begin{subequations}
  \label{eq:FL-F2-R}
  \begin{align}
    \label{eq:FL-FN-R}
    F_L(x,Q^2) &= 
                 F_2(x,Q^2) \left(1 + \frac{4m_p^2 x^2}{Q^2} \right)
                 \frac{R_{L/T}(x,Q^2)}{1 + R_{L/T}(x,Q^2)}\,,
    \\
    \label{eq:F2-from-sigmaTL}
    F_2(x,Q^2) &= \frac{1}{4\pi^2 \alpha} 
                 \frac{Q^2 (1-x)}{1 + \frac{4x^2 m_p^2}{Q^2}}
                 \left(\sigma_T(x,Q^2) + \sigma_L(x,Q^2)\right)\,,
    \\
    \label{eq:R-sigmaTL}
    R_{L/T}(x,Q^2) &= \sigma_L(x,Q^2)/\sigma_T(x,Q^2)\,.
  \end{align}
\end{subequations}
As discussed in App.~\ref{sec:low-Q2-F2-FL}, 
$F_2$ and $F_L$ vanish respectively as $Q^2$ and
$Q^4$ in the limit of small $Q^2$ and fixed $W^2$.
In this limit, as one can see from Eq.~(\ref{eq:Wsq}), $x$ goes to
zero in proportion to $Q^2$.
The quantity $\sigma_T(x,Q^2)$ becomes a function $W^2$ only and it
follows from Eq.~\eqref{eq:FL-F2-R} that $\sigma_L(x,Q^2)$ must vanish
as $Q^2$, leading to
\begin{equation}
  F_2(x,Q^2)  \underset{Q^2\to 0}\Longrightarrow \frac{Q^2}{4\pi^2 \alpha} \sigma_T(W^2)\,, 
  \label{eq:photoprodlim}
\end{equation}
where $\sigma_T(W^2)$ is the photoproduction cross section. 
As a result, in Eq.~(\ref{eq:master}), $Q^2$ scales much below
the proton mass will not give a substantial contribution.

\begin{figure}
  \centering
  \includegraphics[width=0.55\textwidth]{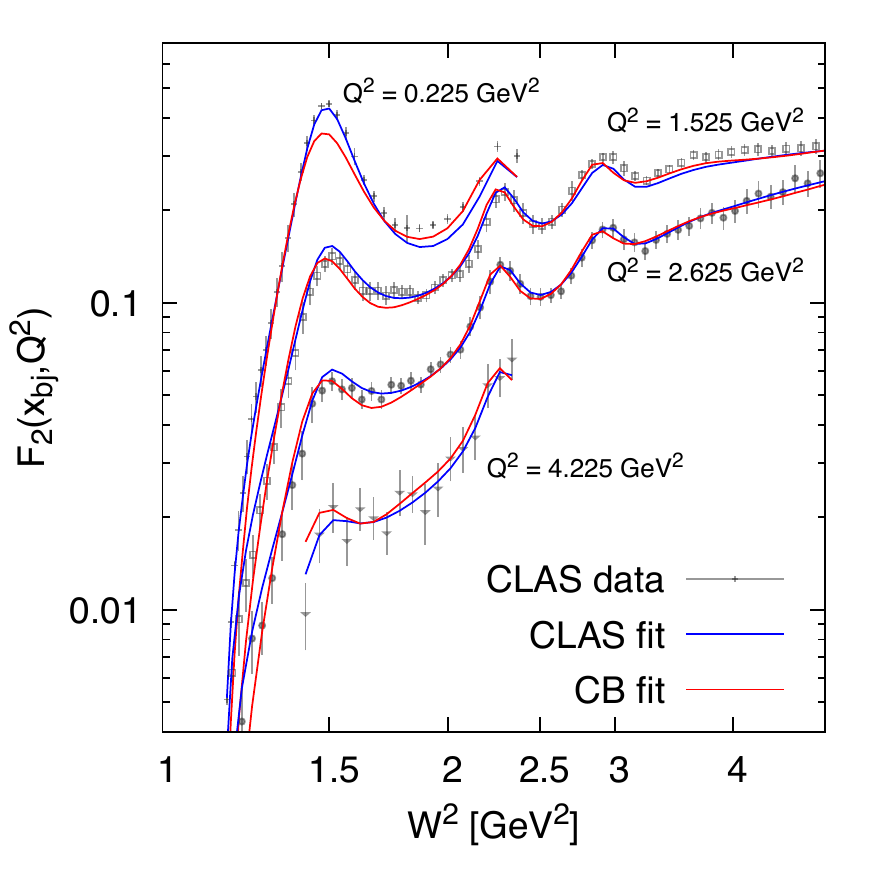}\hfill%
  \includegraphics[width=0.44\textwidth]{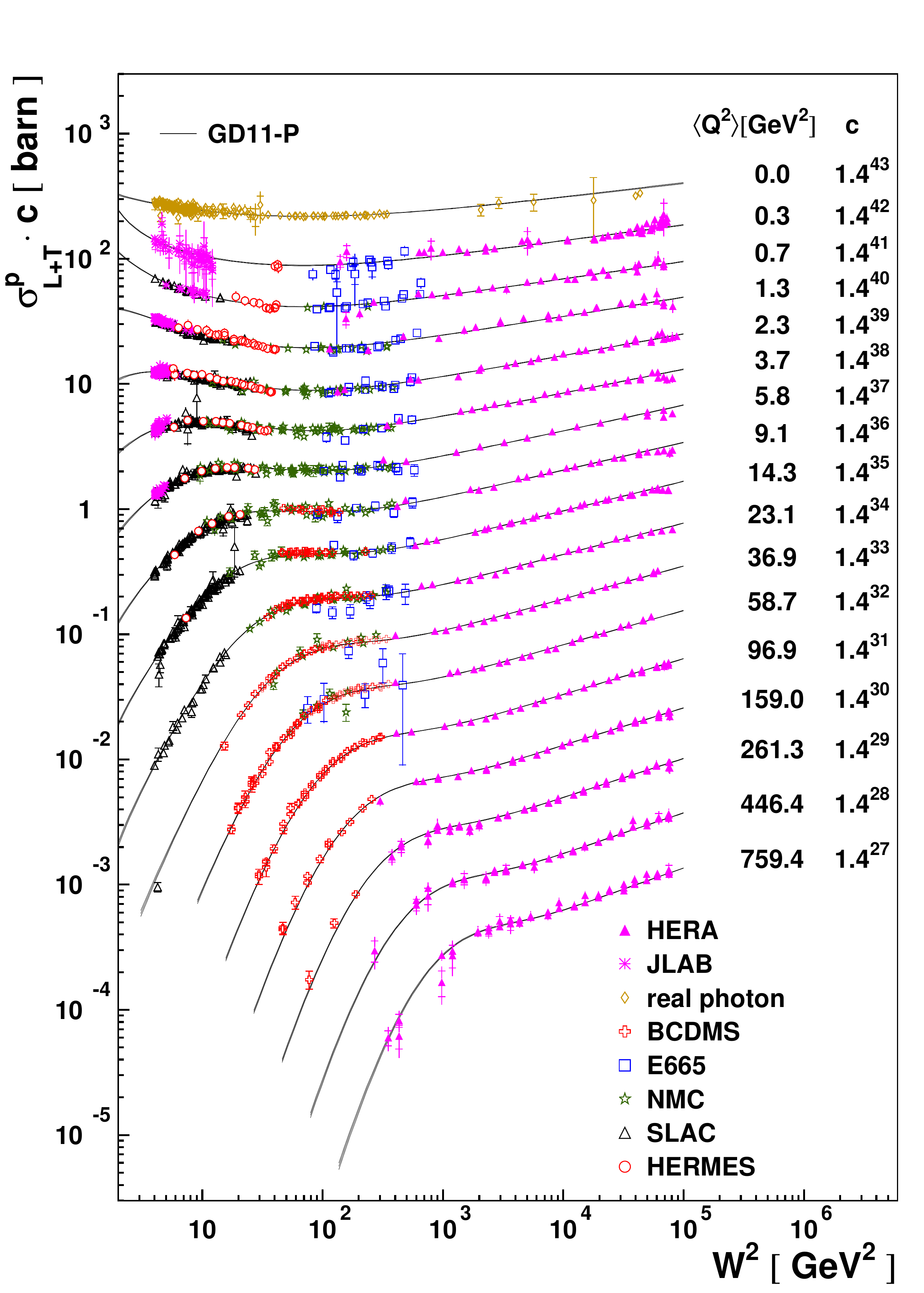}
  \caption{Left: illustration of a subset of the CLAS
    data~\cite{Osipenko:2003bu} in the resonance region, compared to
    two fits, one from the CLAS paper and the other from Christy and
    Bosted (CB)~\cite{Christy:2007ve}.
    The CLAS dataset covers the range $0.225 \le Q^2/\text{GeV}^2 \le 4.725$ in
    steps of $0.05$, with a quality comparable to that shown in the plot across
    the whole $Q^2$ range.
    The errors on the data correspond to the sum in quadrature of
    statistical and systematic components.
    The CLAS data is only a small part of the data that is available
    in the resonance region and used for the fits (see  also Fig.~6 of
    Ref.~\cite{Osipenko:2003bu}).  
    Right: illustration of the GD11-P fit from the HERMES
    collaboration~\cite{Airapetian:2011nu} and corresponding data in
    the continuum region
    (plot reproduced with kind permission of the HERMES
    collaboration). 
}
  \label{fig:clas-data-v-fits}
\end{figure}

There is wealth of data covering the low $Q^2$ region, including also
 photoproduction data through the constraint Eq.~\eqref{eq:photoprodlim}. 
Rather than using these data directly, we will rely on existing fits
of those data.
Fits generally focus upon either the resonance region, $W^2 \lesssim
3\GeV^2$ or the continuum region, $W^2 \gtrsim 4\GeV^2$.
Fig.~\ref{fig:clas-data-v-fits} (left) shows data from the CLAS
experiment~\cite{Osipenko:2003bu}, compared to two global fits to
resonance region data.
The figure (which includes only a small subset of the available data)
illustrates the coverage in $Q^2$ and the quality of the available
data.
The data is shown as a function of $W^2$ in order to clearly
show the resonance peaks starting with the $\Delta$  resonance and beyond.
The CLAS fit is intended for use only for $Q^2 > 0.5\GeV^2$, while
the Christy-Bosted~\cite{Christy:2007ve} fit is intended for use down
to $Q^2 = 0$ and explicitly includes photoproduction data.
Comparing the CLAS and Christy-Bosted fits at $Q^2$ values below the
quoted validity range of the CLAS fit shows, however, that they are
relatively similar.

For the continuum region, the HERMES collaboration has provided a fit,
GD11-P~\cite{Airapetian:2011nu}, using a wide range of data and the
ALLM~\cite{Abramowicz:1991xz} functional form.
Fig.~\ref{fig:clas-data-v-fits} (right), taken from
Ref.~\cite{Airapetian:2011nu}, illustrates the good quality of the
fit. 
Careful inspection of the figure reveals that at each $Q^2$ value the
fit consists of three lines, whose separation represents the
uncertainty. 

\begin{figure}
  \centering
  \includegraphics[width=0.5\textwidth]{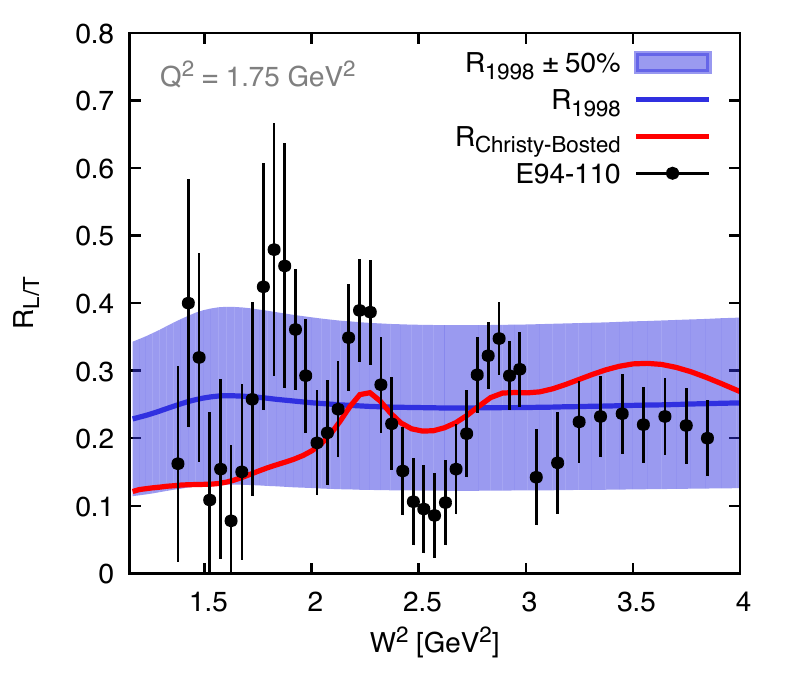}%
  \includegraphics[width=0.5\textwidth]{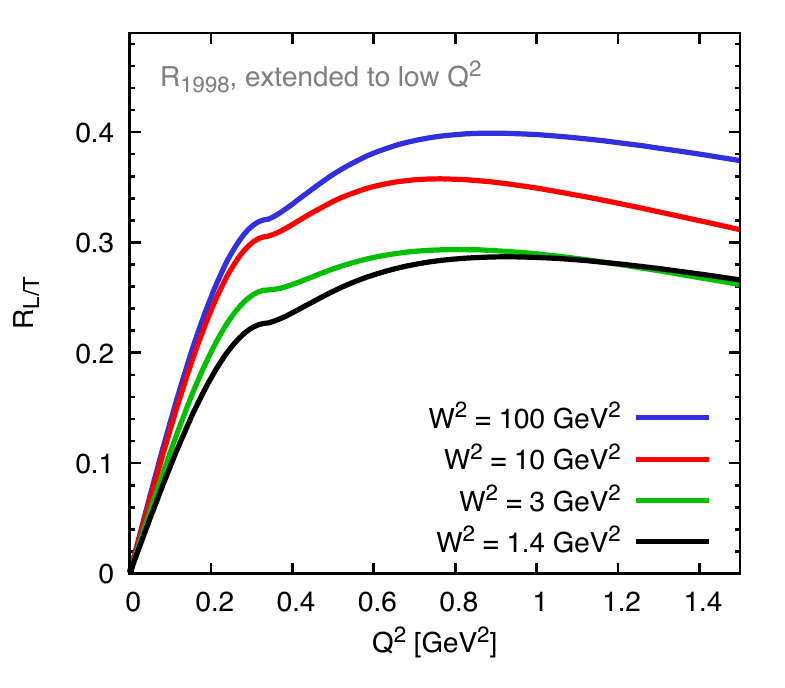}
  \caption{Left: the $R_{1998}$ and Christy-Bosted fits for $R_{L/T}$ as a
    function of $W^2$, compared to data from the E94-110 experiment at
    JLAB~\cite{Liang:2004tj}.
    The data corresponds to version 2 of the arXiv submission of the
    E94-110 article, which postdates the Christy-Bosted fit.
    Also shown is the uncertainty band that we adopt for the
    $R_{1998}$ fit.
    Right: the $Q^2$ dependence for the $R_{1998}$ fit at various
    $W^2$ values, including our extension to low $Q^2$, Eq.~(\ref{eq:R1998mod}).
}
  \label{fig:R-v-W2}
\end{figure}

Electron-proton scattering cross sections give information on both
$F_2$ and $F_L$, with the former often dominating the cross section.
As a result the knowledge of $R_{L/T}$ in Eq.~(\ref{eq:FL-FN-R}) is
often much poorer than the knowledge of $F_2$.
Some of the $F_2$ fits (CLAS, GD11-P) rely on earlier independent fits
for $R_{L/T}$, notably the $R_{1998}$~\cite{Abe:1998ym} fit and that of
Ref.~\cite{Ricco:1998yr}. 
Instead the Christy-Bosted article carried out an independent fit for
$R_{L/T}$. 
Fig.~\ref{fig:R-v-W2} (left)
shows the $R_{1998}$ and Christy-Bosted fits for
$R_{L/T}$, compared to a subsequent extraction of $R_{L/T}$ by the
E94-110 collaboration~\cite{Liang:2004tj}. 
The moderate degree of agreement between data and fits will motivate
us to assign a generous $\pm50\%$ uncertainty to $R_{L/T}$ in the
low-$Q^2$ region.

As a default we will adopt the $R_{1998}$ fit for $Q^2 > Q_b^2$ with
$Q_b^2 = 0.34\GeV^2$.
The fit is, however, not intended for use at low values of $Q^2$, and
so for $Q^2 < Q_b^2$ we will use the form 
\begin{equation}
  \label{eq:R1998mod}
  R_{1998,\text{low-}Q^2}(W^2, Q^2) = R_{1998}(W^2, Q_b^2)\,
  \frac{3u-u^3}{2}\,,\qquad u \equiv \frac{Q^2}{Q_b^2},
\end{equation}
that is continuous and reasonably smooth at $Q_b^2$ and vanishes as
$Q^2$ for $Q^2 \to 0$. This is shown in Fig.~\ref{fig:R-v-W2} (right).

No single low-$Q^2$ fit for $F_2$ is simultaneously adequate in the resonance
and continuum regions.
Accordingly we will combine $F_2$ resonance ($F_2^\text{res}$) and
continuum ($F_2^\text{cont}$) fits using two transition scales
$W^2_\text{lo} = 3\GeV^2$ and $W^2_\text{hi} = 4 \GeV^2$
\begin{equation}
  \label{eq:resonance-continuum-combination}
  F_2(x,Q^2) = \left\{
    \begin{array}{lll}
      F_2^\text{res} 
      & & W^2 < W^2_\text{lo}\,,\\[5pt]
      (1-\rho(W^2)) F_2^\text{res} + \rho(W^2) F_2^\text{cont} 
      & & W^2_\text{lo} < W^2 < W^2_\text{hi}\,,\\[5pt]
      F_2^\text{cont} 
      & & W^2 > W^2_\text{hi}\,,\\
    \end{array}
  \right.
\end{equation}
where $W^2$ is evaluated as in Eq.~(\ref{eq:Wsq}) and $\rho(W^2)$ is 
\begin{equation}
  \label{eq:rho_Wsq}
  \rho(W^2) = 2\omega^2 - \omega^4\,,\qquad 
  \omega = \frac{W^2 - W^2_\text{lo}}{W^2_\text{hi} - W^2_\text{lo}}\,.
\end{equation}
This ensures a continuous and smooth transition between the low- and
high-$W^2$ regions.

We will consider two combinations of low and high-$W^2$ fits for
$F_2$: our main one will be CLAS with GD11-P, using $R_{1998}$ for
both (including the modification Eq.~(\ref{eq:R1998mod})).
As a cross-check of uncertainties we will have a subsidiary
combination, composed of Christy-Bosted and GD11-P for $F_2$, and
Christy-Bosted $R_{L/T}$ at low $W^2$ and $R_{1998}$ at high-$W^2$,
with the two $R_{L/T}$ fits combined in a manner identical to
Eq.~(\ref{eq:resonance-continuum-combination}).

\begin{figure}
  \centering
  \includegraphics[width=0.49\textwidth,page=1]{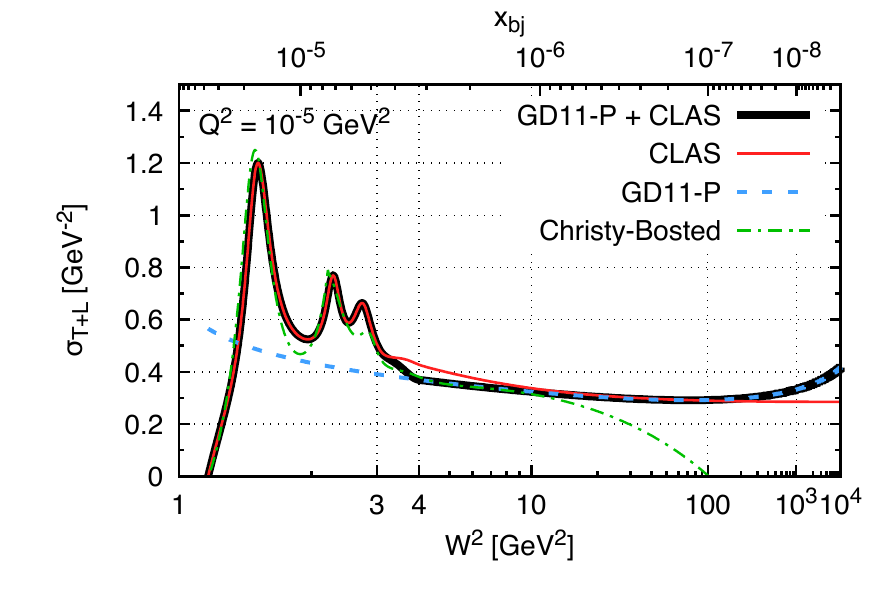}\hfill%
  \includegraphics[width=0.49\textwidth,page=2]{Figs/check-low-Q2-paper.pdf}\\
  \includegraphics[width=0.49\textwidth,page=3]{Figs/check-low-Q2-paper.pdf}\hfill%
  \includegraphics[width=0.49\textwidth,page=4]{Figs/check-low-Q2-paper.pdf}
  \caption{The sum of $\sigma_T+\sigma_L$,
    cf.\ Eq.~(\ref{eq:F2-from-sigmaTL}), shown as a function of $W^2$
    (lower axis) and $x$ (upper axis) for various $Q^2$ values.
    Our default prescription corresponds to the curve labelled
    ``GD11-P + CLAS'', which transitions smoothly between the GD11-P
    and CLAS fits in the region of $W^2$ between $3$ and $4\GeV^2$.  }
  \label{fig:low-Q2-check}
\end{figure}

Our main combination is shown in Fig.~\ref{fig:low-Q2-check}, compared
to the individual CLAS, Christy-Bosted and GD11-P results.
The need for a combination of more than one fit is evident from the
fact that the GD11-P fit misses the resonance structures below
$4\GeV^2$, while the resonance fits miss the rise at large-$W^2$.
For all $Q^2$ values shown, the transition region between $3$ and
$4\GeV^2$ is reasonably covered.
Note also the reduced significance of the resonance peaks at
high-$Q^2$, though the Christy-Bosted fit appears to show artefacts in
this regard near $W^2=3\GeV^2$.
For this reason, when we use the Christy-Bosted fit in the region of
$Q^2 > Q_{0,\text{CB}}^2 = 8\GeV^2$ we will modify its large-$Q^2$
behaviour as follows\\
\begin{subequations}
  \begin{align}
    \label{eq:modified-CB}
    F_2^\text{CB,mod}(W^2,Q^2) &= 
          F_2^\text{CB}(W^2,Q^2_\text{mod}) \times
                                 \left(
                                 \frac{Q_{1,\text{CB}}^2}{
                                 Q_{1,\text{CB}}^2 + \Delta Q^2}
                                 \right)
                                 \,,
    \\
    Q^2_\text{mod} &= Q_{0,\text{CB}}^2 +
                     \frac{\Delta Q^2}{1 + \frac{\Delta Q^2}{Q_{1,\text{CB}}^2 -
                     Q_{0,\text{CB}}^2}}\,,
    \\
    \Delta Q^2 &= Q^2 - Q_{0,\text{CB}}^2\,.
  \end{align}
\end{subequations}
with $Q_{1,\text{CB}}^2 = 30\GeV^2$. 
The Christy--Bosted curves shown in Fig.~\ref{fig:low-Q2-check} do not
include this modification.

\begin{figure}
  \centering
  \includegraphics[width=0.49\textwidth]{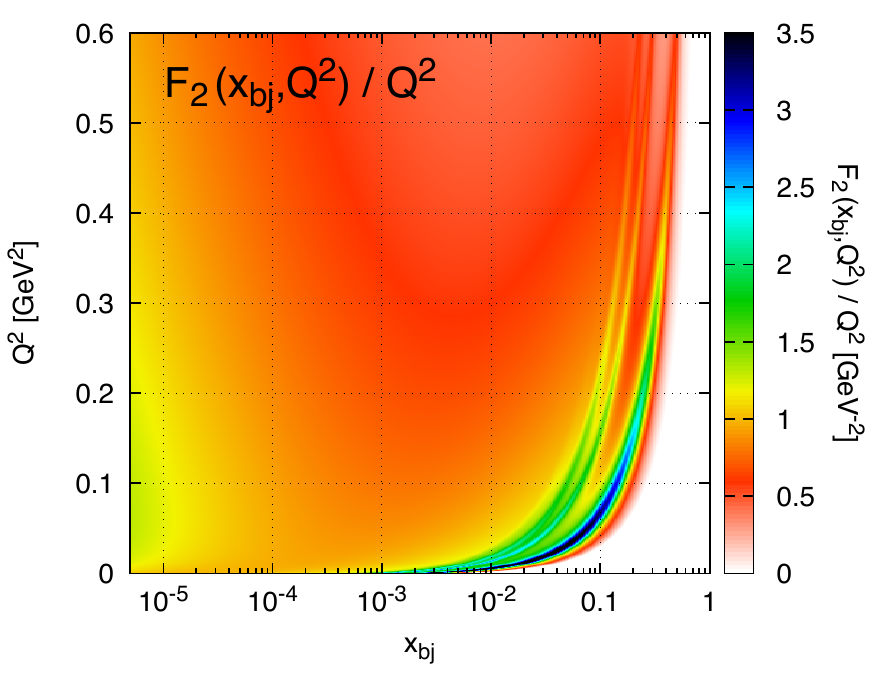}
  \hfill
  \includegraphics[width=0.49\textwidth]{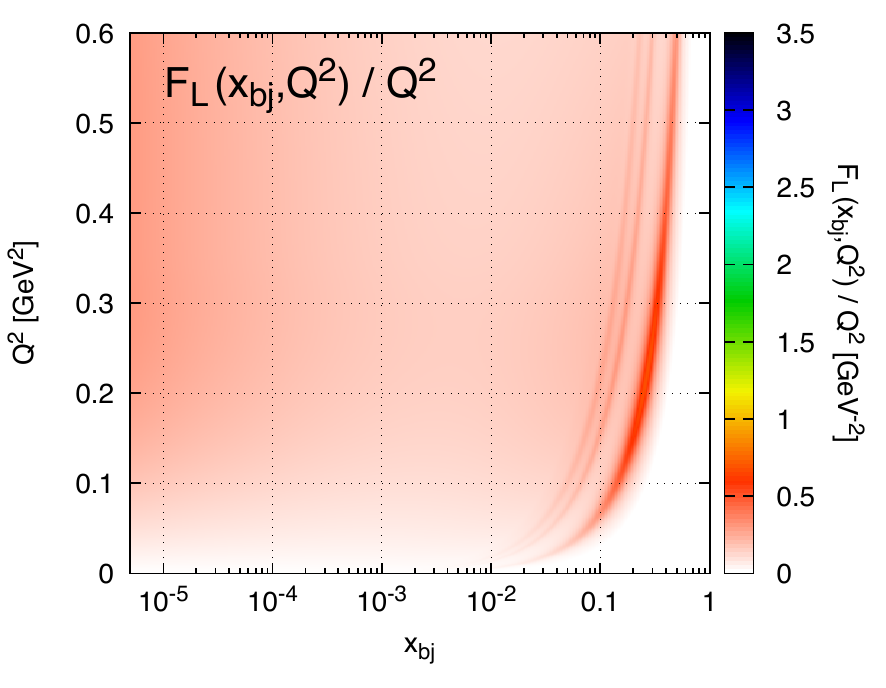}
  \caption{Value of the structure functions $F_2$ (left) and $F_L$
    (right) divided by $Q^2$ as a function of $\xbj$ and $Q^2$, using
    our CLAS+GD11-P prescription.
    They are divided by $Q^2$ to reflect the structure of the
    integrand for $f_{\gamma}(\xph,\mu^2)$, Eq.~\eqref{eq:master}.
    The latter receives contributions only from the region from $\xbj > \xph$.
    For a full understanding of the contribution of different
    regions of $\xbj$ and $Q^2$ to the integral, one must also take
    into account the $z=\xph/\xbj$-dependent factors that multiply the
    structure functions and the exact limits in $\xbj$ and $Q^2$.
  }
  \label{fig:F2FL-plane}
\end{figure}

To close our discussion of the low-$Q^2$ region, we show in
Fig.~\ref{fig:F2FL-plane} $F_2$ (left) and $F_L$ (right) divided by
$Q^2$, as a function of $\xbj$ and $Q^2$.
This gives an indication of the size of the contribution of different
$\xbj$ and $Q^2$ values in Eq.~\eqref{eq:master}.
The main features to note are the importance of the resonance region
and the relevance of $Q^2$ values down to zero over nearly the whole
$\xbj$ range.
One also sees that $F_L$ brings a substantially smaller contribution
than $F_2$, and tends to vanish for small $Q^2$, as dictated by our
use of Eq.~(\ref{eq:R1998mod}).
The figure focuses on the region $Q^2 < 0.6\GeV^2$.
For larger values of $Q^2$, the usual scaling behaviour of $F_2$ sets in, 
with, to first approximation, a uniform contribution in $\ln Q^2$.

\subsection{High-$Q^2$ continuum}
\label{sec:Q2cont}
For sufficiently large $Q^2$ and $W^2$ one can calculate $F_2$ and
$F_L$ from parton distribution functions (PDFs) using the known
perturbative expansion of the DIS coefficient functions.
This is more reliable than using a fit to available data (e.g.\ GD11-P
also includes some high-$Q^2$ data), because the extension to
arbitrarily large $Q^2$ is provided by DGLAP evolution rather than the
extrapolation provided by some a priori arbitrary parametrisation.
Furthermore, in recent years there has been extensive progress in the
extraction of PDFs from DIS and collider data, including detailed and
well-tested uncertainty estimates.

Our prescription for evaluating $F_2$ and $F_L$ is as follows.
Our choice of PDF and associated uncertainties will be
PDF4LHC15\_nnlo\_100~\cite{Butterworth:2015oua}.
This is based on a combination of the CT14nnlo~\cite{Dulat:2015mca},
MMHT2014nnlo~\cite{Harland-Lang:2014zoa} and
NNPDF30\_nnlo\_as\_0118~\cite{Ball:2014uwa} global PDF fits.
We will use NNLO coefficient
functions~\cite{vanNeerven:1991nn,Zijlstra:1992qd,vanNeerven:1999ca,vanNeerven:2000uj},
implemented in
\texttt{HOPPET}~\cite{Salam:2008qg,Cacciari:2015jma,Dreyer:2016oyx}. 
All quark flavours will be treated as massless and we will
correspondingly use a zero-mass variable flavour-number scheme
(ZM-VFNS).
The heavy flavour contribution to $F_2$ and $F_L$ is of order $\as$
and by taking the massless approximation for a quark of mass $m_q$ we
mistreat that order $\as$ contribution in a region of $Q^2 \sim
m_q^2$.
Examining Eq.~(\ref{eq:master}), one sees that this will translate to
an inaccuracy in $f_{\gamma}$ of order $\alpha \as$, i.e.\ beyond
the order that we aim to reproduce in our analysis with data.

Determinations of structure functions from PDFs are potentially
subject to corrections from higher-twist effects.
Recent studies of DIS
data~\cite{Harland-Lang:2016yfn,Cooper-Sarkar:2016foi} suggest that
higher-twist effects could be substantial for $F_L$, at least at small
values of $\xbj$.
In line with the observations of this study, we will account for such
a possibility when evaluating our uncertainties, using a
multiplicative correction
\begin{equation}
  \label{eq:FL-withHT}
  F_L^\text{with HT}(\xbj,Q^2) = \left(1 + \frac{5.5\GeV^2}{Q^2}\right)
  F_L^\text{NNLO}(\xbj,Q^2)\,,
\end{equation}
taking the larger of the two normalisations in
Refs.~\cite{Harland-Lang:2016yfn,Cooper-Sarkar:2016foi}. 

\begin{figure}
  \centering
  \includegraphics[width=0.49\textwidth,page=1]{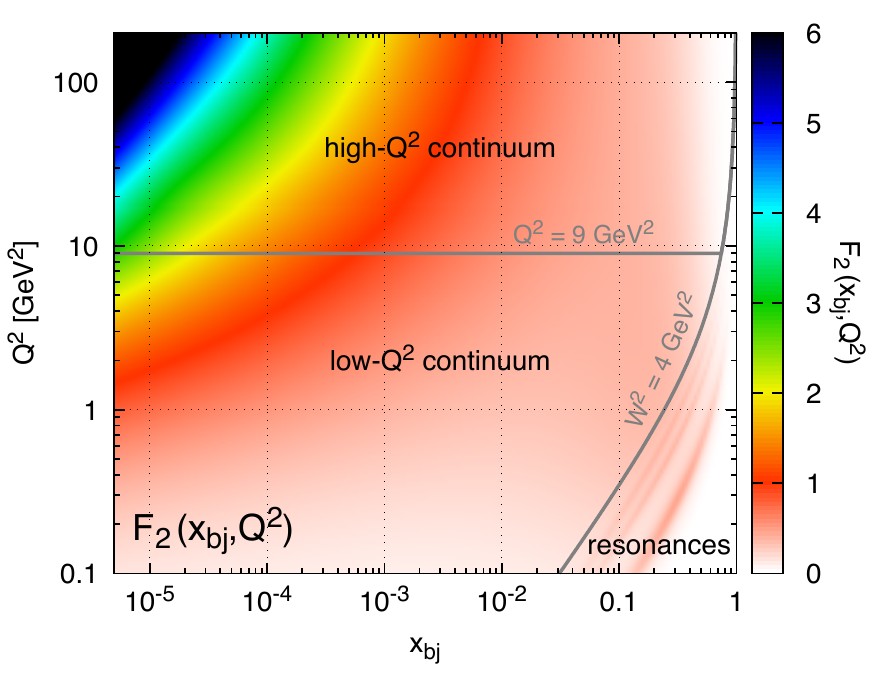}%
  \hfill
  \includegraphics[width=0.49\textwidth,page=2]{Figs/F2-plane.pdf}
  \caption{Values of the structure functions $F_2$ and $F_L$ as a
    function of $\xbj$ and $Q^2$, using a PDF4LHC15\_nnlo\_100-based
    NNLO ZM-VFNS prescription for $Q^2>9\GeV$ and $W^2>4\GeV^2$, and
    the CLAS+GD11-P combination elsewhere.}
  \label{fig:F2FL-plane-full}
\end{figure}

Our default domain for using a PDF-based evaluation of the structure
functions will be $Q^2 > Q^2_\text{PDF}  = 9\GeV^2$ and $W^2 > 4\GeV^2$.
In the rest of the kinematic plane we will use the resonance and low
$Q^2$ continuum fits.
The breakup of the $\xbj{-}Q^2$ plane is summarised in
Fig.~\ref{fig:F2FL-plane-full}, analogous to
Fig.~\ref{fig:F2FL-plane}, but showing a larger range of $Q^2$.
The $Q^2$ scale is now logarithmic and we no longer divide $F_2$ and
$F_L$ by $Q^2$. The colour-coding once again provides a
visualisation of the density of the integrand in
Eq.~(\ref{eq:master}).
At large $\xbj$ one sees the gradual reduction with increasing $Q^2$
of the structure functions.
A consequence of this is that the resonance part of the low-$Q^2$
region plays an especially important role in the determination of the
large-$x$ photon distribution.
At moderate $\xbj$, the structure functions are largely independent of
$\xbj$, i.e.\ they display Bjorken scaling, while at small $\xbj$ the
structure functions increase rapidly with $Q^2$, a consequence mainly
of double logarithms of $\xbj$ and $Q^2$ in the scaling violations.

Careful inspection of Fig.~\ref{fig:F2FL-plane-full} reveals that
$F_2$ and $F_L$ are not perfectly continuous at the transition scale
of $Q_{\rm trans}^2=9\GeV^2$, i.e.\ the results of the GD11-P fit and
PDF-based structure function evaluations do not quite match up.
This is most visible if one inspects the band of yellow colour in the
plot for $F_2$, around $\xbj=2\cdot10^{-5}$.
To probe the impact of this discontinuity in our $f_{\gamma}$
determinations we will introduce, as one of our uncertainty sources, a
variation of the $Q^2$ threshold for switching between the low-$Q^2$
GD11-P fit and the high-$Q^2$ PDF-based determination.
The alternative $Q^2$ threshold that we will use is $Q_{\rm
  trans}^2=5\GeV^2$.

\section{Uncertainty from missing higher orders}
\label{sec:scale}

The integral in our photon PDF was broken up into two pieces, an
all-$Q^2$ part which gave the ``physical'' PDF, and a high-$Q^2$ which
gave the $\MSbar$-conversion piece, cf.\ Eqs.~(\ref{6.12a}, \ref{9.11}).
In this section, we examine the uncertainty in the final photon PDF
result depend on the precise way in which the PDF integral is split
up, i.e.\ on the upper limit of the $Q^2$ integral in
Eq.~(\ref{6.12a}).

\subsection{Dependence on physical factorisation separation scale}

To probe the impact of the choice of separation between the physical
PDF and the $\MSbar$-conversion,
we use the following alternative definition of the physical and
$\MSbar$-conversion terms
 \begin{align}
f^{\text{PF}}_\gamma(x,\mub^2,[\muzfunNoSqr]) &= \frac{1 }{2 \pi
  \alpha(\mub^2) x} \int_x^1 \frac{\mathd z}{z} \int_{\frac{m_p^2
    x^2}{1-z}}^{\muzfun(z)} \frac{\mathd Q^2}{Q^2} \aph^2(q^2) \nn &
\biggl\{- z^2 F_L(x/z,Q^2) + \left[ 2-2z + z^2 + \frac{2 m_p^2
    x^2}{Q^2}\right] F_2(x/z,Q^2)\biggr\} \,,
\label{6.12_alt}
 \end{align}
\begin{align}
 f_\gamma^{\text{con}}(x,\mub^2,[\muzfunNoSqr])
& =   \frac{8 \pi }{x \alpha(\mub^2) \smu^{2\epsilon}} \frac{1}{(4\pi)^{D/2}} \frac{1}{\Gamma(D/2-1)}  \int_x^1 \frac{\mathd z}{z}  (1-z)^{D/2-2}  \int_{\muzfun(z)}^\infty  \frac{\mathd Q^2}{Q^2}  \nn
 & \left(Q^2\right)^{D/2-2}  \aphD^2(q^2)  \bigg\{ - z^2 (1-\epsilon) F_{L, D}(x/z,Q^2)  \nn
& +\left[ 2-2z  +  z^2 -\epsilon z^2  \right] F_{2, D}(x/z,Q^2) \bigg\}\,, 
\label{6.13_alt}
\end{align}
where $[\muzfunNoSqr]$ denotes the functional dependence on $\muzfunNoSqr(z)$. 
We easily find that
\begin{eqnarray}
 f_\gamma^{\text{con}}(x,\mub^2,[\muzfunNoSqr]) 
  &=&\, f_\gamma^{\text{con}}(x,\mub^2)
+ \frac{1}{x \alpha(\mub^2)} \frac{1}{2\pi}  \int_x^1 \frac{\mathd z}{z}  \int_{\muzfun(z)}^{\frac{\mu^2}{(1-z)}}  \frac{\mathd Q^2}{Q^2}\times  \nn
 &&  \aph^2(q^2)  \biggl\{ - z^2 F_{L}(x/z,Q^2) + \left[ 2-2z  +  z^2   \right] F_{2}(x/z,Q^2) \biggr\} \nn
  &=&\, f_\gamma^{\text{con}}(x,\mub^2)
+ \frac{1}{x \alpha(\mub^2)} \frac{1}{2\pi}  \int_{zv,x} \int_{\muzfun(z)}^{\frac{\mu^2}{(1-z)}}
\frac{\mathd Q^2}{Q^2}\times  \nn
 &&  \aph^2(q^2)  \biggl\{ zp_{\gamma q}(z) F_{2}(v,Q^2) - z^2 F_{L}(v,Q^2)  \biggr\}\;.
\label{6.13_alt2}
\end{eqnarray}
In the following we will consider two alternative choices for $\muzfunNoSqr(z)$:
\begin{subequations}
  \label{eq:mu_both_choices}
  \begin{equation}
    \muzfun(z) = \frac{\muz^2}{1-z}\,,\qquad \frac{1}{2} \le \frac{\muz}{\mu} \le 2\,,   
    \label{eq:mu_1stchoice}
  \end{equation}
  and
  \begin{equation}
    \muzfun(z) = \muz^2 \,,\qquad \frac{1}{2} \le \frac{\muz}{\mu} \le 2\,.  
    \label{eq:mu_2ndchoice}
  \end{equation}
\end{subequations}
The variation of the answer as a function of $\muz$ will provide us with
an estimate of uncertainties from missing higher-order (QCD)
contributions.  
This is similar in spirit, but different in its details, from standard
scale variation.

$M(z)$ is a large scale, where we can use QCD expressions for the
structure functions in terms of parton distributions. We have
\begin{align}
\frac{F_2(v,Q^2)}{v} &= \sum_{\qg\in\qgset} \int_{yw,v} \mathscr{F}_{2,\qg} (y, Q^2,\mub^2) f_\qg(w,\mub^2) , \nn
\frac{F_L(v,Q^2)}{v} &= \sum_{\qg\in\qgset} \int_{yw,v} \mathscr{F}_{L,\qg} (y, Q^2,\mub^2)  f_\qg(w,\mub^2), 
\label{9.3b}
\end{align}
where, up to NLO QCD accuracy for the coefficient functions,
\begin{align}
  \mathscr{F}_{2,a\in\qset}(y,Q^2,\mu^2) &=  C^{(0,0)}(y) + \frac{\as(\mu^2)}{2\pi} \left( \ln\frac{Q^2}{\mu^2} \Ptwoqa(y) +C_{2,a}^{(1,0)}(y) \right)\,, \nn
    \mathscr{F}_{2,g}(y,Q^2,\mu^2) &=  \frac{\as(\mu^2)}{2\pi} \left( \ln\frac{Q^2}{\mu^2} \Ptwog(y) +C_{2,g}^{(1,0)}(y) \right)\, , \nn
    \mathscr{F}_{L,a\in\qset}(y,Q^2,\mu^2) &=  \frac{\as(\mu^2)}{2\pi} C_{L,a}^{(1,0)}(y) \,,\nn 
      \mathscr{F}_{L,g}(y,Q^2,\mu^2) &= \frac{\as(\mu^2)}{2\pi} C_{L,g}^{(1,0)}(y) \,,
  \label{eq:F2L_def}
\end{align}
where $\Ptwoqa$ and $\Ptwog$ are defined in Eq.~(\ref{G:9.51}).
We find the following $[\muzfunNoSqr]$ dependent additions to the 
$f_\gamma^{(0,1)}$ and $f_\gamma^{(1,1)}$ $\MSbar$-conversion terms: 
\begin{align}
  \label{eq:scale-dep-01}
 f_\gamma^{(0,1)}(x,\mub^2,[\muzfunNoSqr]) 
  &= f_\gamma^{(0,1)}(x,\mub^2)  
    + \sum_{\qg\in\qset} \int_{w z, x} 
    \log\frac{\mu^2}{(1-z)M^2(z)} p_{\gamma q}(z) f_\qg(w, \mu^2)\,,
  \\
 f_\gamma^{(1,1)}(x,\mub^2,[\muzfunNoSqr]) 
  &= f_\gamma^{(1,1)}(x,\mub^2)  
    +
    \int_{y w z, x} 
    \Bigg\{ \nn
  &  \log\frac{\mu^2}{(1-z)M^2(z)}\sum_{\qg\in\qgset} \Big[
    p_{\gamma q}(z) C_{2,\qg}^{(1,0)}(y) f_\qg(w, \mu^2)
    -  z \, C_{L,\qg}^{(1,0)}(y) f_\qg(w, \mu^2) \Big] \nn
  & +\frac{1}{2}\left(\log^2\frac{1}{1-z}-
    \log^2\frac{M^2(z)}{\mu^2}\right) p_{\gamma q}(z)
     \sum_{\qg \in \qgset}\Ptwoqa(y) f_\qg(w,\mu^2)
  \label{eq:scale-dep-11}
\Bigg\}\,.
\end{align}
We omit explicit analytic results for the convolutions, insofar as
these are trivial to perform numerically with the
\texttt{HOPPET}~\cite{Salam:2008qg}
package.

\subsection{Numerical impact of the separation scale}
\label{sec:scale-uncert-res}

\begin{figure}
  \centering
  \includegraphics[width=0.5\textwidth,page=6]{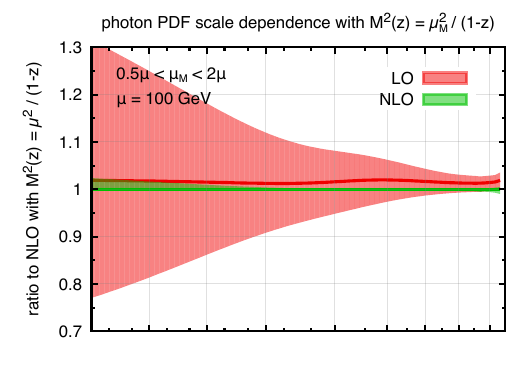}%
  \includegraphics[width=0.5\textwidth,page=8]{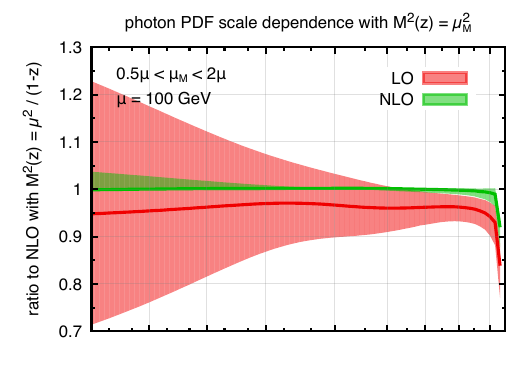}%
  \\[-35pt]
  \includegraphics[width=0.5\textwidth,page=2]{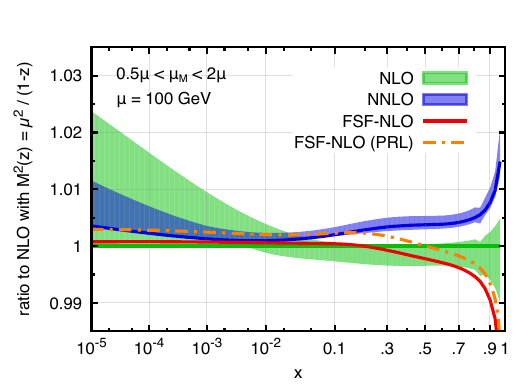}%
  \includegraphics[width=0.5\textwidth,page=4]{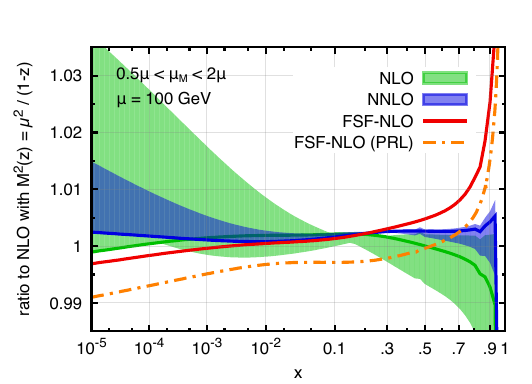}
  \caption{Scale dependence of the photon PDF at LO (top only),  NLO
    and NNLO (bottom only).
    The left and right plots show results with $\muzfun(z)$, the upper
    scale in the physical factorisation PDF, respectively as in
    Eq.~(\ref{eq:mu_1stchoice}) and Eq.~(\ref{eq:mu_2ndchoice}).
    All result are normalised to the NLO answer with our default choice,
    $\muzfun(z) = \mu^2/(1-z)$. 
    Lines correspond to a central scale choice $\mu_M^2 = \mu^2$.
    Further details, including the specification of the FSF-NLO
    curves, are given in the text.  }
  \label{fig:scale-dep}
\end{figure}

Fig.~\ref{fig:scale-dep} shows the numerical consequences of the above
results.
Most of the details of our numerical implementation will be discussed
below in Sec.~\ref{sec:practical-implementation}, and here we
concentrate just on the scale dependence.
The figure shows results with both $\muzfun(z)$ choices in
Eq.~(\ref{eq:mu_both_choices}).
LO corresponds to Eq.~(\ref{9.11}), with no
$\MSbar$-conversion term, i.e.\ $f_\gamma^\text{PF}$ of
Eq.~(\ref{6.12_alt}).
NLO corresponds to Eq.~(\ref{9.11}), where
$f^{\text{con}}_\gamma(x,\mub^2)$ includes just the
$f_\gamma^{(0,1)}(x,\mu^2)$ term, Eqs.~(\ref{9.6a}) and
(\ref{eq:scale-dep-01}), in its series expansion,
Eq.~(\ref{eq:FconExpansion}).
The NNLO result additionally includes the $f_\gamma^{(1,1)}(x,\mu^2)$
term, Eqs.~(\ref{eq:fcon11}) and (\ref{eq:scale-dep-11}).
Higher-order QED corrections are not included.
The high-$Q^2$ contribution to the $f^{\text{PF}}$ term is always
evaluated including a NNLO PDF
(PDF4LHC15\_nnlo\_100~\cite{Butterworth:2015oua}), NNLO splitting
kernels~\cite{Moch:2004pa,Vogt:2004mw} and massless NNLO QCD
coefficient
functions~\cite{vanNeerven:1991nn,Zijlstra:1992qd,vanNeerven:1999ca,vanNeerven:2000uj}
as implemented~\cite{Cacciari:2015jma,Dreyer:2016oyx} in
\texttt{HOPPET}~\cite{Salam:2008qg}.

The upper plot shows the substantial reduction in the scale
uncertainty in going from LO to NLO with both scale choices. There is
an especially large uncertainty in the LO photon PDF for small $x$. In
this region the photon PDF is dominated by the high-$Q^2$ integration
due to the DGLAP evolution of the quark distribution.  Changing the
scale by a factor of two around our central value of $\mu=100$\,GeV
causes a 20\% shift in $\ln(\mu/Q_{\rm trans})$.
This helps explain the size of the uncertainty in the upper figures.
Recall that $Q_{\rm trans} = 3\GeV$ is the scale above which we use
PDFs and coefficient functions to evaluate the structure functions, as
explained at the end of Sec.~\ref{sec:Q2cont}.
Previous estimates of the photon PDF were only accurate to LO, and so
had an intrinsic uncertainty comparable to that of our LO band.%
\footnote{One should keep in mind that many current predictions for
  processes involving an incoming photon are only available at LO for
  the photon-induced process.
  They will therefore have a substantial relative scale uncertainty.
  This uncertainty should mostly be cancelled in calculations that
  include NLO QED corrections to the photon-induced processes,
  e.g.~Refs.~\cite{Dittmaier:2009cr,Kallweit:2017khh}.}

In the lower left-hand plot, the curve labelled FSF-NLO (``Full
Structure Function") corresponds to Eq.~(\ref{eq:master}), with $F_2$
computed at NNLO in both the physical-factorisation and the
$\MSbar$-conversion terms.
The PRL variant of the FSF-NLO curve corresponds to the result of
Ref.~\cite{Manohar:2016nzj}, whose implementation had a formally
subleading bug in the scale of $\alpha$ that multiplied the
$\MSbar$-conversion term.
The corresponding curves in the lower right-hand plot are analogous
but with $M^2(z) = \mu^2$ as the upper limit of the
physical-factorisation component.

For our final central result, to be shown in the next section, we will
take the NLO curve with $M^2(z) = \mu^2/(1-z)$, i.e.\ using
$f^{\text{con}}_\gamma(x,\mub^2)$ that includes just the
$f_\gamma^{(0,1)}(x,\mu^2)$ term, Eq.~(\ref{9.6a}).
As an estimate of the uncertainty from missing higher-order (MHO)
contributions, we will take the largest deviation of any of the NNLO
$M^2(z) = \mu^2_M/(1-z)$ scale choices and symmetrise it with respect
to the central NLO result.

\begin{figure}
  \centering
  \includegraphics[width=0.6\textwidth,page=1]{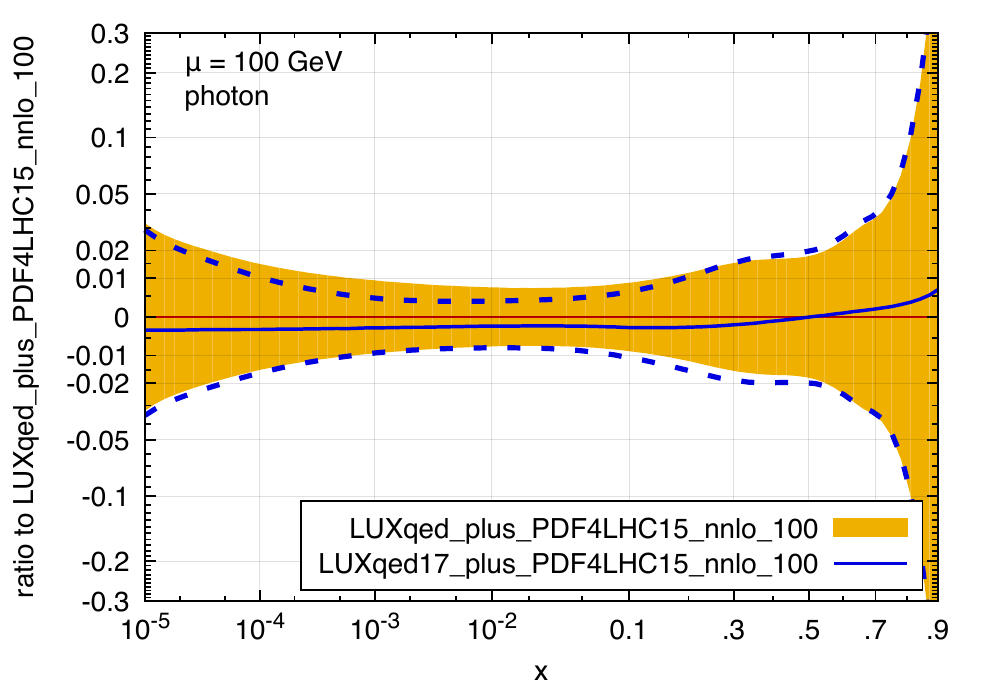}%
  \caption{%
    Ratio of the photon PDF in this paper (LUXqed17), shown
    as a blue solid line, to our previous
    result~\cite{Manohar:2016nzj} (LUXqed) at $\mu=100$\,GeV.
    The solid orange area is the original error band, and the dashed
    blue lines indicate the new error band.}
  \label{fig:compare}
\end{figure}

This differs from the prescription adopted in
Ref.~\cite{Manohar:2016nzj}, where we used the FSF-NLO evaluation with
$M^2(z) = \mu^2/(1-z)$.
There the MHO uncertainty was estimated from the difference between
FSF-NLO evaluations with $M^2(z) = \mu^2/(1-z)$ and $M^2(z) = \mu^2$,
corresponding to the difference between the orange (dot-dashed) curves
in the left and right-hand plots of Fig.~\ref{fig:scale-dep}.
The difference between $M^2(z) = \text{constant}$ and
$M^2(z) \propto 1/(1-z)$ is a $\log(1-z)$ contribution, which gets
large as $z \to 1$. 
These ``endpoint'' logarithms arise because a new scale $Q^2(1-z)$,
the invariant mass of the final state hadronic system, enters the
problem. 
While formally, this spurious logarithm is cancelled by a
corresponding one in the \MSbar-conversion  
term, there is a left-over higher-order $\log(1-z)$ with the choice
$M^2(z) = \text{constant}$. 
For our default photon-PDF prediction, we prefer not to add a spurious
endpoint logarithm to $f_\gamma^\text{con}$ via our choice of
$M^2(z)$. 
Figure~\ref{fig:compare}
compares the old and new versions of our photon PDF.

Note that we have chosen not to use a full NNLO result. 
While this is in principle straightforward to achieve, several further
practical elements would be needed beyond those we have implemented so
far. 
The one that is potentially most problematic relates to the fact that
one must take into account QED corrections to the structure functions.
This would call for a detailed assessment of the way in which QED
corrections have been removed from published data.

\section{Evaluation of the photon distribution and
  prescription for  other partons}
\label{sec:practical-implementation}

We evaluate the photon distribution using Eqs.~(\ref{6.12a},
\ref{9.6a}, \ref{9.11}), keeping only the $f_\gamma^{(0,1)}$ term in
Eq.~(\ref{9.11b}).
The elastic component is given by Eq.~(\ref{eq:elastic-component}).
A number of choices need to be made in the evaluation, both for the
central value of the photon distribution and its uncertainties.
For practical usage one also requires a consistent set of parton
distribution functions for the other flavours.
In this section we describe the approach we take and considerations
for future evaluations.

We recall that, as in Sec.~\ref{sec:input-data}, our accuracy aim
for the photon distribution will be to control terms up to $\alpha
(\as L)^n$ and $\alpha^2 L^2 (\as L)^n$.
%

\subsection{Evaluation scale and evolution}
\label{sec:evaluation-scale}

A first choice relates to the question of the $\mu^2$ value(s) where
the photon distribution is evaluated and the issue of higher-twist
corrections.
When $\mu^2$ is not very large, the upper limit of
Eq.~(\ref{6.12a}) is in a region where both the inelastic and
elastic parts of structure functions may themselves contain
higher-twist corrections.

For the elastic part, in Eq.~(\ref{eq:elastic-component}) we have
deliberately set the upper limit in the $Q^2$ integration to infinity.
This is possible because the contribution to the
$\mu^2$ dependence of the elastic part comes from the running of the
QED coupling in front of Eq.~(\ref{eq:elastic-component}), which is
not affected by the choice of upper limit in $Q^2$.
Leaving the upper limit as $\mu^2/(1-z)$ would instead have resulted
in higher twist contributions associated with the $1/Q^4$ scaling of
the form factors.

In the inelastic part, there is no easy way, at low $\mu^2$ scales, of
separating leading and higher twist components.
This is because at the corresponding upper limit of the $Q^2$
integration, the inelastic structure functions mix both leading and
higher-twist effects.
Accordingly we choose to evaluate Eq.~(\ref{9.11}) at a scale
$\mu^2_\text{eval} \gg \mpr^2$ and then determine the photon
distribution at other scales through DGLAP evolution.
We take $\mu_\text{eval} = 100\GeV$ so as to ensure that potential
residual higher-twist effects, of order $\mpr^2/\mu^2_\text{eval}$,
are much smaller than the precision we will be seeking, which will be
roughly at the $1\%$ level.

For the DGLAP evolution we will include the $\order{\alpha}$ and
$\order{\alpha \as}$ splitting kernels~\cite{deFlorian:2015ujt}, as
well as the QCD kernels up to
$\order{\as^3}$~\cite{Moch:2004pa,Vogt:2004mw}.
Given the precision that we use in the evaluation of the photon
distribution, and our assumption $\as^2 \sim \alpha$, this forms a
consistent set of evolution terms.

\subsection{Uncertainties}
\label{sec:uncertainties}

\begin{figure}
  \centering
  \includegraphics[width=0.6\textwidth,page=2]{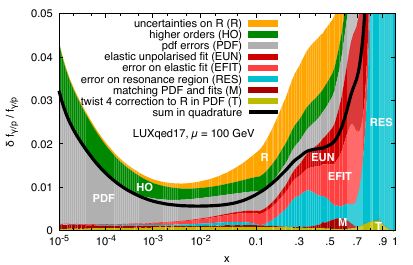}
  \caption{Breakdown of uncertainties on the photon distribution. 
    The uncertainties are shown stacked linearly, while the sum in
    quadrature, i.e.\ our final uncertainty, is represented by the
    thick black line. See the text for a detailed description of the
    various contributions.
  }
  \label{fig:uncertainty-breakdown}
\end{figure}

The final uncertainty on our PDF distribution is taken to be the sum
in quadrature of many individual uncertainty sources, because they are
uncorrelated. The individual contributions have already been discussed
in Sec.~\ref{sec:input-data}, where we explained the input data to our
analysis. The various uncertainties with labels as in
Fig.~\ref{fig:uncertainty-breakdown} are:
\begin{itemize}
\item[(EFIT)] The uncertainty on the elastic contribution that comes
  from the uncertainty on the A1 world polarised form factor fits, as
  shown in Fig.~\ref{fig:elastic-component-ratios}.
  This band is asymmetric and we symmetrise it using the largest
  deviation to obtain a (more conservative) symmetric band.

\item[(EUN)] The uncertainty that comes from replacing the A1 world
  polarised fit (which includes a two-photon-exchange correction) with
  just the world unpolarised data (which does not).
  This provides a one-sided uncertainty, which we again symmetrise. 

\item[(RES)] We replace the CLAS resonance-region fit with the
  Christy-Bosted fit (modified as in Eq.~(\ref{eq:modified-CB}).
  This replacement gives a one-sided uncertainty, which we once again
  symmetrise. 

\item[(R)] A modification of $R^{1998}_\text{L/T}$ by $\pm 50\%$
  around its central value, as shown in Fig.~\ref{fig:R-v-W2}.
  Recall that this $R$ choice is only used in the regions where we
  take $F_2$ from one of the GD11-P, CLAS or Christy-Bosted fits.
  Of the $\pm50\%$ variations of $R$, the one with the larger impact
  on the photon distribution is identified and the resulting
  uncertainty symmetrised.

\item[(M)]  A modification of the $Q^2_\text{PDF}$ scale which governs the
  transition from the GD11-P structure function fit to a PDF-based
  evaluation.
  The default choice of $Q^2_\text{PDF} = 9\GeV^2$ is reduced to
  $5\GeV^2$ and since this is a one-sided uncertainty, the resulting
  effect is symmetrised.

\item[(PDF)] The input PDF uncertainties for $Q^2 > Q_\text{PDF}^2$ according
  to the default prescription for the PDF (PDF4LHC15\_nnlo\_100).

\item[(T)] A twist-4 modification of $F_L$ as in Eq.~(\ref{eq:FL-withHT}). 
  This is a one-sided modification that is then symmetrised.

\item[(HO)] An estimate of missing higher-order effects obtained by taking
  the largest deviation of any of the NNLO results with the
  $\mu^2/(1-z)$ scale choice in Fig.~\ref{fig:scale-dep} (left)
  relative to the NLO result with scale choice $\mu^2/(1-z)$.
  The resulting uncertainty is symmetrised.
\end{itemize}
Note that we do not include the quoted uncertainty from the GD11-P fit. 
When studying that uncertainty we found on one hand that its impact
was negligible compared to the other uncertainties, and on the other
hand that the resulting uncertainty band did not always overlap with
$F_2$ as calculated from PDFs in regions at small $x$ and high $Q^2$
that lack direct $F_2$ data.
This observation motivated our choice to restrict the use of the
GD11-P fit to $Q^2$ values with sufficient data coverage and to vary
the $Q_\text{trans}^2$ transition scale (and $R$) for the uncertainty
estimate. We also did not include any uncertainty associated with the
prescription for matching other parton flavours.

The impact of the different sources of uncertainty is shown in
Fig.~\ref{fig:uncertainty-breakdown}, and our final uncertainty, shown
by the black line, is given by adding the contributions in
quadrature.\footnote{There are correlations between high and low $Q^2$
  that have not been included in our analysis. For example, low $Q^2$
  values for $F_2(x,Q^2)$ are correlated with quark and gluon PDFs at
  high $Q^2$, via DGLAP evolution of $F_2$.}
The overall uncertainty is less than 2\% for $10^{-4} < x < 0.1$. For
small values of $x$, the uncertainty is dominated by the uncertainties
in the parton distributions of quarks (and gluons), which enter the
high-$Q^2$ part of the photon PDF integral. 
As $x \to 1$, the
uncertainties are dominated by the low-$Q^2$ part of the photon PDF
integral from elastic form factors, the resonance contribution, and
$\sigma_L$. 
This is a reflection of the fact that non-perturbative effects (such
as higher twist corrections) grow like
$\Lambda_\text{QCD}^2/[Q^2(1-x)]$ as $x \to 1$, and that, for $x$
close to $1$, quark PDFs fall off rapidly as $Q^2$ increases, so the
low-$Q^2$ region contributes significantly to $f_\gamma$ as $ x\to 1$.

\subsection{Matching to other partons}
\label{sec:match-other-part}

The introduction of a photon component of the proton necessarily
implies modifications of other partons relative to a set that has been
determined without QED corrections.
QED corrections to the quarks start at $\order{\alpha L (\as L)^n}$,
i.e.\ the same order as the leading photon contribution.
These terms are generated by the order $\alpha$ QED contribution in
DGLAP evolution.
If one wishes to know the full set of partons to the same accuracy as
the photon, i.e.\ $\order{\alpha (\as L)^n}$ relative corrections,
then it is effectively necessarily to have a reliable estimate of the
QED corrections to the initial conditions for the DGLAP evolution.
This would require that one repeat a global PDF fit with QED
corrections, e.g.\ the order ${\alpha}$ corrections to the DIS
coefficient functions, and with information about the photon PDF as an
input (e.g.\ because it affects the momentum sum rule at order
${\alpha}$ at the starting scale).
Such a fit is beyond the scope of our article, though below we outline
a procedure for how it might be performed.

\subsubsection{Our procedure}
\label{sec:our-photon+quarks}

The prescription that we adopt is as follows.
At a scale $\mu^2_\text{match}$, we assume that the quarks are
unchanged relative to a global fit without QED contributions.
At this scale we rescale the gluon by a factor as follows:
\begin{equation}
  \label{eq:gluon-rescale}
  f_{g}^\text{rescaled}(x,\mu^2_\text{match}) = 
  \left[ 1 - \frac{
    \omega_{\gamma}(\mu^2_\text{match})}{\omega_{g}(\mu^2_\text{match})}\right]
  \times f_{g}(x,\mu^2_\text{match})\,,
\end{equation}
where $\omega_{i}(\mu^2)$ is the momentum fraction carried by
parton flavour $i$ at scale $\mu^2$, 
\begin{equation}
  \label{eq:momentum-definition}
  \omega_{i}(\mu^2) = \int_0^1 dx\, x f_{i}(x,\mu^2).
\end{equation}
As we shall see below, Fig.~\ref{fig:mom}, the photon momentum
fraction, $\omega_\gamma$, is a fraction of a percent.
The above procedure ensures that the momentum sum rule, including the
photon contribution, is satisfied.
The reason for absorbing the momentum into an adjustment of the gluon
is that the gluon is the parton least directly constrained from DIS
data. 

The choice of $\mu_\text{match}^2$ is somewhat arbitrary.
Ideally, it should be close to
the $Q^2$ scales in DIS that provide the greatest constraint on the
quark PDFs,
given that our procedure leaves the quark distribution unchanged at
scale $\mu^2_\text{match}$.
Since we use a PDF fit that was determined without QED corrections,
when we convolute with the QCD coefficient functions (without QED
corrections), we should reproduce the true experimental $F_{2/L}$ at
the scales where the DIS data is most constraining.
There is however a practical problem that we should consider.
The PDF4LHC15\_nnlo\_100 set, which we take as our base PDF set, is
designed for use at large $\mu^2$ values.
At low $\mu^2$ we encountered two issues,
discussed in detail in App.~\ref{sec:PDF4LHC15-issues}:
one is that it is a merger
of sets with different underlying heavy-quark thresholds, and as a
result does not strictly satisfy DGLAP evolution.
The second relates to the way the PDF4LHC15\_nnlo\_100 underlying PDF
sets are encoded in LHAPDF~\cite{Buckley:2014ana} files.
These issues prevent us from starting a 
DGLAP evolution from scales below about $6\GeV$.
Putting together the various considerations, we opted for the choice
$\mu_\text{match} = 10\GeV$.
The resulting set, LUXqed17\_plus\_PDF4LHC15\_nnlo\_100, is valid only
for scales $\mu \ge 10\GeV$ (below this scale, LHAPDF interpolates the
LUXqed17 partons to zero).

\begin{figure}
  \centering
  \includegraphics[width=0.5\textwidth,page=2]{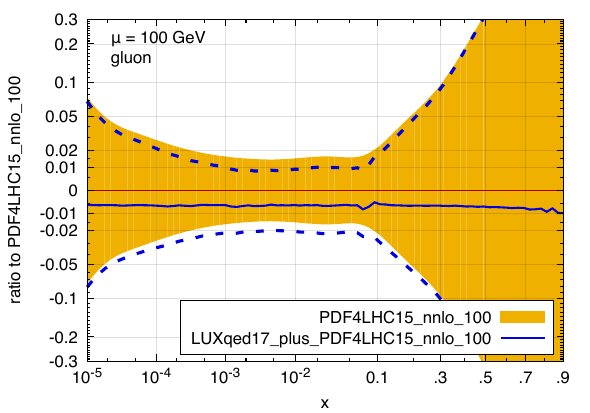}%
  \includegraphics[width=0.5\textwidth,page=5]{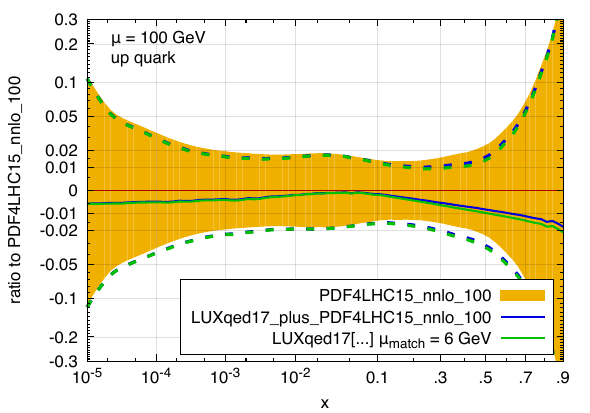}%
  \caption{%
    Ratio of the gluon and $u$-quark PDF in this paper
    (LUXqed17\_plus\_PDF4LHC15\_nnlo\_100) to the corresponding
    PDF4LHC15\_nnlo\_100 \cite{Butterworth:2015oua} distributions at
    $\mu=100$\,GeV, shown as the solid blue line.
    The solid orange region is the original PDF4LHC15\_nnlo\_100
    error band, and the dashed blue lines represent our error band.
    In the case of the up-quark distribution, we also show in green
    the impact of using $\mu_\text{match} = 6\GeV$ instead of our
    default of $10\GeV$.  }
  \label{fig:compareparton}
\end{figure}

Since we include QED effects in the DGLAP evolution, at scales other
than $\mu_\text{match}$ all partons acquire QED-induced modifications,
of order $(\alpha L)^n (\as L)^m$ for the quarks, with $n \ge 1$,
$m \ge 0$.
The changes in the gluon and up-quark PDFs at scale
$\mu = 100\GeV$ are illustrated in Fig.~\ref{fig:compareparton}.
In the case of the up-quark PDF, we also show the effect of reducing
$\mu_\text{match}$ from our default of $10\GeV$ to $6\GeV$,
demonstrating that the impact is minimal compared to the overall
uncertainty on the up-quark distribution.

\begin{figure}
  \centering
  \includegraphics[width=0.6\textwidth]{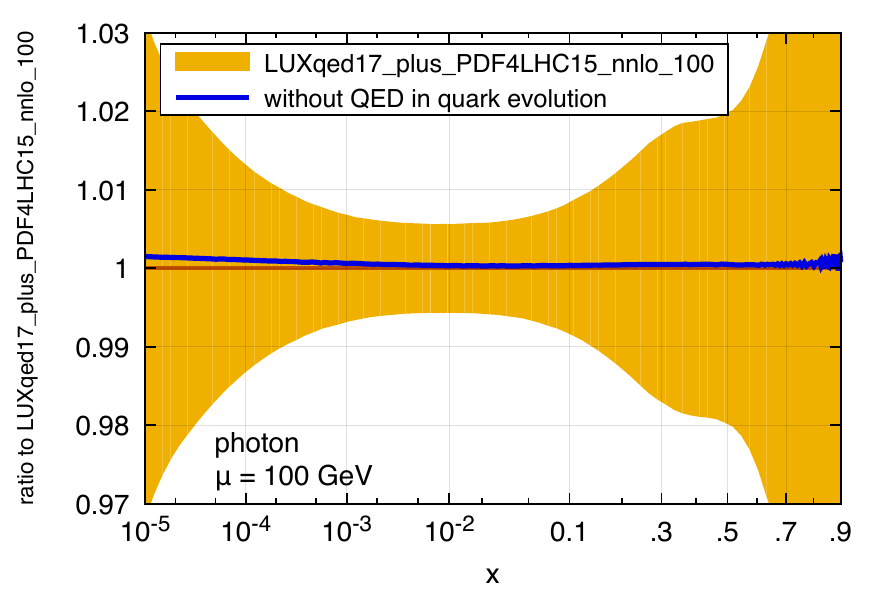}
  \caption{The impact on the photon distribution from leaving out QED 
    effects in 
    the quark evolution (blue curve), compared to the
    LUXqed17\_plus\_PDF4LHC15\_nnlo\_100 uncertainty.
    The irregularities in the curve at large $x$ values are an
    artefact associated with the use of different underlying $x$ grids
    between the results with and without QED effects in the quark evolution.
  }
  \label{fig:quark-qed-on-photon}
\end{figure}

As discussed in Sec.~\ref{sec:evaluation-scale}, it is not
advisable to directly use Eqs.~(\ref{6.12a}, \ref{9.6a}, \ref{9.11})
to evaluate the photon PDF at scales as low as
$\mu_\text{match}$, because of the presence of higher twist effects
that are difficult to control.
On the other hand, if we evaluate the photon PDF at a scale
$\mu_\text{eval} \gg \mu_\text{match}$, the $\order{\alpha L}$
modifications to the quark distributions from DGLAP evolution must be
taken into account in order to correctly treat contributions of order
$(\alpha L)^2(\as L)^n$ ($n \ge 0$) in the photon distribution.
We do so technically as follows, keeping in mind that our base
PDF4LHC15\_nnlo\_100 PDF is based on fits that have neither QED
evolution, nor QED corrections in coefficient functions.
First we evaluate the photon distribution at scale
$\mu_\text{eval} \gg \mu_\text{match}$, using a high-$Q^2$ $F_2$
calculated without QED effects.
We then evolve the photon distribution down to the scale
$\mu_\text{match}$, using a special variant of DGLAP evolution in
which all QED contributions to $P_{qi}$ and $P_{gi}$ are set to zero,
for all parton flavours $i$. 
With this procedure, the quark and gluon densities remain
identical to those of the original PDF set. These unchanged distributions are then used in the evaluation of
$P_{\gamma q}$ terms for the photon evolution.
We perform the matching at scale $\mu_\text{match}$, as described
above, and finally evolve back up in scale using DGLAP evolution
including the full set of QED contributions.
We stress that when evolving back up to the $\mu_\text{eval}$
scale, all partons will acquire corrections of
relative order $(\alpha L)^n$, with $n\ge 1$. In particular the
photon PDF will acquire relative correction of order $\alpha L$,
that are required by our aimed accuracy.
As illustrated in Fig.~\ref{fig:quark-qed-on-photon}, without the QED
effects in the quark evolution (which tend to reduce the quark
distribution), the photon distribution comes out slightly higher.
However, the effect is minimal compared to the overall
LUXqed17\_plus\_PDF4LHC15\_nnlo\_100 uncertainty.
Note that an $\order{1}$ change in the choice of $\mu_\text{match}$
corresponds to a NNLO, $\order{\alpha^2 L (\as L)^n}$, effect on the
photon distribution.

The smaller effect in the photon density in
Fig.~\ref{fig:quark-qed-on-photon} relative to that in the quark
densities in Fig.~\ref{fig:compareparton} (right) is because the
higher-order QED modification of the photon is driven by the average
of the QED modification of the quarks across all scales from
$\mu_\text{match}$ up to $\mu$.
However the quarks are modified mostly at scales close to $\mu$, i.e.\
the average modification is suppressed.

A final comment concerns lepton distributions. 
For consistency with the running of the QED coupling, photon splitting
to leptons should be included, which generates non-zero lepton
distributions. 
These are of order $(\alpha L)^2$.
Their feedback on the photon distribution is an $(\alpha L)^3$ effect
and so beyond our accuracy.\footnote{Leptons also affect the running
  of the QED coupling, with an order $(\alpha L)^2$ effect on the
  photon distribution, which we do include.}
In practice we set the lepton distributions to zero at scale
$\mu_\text{match}$ and include the full set of $P_{l \gamma}$,
$P_{ll}$ and $P_{\gamma l}$ splitting functions in the evolution.
A more complete approach would determine the lepton distributions
directly from $F_2$ and $F_L$, in a manner analogous to that used for
the photon distribution in this paper.
Note that while we include leptons in our DGLAP evolution, their
distributions are not included in the LHAPDF files that we make
available.

\subsubsection{Iterative procedure}
\label{sec:iterative-photon+quarks} 

The above procedure allows us to obtain NLO QED accuracy for the
photon distribution and LO QED accuracy for the other partons.
Recall that LO corresponds to $\alpha L$ and NLO to an extra
suppression of $\alpha L$ or $\as$, with $\as L$ considered to be of
order $1$.
For compactness, below we will use $\delta$ to refer to any quantity
of order of $\alpha L$ or $\as$.
The photon distribution and QED effects in the quarks both start at
order $\delta$.

To go beyond the NLO QED accuracy for the photon, and to achieve
better than LO QED accuracy for the other partons, requires a more
sophisticated procedure than that discussed in the previous section.

One conceivable difficulty in constructing a general approach is that
the photon evaluation depends on one's knowledge of the quark
distributions, while the extraction of the quark distributions from
data is itself affected by the photon distribution.

This apparent circularity can be circumvented using an iterative
procedure that exploits the smallness of the QED coupling, or
equivalently of $\delta$.
First one determines a photon distribution, $f_{\gamma}^{(0)}$ as in
Sec.~\ref{sec:our-photon+quarks}, based on QCD parton distributions
obtained from a PDF fit without QED effects,
$f_{q}^{\text{no-QED}}$.
As discussed in the previous subsection,
the resulting photon PDF includes terms up to the order
$(\alpha L)^2 (\as L)^n \sim \delta^2$ and $\alpha (\as L)^n \sim \delta^2$, 
i.e.\ of relative order $\delta$ with respect to the leading term.
One then carries out a global fit with QED effects in the evolution
and coefficient functions,\footnote{We assume, for the principle of the
  demonstration, that the evolution and coefficient functions are
  known to all perturbative orders.
  We also assume that lepton distributions are treated correctly to
  the relevant order.}
and a photon distribution $f_{\gamma}^{(0)}$ that is not changed
during the fit.
This gives a zeroth iteration of the QCD parton distributions with QED
effects, $f_{q}^{(0)}$.
Their accuracy depends on the hard processes being used in the QCD
fit:
\begin{itemize}
\item For Drell-Yan production, i.e.\ $q\bar q \to \ell^+\ell^-$ and
  $\gamma \gamma \to \ell^+\ell^-$, the relative contribution from the
  photon distribution starts at order
  $(f_{\gamma}/f_{q})^2 \sim (\alpha L)^2 \sim \delta^2$ (the
  $\gamma \gamma$ and $q\bar q$ Drell-Yan production channels differ
  only by the sizes of the incoming parton distributions).
  Since the photon distribution is known up to relative order $\delta$, the
  QED corrections to the quark distribution are known up to order
  $\delta^{3}$.
  In general if the photon distribution is known up to relative order
  $\delta^n$, the quark distribution can be extracted to order
  $\delta^{n+2}$.

\item For DIS structure functions, the relative contribution from the
  photon distribution starts at order $\alpha f_{\gamma} / f_{q}
  \sim \alpha^2 L \sim \delta^3$.
  Since the photon distribution is known up to relative order $\delta$, the
  QED corrections to the quark distribution are known up to order
  $\delta^{4}$.
  In general if the photon distribution is known up to relative order
  $\delta^n$, the quark distribution can be extracted to order
  $\delta^{n+3}$.
\end{itemize}
It is the Drell-Yan process, an important input in all modern global PDF
fits, that will limit the accuracy of the iteration, so we concentrate
on that process.
We now use the QCD partons $f_{q}^{(0)}$ to determine an improved
approximation to the photon distribution, $f_{\gamma}^{(1)}$.
Since the QCD partons are known to QED accuracy $\delta^3$, this new
estimate of the photon will be accurate to order
$\alpha L \times \delta^3 \sim \delta^4$.
One can then insert $f_{\gamma}^{(1)}$ into a renewed determination
of the QCD partons to obtain $f_{q}^{(1)}$, which will be accurate
to order $\delta^5$.
In general, at iteration $i$, $f_{\gamma}^{(i)}$ will be accurate up
to and including order $\delta^{2+2i}$, while $f_{q}^{(i)}$ will be
accurate up to and including order $\delta^{3+2i}$.
The convergence of this procedure is therefore fast, and the limiting
consideration will be the availability of coefficient and splitting
functions of the required order.\footnote{As discussed in
  Sec.~\ref{sec:higher-order}, starting from order $\alpha^2 =
  \delta^4$, the photon distribution is itself a direct input to its
  own determination; an iterative type approach also addresses this case.}$^,$%
\footnote{The above discussion assumes that PDF fits for the QCD
  partons are carried out without an imposed momentum sum rule and are
  constrained exclusively by data.
  However, it is quite common for the momentum sum to be imposed.
  In this case the momentum carried by the photon acts as an effective
  input to the fit, since it determines the momentum allowed for the
  QCD partons.
  For the $f_q^{(0)}$ determination, the highest known QED term in the
  momentum sum is $\delta^2$.
  Thus, insofar as the overall accuracy of the fit is limited by the
  input(s) with worst accuracy, after one iteration the absolute accuracy
  for both $f^{(i)}_{\gamma}$ and $f^{(i)}_{q}$ would be
  $\delta^{2+i}$.}

\section{Numerical results}
\label{sec:photon-pdf-results}

In this paper, we have made some small changes to the procedure used
to evaluate the photon PDF compared to our previous paper, as
discussed in Sec.~\ref{sec:scale-uncert-res}. Figure~\ref{fig:compare}
in that section showed the difference in the PDF using the two methods. The change is very small, and well within our errors.

\begin{figure}
 \includegraphics[width=0.48\textwidth,page=1]{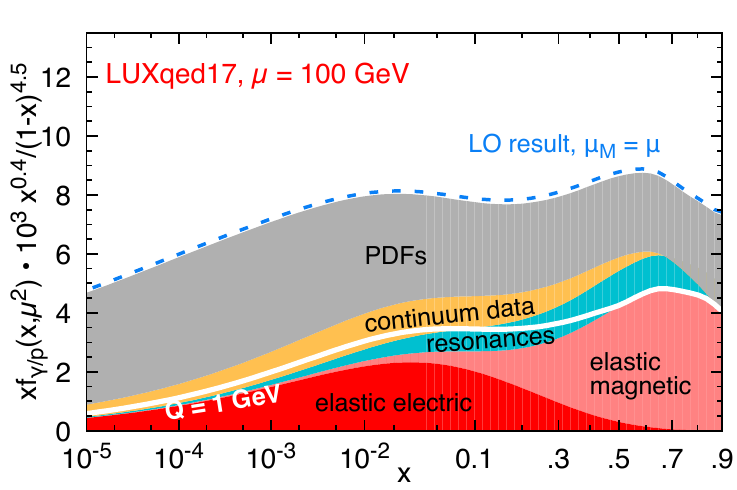}\hfill
  \includegraphics[width=0.48\textwidth,page=8]{Figs/contribs.pdf}
  \caption{The left panel shows the contributions to the photon PDF at
    $\mu=100$~GeV,
    Eqs.~(\ref{eq:fgamma-split}, \ref{6.12a}, \ref{eq:FconExpansion}, \ref{9.6a}),
    multiplied by $10^3 x^{0.4}/(1-x)^{4.5}$, with a breakdown into the various
    components discussed in the text. 
    The white line is the sum of the inelastic contribution from $Q^2 \le 1\,
    \text{(GeV)}^2$ in Eq.~(\ref{6.12a}) and the full
    elastic contribution. 
    The full physical factorisation result of Eq.~(\ref{6.12a}), which is equivalent to a
    LO result, is given by the
    dashed blue line. 
    The right panel shows the same plot for $\mu=500$\,GeV, with the
    scale $\mu_M$ in Eq.~(\ref{eq:mu_1stchoice}) for the LO results
    set to $\mu/2$ or $2\mu$. 
    The total PDF (edge of grey region) is shown for $\mu_M=\mu$. The
    $\MSbar$-conversion term (difference between grey region and
    dashed blue curve) 
    has a significant impact with scale choices other than $\mu_M=\mu$.
}
  \label{fig:breakup}
\end{figure}

The photon PDF computed using the method of this paper is given in the left-hand
panel of Fig.~\ref{fig:breakup} for $\mu=100$\,GeV.
It is rescaled by a factor $10^3 x^{0.4}/(1-x)^{4.5}$ to facilitate
the simultaneous study of different $x$ regions.
The plot includes a breakup into the different contributions discussed in
Sec.~\ref{sec:input-data}. 
There is a sizeable elastic contribution, with an important magnetic
component at large values of $x$.
The resonance and continuum regions are also quantitatively relevant.
The white line represents contributions arising from the $Q^2<1$
region of all the structure functions, including the full elastic
contribution, and this serves to illustrate the importance of a proper
inclusion of the low $Q^2$ region, given the accuracy we aim for.
The PDF contribution, in the physical factorisation scheme, is from
the bottom of the grey region to the blue dashed curve.
The $\MSbar$-conversion term, Eqs.~(\ref{eq:FconExpansion},
\ref{9.6a}), is negative and corresponds to the difference between 
the blue dashed curve (PF result) and the top edge of the grey region
(final full result in the $\MSbar$ scheme).

The right-hand plot of Fig.~\ref{fig:breakup} illustrates how the
components evolve when increasing the factorisation scale $\mu$.
The main change is associated with the $\log Q^2$ growth of the ``PDF''
contribution and is most important at small $x$ values, a consequence
of the fact that the quarks distributions themselves increase rapidly
with $Q^2$ at small $x$.
The elastic, resonance and (low-$Q^2$) continuum contributions to the
photon PDF all depend slowly on $\mu$ via the overall
$1/\alpha(\mu^2)$ factor in Eq.~(\ref{6.12a}).
These components, though formally NLO, remain a significant fraction
of the overall photon PDF, even at this large value of $\mu$.

The right panel in Fig.~\ref{fig:breakup} also shows the impact of
scale variation on the contributions at $\mu=500$\,GeV. 
The blue dashed curves are for $\mu_M=\mu/2$ and $\mu_M=2\mu$ in
Eqs.~(\ref{6.12_alt},\ref{eq:mu_1stchoice}) for the PF photon.
The total $\MSbar$ photon PDF (the top edge of the grey band) uses our
central choice of $\mu_M=\mu$ in Eq.~(\ref{eq:mu_1stchoice}).
The impact 
of a change of $\mu_M$ on the $\MSbar$ photon PDF would be
barely visible in the plot, because the substantial scale dependence
of the PF result is largely cancelled by that of the
$\MSbar$-conversion term, as was noted earlier in the discussion of
Fig.~\ref{fig:scale-dep}.
Previous photon PDFs were at best accurate to leading order, and hence
had much larger scale uncertainties than LUXqed.

\begin{figure}
  \centering
  \includegraphics[width=0.5\textwidth,page=2]{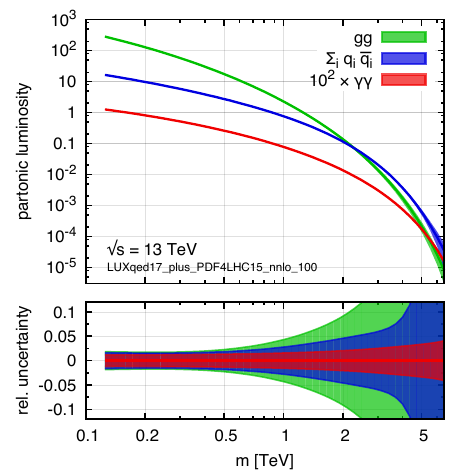}%
  \includegraphics[width=0.5\textwidth,page=6]{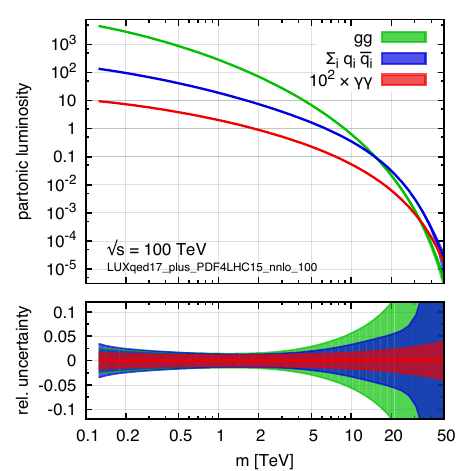}
  \caption{Upper panels:  partonic luminosities in
    $pp$ collisions, as a function of the 
    partonic invariant mass $m$, at centre-of-mass energies of
    $13\TeV$ (left) and $100\TeV$ (right).
    The $\gamma\gamma$ (scaled by $10^2$), $q\bar q$
    and $gg$ luminosities appear from bottom to top.
    Lower panels: the relative uncertainties of the luminosities.
    Our luminosity definition is given in Eq.~(\ref{eq:lumi-def}).  }
  \label{fig:lumi}
\end{figure}

Figure~\ref{fig:lumi} shows the $\gamma\gamma$ luminosity compared
with the $gg$ and total $q \overline q$ luminosities for
$\sqrt{s}=13$\,TeV and 100\,TeV, where the luminosity $\rd L_{ij}/\rd
\ln m^2$ for partons $i$ and $j$ in $pp$ collisions is defined by
\begin{align}
  \label{eq:lumi-def}
  \frac{\rd L_{ij}}{\rd\ln m^2} =
  \frac{m^2}{s} \int \frac{dz}{z}\, f_{i}(z,m^2)\, 
  f_{j}\left( \frac{m^2}{z s}, m^2\right) \,.
\end{align}
For the $\sum_i q_i \overline q_i$ luminosity, we have included a
factor of two in the sum, since either quarks or antiquarks can come
from each beam.  The $\gamma \gamma$ luminosity is about three orders
of magnitude smaller than the $gg$ and $q \overline q$ luminosities
over a wide range of masses. The impact of the $\gamma \gamma$
luminosity is however enhanced in processes with leptons and
electroweak gauge bosons, such as $p p \to WW$, where the $\gamma
\gamma$ process has a $t$-channel exchange diagram which is enhanced
in some kinematic regions.
\begin{figure}
  \centering
  \includegraphics[width=0.5\textwidth,page=2]{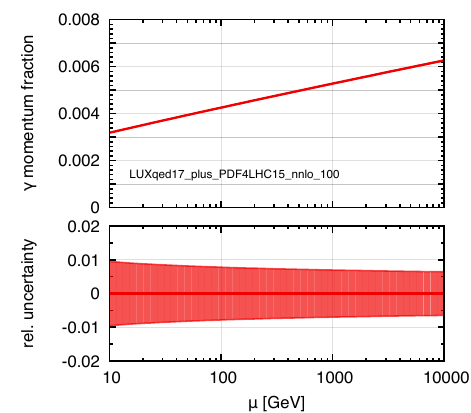}
  \caption{Momentum fraction (top) and its relative uncertainty (bottom)
    carried by the photon, as a function of the factorisation scale
    $\mu$.  
   }
  \label{fig:mom}
\end{figure}

Figure~\ref{fig:mom} shows the photon momentum fraction of the proton
as a function of $\mu$. The momentum fraction is $\sim 0.43\%$ at
$\mu=100 \GeV$, and increases by about 0.1\% for each factor of 10
increase in $\mu$. 
Eventually, neglecting electroweak corrections, the
photon momentum fraction saturates at a value that is independent of
$\alpha$ and $\alpha_s$, however this occurs at trans-Planckian
scales. 
%

\section{Conclusions}
\label{sec:conclusions}

In this paper, we have provided a detailed derivation and explanation
of the results in Ref.~\cite{Manohar:2016nzj}. In addition, we have discussed
several extensions of the results. We have computed the polarised photon
PDFs and the photon TMDPDF using the same method, and given the
corresponding formul\ae.

We have given an alternative derivation of the photon PDF directly
from the operator definition of the photon parton density, and we have
used this definition to compute the $\MSbar$ conversion term in the
photon PDF to one higher order in $\alpha_s$ and $\alpha$ than our
previous result.
Our formalism allows for the computation of the $\MSbar$ conversion
term to yet higher orders using $\MSbar$ DIS coefficient function
expansions in $D$ dimensions.
On the phenomenological side, we have used the higher-order $\alpha_s$
correction to give a more detailed analysis of the theoretical error.

We have highlighted what is needed as experimental input for the
proton structure functions in order to make phenomenological use of
our new higher-order calculations, in particular the QED contributions.
Specifically, our photon PDF formula requires
experimental data on DIS structure functions without removing
the QED corrections on the hadronic side. 
Two-photon
exchange contributions, which couple the hadronic and leptonic sides, 
must still be removed. 
We also stress that,
rather than the
measurement of the proton form factor for the elastic contribution, it
would be useful to have a measurement of the structure functions also
below the hadronic inelastic threshold $W^2=(m_p+m_{\pi})^2$, in order to account
for final states consisting of a proton accompanied by an arbitrary
number of photons and electron-positron pairs.

The evolution of the photon PDF can also be used to obtain
higher order $P_{\gamma x}$ DGLAP splitting functions.
We verified that the resulting
$\alpha \alpha_s$ and $\alpha^2$ $P_{\gamma x}$ unpolarised splitting
functions agree with recent computations~\cite{deFlorian:2015ujt,deFlorian:2016gvk}. 
From our results it is also possible to derive the $\alpha \alpha_s$
polarised splitting function, as well as the unpolarised splitting
functions to one higher order.
We have not, however, given explicit results for them.

We have given a detailed explanation of the data inputs for our photon
PDF, and the resulting uncertainties.  The detailed treatment of
the higher-order contributions used to
evaluate the PDF and estimate uncertainties is slightly different from
the previous version~\cite{Manohar:2016nzj}, as explained in the text,
cf.\ also Fig.~\ref{fig:compare}.
To distinguish this new set from our previous determination, we call
it LUXqed17\_plus\_PDF4LHC15\_nnlo\_100 (or LUXqed17 for brevity),
and it is valid for scales $\mu \ge 10\GeV$.

The few percent precision that we have achieved for the photon PDF is
more than adequate for the computation of photon induced corrections
to Standard Model processes at present. It is possible and natural to
adopt this method in the context of global PDF fits.\footnote{This is
  currently being done by the NNPDF collaboration.}

Since we have shown that a proton collider can be viewed as a broad beam photon
collider, and that the broad-beam distribution can be computed with high precision,
the question remains whether such precision can be fully exploited in
experimental measurements. The first thing that comes to mind is the possibility
to search for totally hadrophobic BSM particles, whose cross section could be
computed with very high precision using the results presented here.
It is tempting to accompany such searches or measurements with a veto on the
event hadronic activity, in view of the large elastic component that we found.
We must point out, however, that even for the elastic component one cannot
guarantee the absence of hadronic activity. 
While this is certainly the case
for lepton-proton collisions, one should remember that in the proton-proton
case the colliding protons are likely to pass close, or even to cross each other,
sometimes giving rise to the production of secondary
hadrons~\cite{Harland-Lang:2016apc}.  
The precision of the photon PDF input would then be spoiled by a less
well known survival probability factor for the protons.

The discussion in this paper has been for the photon PDF of the
proton, but the method can be applied to derive a photon PDF formula
for any hadron. 
Eqs.~(\ref{6.12a}, \ref{9.11}) for the
unpolarised PDF hold without change for any hadron. 
The corresponding polarised formul\ae\ hold only for hadrons of spin
$1/2$.
Evaluating the PDF from the formula requires, of course,
experimental data on the hadronic form factors and structure
functions. For the neutron, the high $Q^2$ data is available in terms
of quark and gluon PDFs, but low-$Q^2$ data is not as accurate as for
the proton. It is also important to keep in mind that at low $Q^2$, a
nuclear target such as a deuteron is not simply the sum of a neutron
and proton. 
Finally, we remark that the methods developed here can also be
extended to derive the lepton distributions in the proton. 

Data files corresponding to the figures in this paper are available
through Zenodo~\cite{ZenodoUpload}.
We will provide LHAPDF~\cite{Buckley:2014ana} files for the
LUXqed17\_plus\_PDF4LHC15\_nnlo\_100 and
LUXqed17\_plus\_PDF4LHC15\_nnlo\_30 sets through LHAPDF.

\section*{Acknowledgements}

We would like to thank Silvano Simula who provided us with a code for
the CLAS parametrisation, Jan Bernauer for discussions of the A1
results and fits, Gunar Schnell for bringing the HERMES \mbox{GD11-P}
fit to our attention and providing the corresponding code, and Gabriel
Lee and Richard Hill for their preliminary form factor fits.
We also thank Stefan Dittmaier, Claude Duhr and Cynthia Keppel for
helpful discussions.
This work was supported in part by ERC Consolidator Grant HICCUP
(No.\ 614577), ERC Advanced Grant Higgs@LHC (No.\ 321133), a grant
from the Simons Foundation (\#340281 to Aneesh Manohar), by DOE grant
DE-SC0009919, and NSF grant NSF PHY11-25915.
AM and PN acknowledge CERN-TH for hospitality during part of this
work.  AM, GPS and GZ thank MIAPP for hospitality while this work was
being completed.

\appendix

\section{QED corrections}\label{app:emcurrent}

In this Appendix, we summarise some known results on the
renormalisation of the electromagnetic current~\cite{Lurie:1968zz,Beneke:2003zv,Collins:2005nj}
which will be useful for the subsequent discussion. The formul\ae\ are
given for the case of QED for simplicity, but they are trivially
generalised to the case of QCD with electromagnetic interactions.

The QED Lagrangian, including the gauge-fixing term is
\begin{align}
  \mathcal{L} ={} &
    \bar\psi^{(0)}  \left( i \slashed{\partial} 
                            +e_0 \slashed{A}^{(0)}
                            -m_0 
                     \right)
         \psi^{(0)}  
   -\frac{1}{4} \left( F_{\mu\nu}^{(0)}  \right)^2
   - \frac{1}{2\xi} (\partial\cdot A)^2 \nn
    ={}&
    Z_\psi \bar\psi  \left( i \slashed{\partial} 
                            +e\smu^\epsilon \slashed{A}
                            -Z_m m 
                     \right)
         \psi
   -\frac{Z_A}{4} \left( F_{\mu\nu}  \right)^2    - \frac{1}{2\xi} (\partial\cdot A)^2.
   \label{A3.1}
\end{align}
Bare quantities are labelled with a subscript $0$.
The renormalisation constants are defined through 
\begin{align}
e_0 &= Z_e e\, \smu^\epsilon, &
m_0 &= Z_m m , &
A_\mu^{(0)} &= Z_A^{1/2} A^\mu, &
\psi^{(0)} &= Z_\psi^{1/2} \psi \,,
\label{A3.2}
\end{align}
where
$  {\cal S}^2={e^{\gamma_{\rm E}}}/{4\pi}$, and the QED Ward identity implies that 
\begin{align}
Z_eZ_A^{1/2}=1\,.
\label{zeza}
\end{align}
The gauge-fixing parameter $\xi$ is not renormalised, by BRST invariance.

\begin{figure}
\centering
\includegraphics[height=1.5cm]{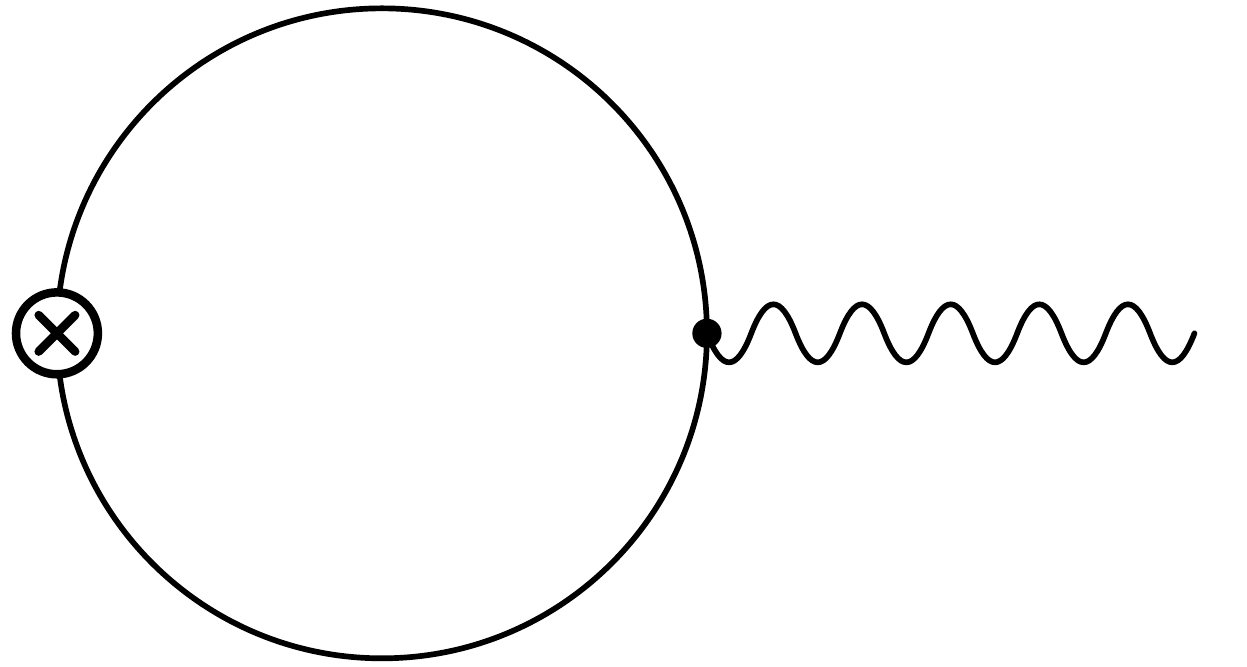}
\caption{\label{fig:03:1} Penguin graph which renormalises the electromagnetic current.}
\end{figure}

The QED $\beta$ function is given in terms of the anomalous dimension
$\gamma_A$ of the photon field,
\begin{align}
\mub \frac{\mathd e}{\mathd \mub} &= - \epsilon e + \beta(e) = - \epsilon e + e \gamma_A (e)\,, &
\mub^2 \frac{\mathd \alpha }{\mathd \mub^2} &= - \epsilon \alpha + \alpha \gamma_A (\alpha)\,,
\label{A3.3}
\end{align}
where $e(\mub)$ is the $\MSbar$ coupling, and
\begin{align}
\gamma_A(e) &= \gamma_0 \frac{e^2}{16\pi^2} + \gamma_1 \left(\frac{e^2}{16\pi^2}\right)^2 + \ldots\,, &
\gamma_A(\alpha) &= \gamma_0 \frac{\alpha}{4\pi} + \gamma_1 \left(\frac{\alpha}{4\pi}\right)^2 + \ldots\,.
\label{A3.4}
\end{align}
The Noether current is
\begin{align}
j^\mu_N &= Z_\psi \overline \psi \gamma^\mu \psi = \overline \psi^{(0)} \gamma^\mu \psi^{(0)}\,.
\label{A3.5}
\end{align}
The usual (incorrect) textbook statement is that the electromagnetic
current is a conserved current, and not renormalised, so that
$j^\mu_N$ has finite matrix elements. This is false, because of the
well-known penguin graph Fig.~\ref{fig:03:1}.
As shown in Ref.~\cite{Lurie:1968zz,Beneke:2003zv,Collins:2005nj}, the renormalised current in the $\MSbar$ scheme is
\begin{align}
  j^\mu_{\MSbar}
  =
  \bar \psi^{(0)} \gamma^\mu \psi^{(0)} 
              + \frac{ 1 - Z_A^{-1} }{ e_0 }
                \partial_\nu F^{(0)\, \nu \mu}\,,
 \label{A3.6}
\end{align}
which satisfies the renormalisation group equation
\begin{align}
\mub \frac{ {\rm d} }{ {\rm d} \mub } j^\mu_{\MSbar} =-2\gamma_A\  j^\mu_{\MSbar}
\qquad\mbox{(in physical matrix elements)},
\label{A3.7}
\end{align}
where the Gupta-Bleuler condition $\partial \cdot A =0$ has been used
for physical matrix elements.

There is a unique \emph{local} electromagnetic
current
\begin{align}
  j^\mu ={}& j^\mu_{\MSbar} 
          + \frac{ \Pi(0,\mub) \,  \partial_\nu F^{\nu\mu} }
                 { e\, \smu^\epsilon }\,,
\label{A3.8}
\end{align}
which satisfies the usual QED Ward identities, and is not renormalised,
\begin{align}
\mub \frac{ {\rm d} }{ {\rm d} \mub } j^\mu =0\,.
\label{A3.9}
\end{align}
Here $\Pi(q^2,\mub)$ is the vacuum polarisation,\footnote{The sign
  convention for $\Pi$ used here is the opposite of that in
  Ref.~\cite{Collins:2005nj}.} defined so that the renormalised photon
propagator is
\begin{equation}
\label{A3.10}
    \frac{i \left( -g_{\mu\nu} + q_\mu q_\nu/q^2\right) }
         { q^2 \, [1 - \Pi(q^2,\mub) ] }
  - \frac{ i q_\mu q_\nu \xi }{ (q^2)^2 }.
\end{equation}

\def\jir{\ensuremath{G_{1\gamma{\rm I}}}}
\def\jtot{\ensuremath{G_{\rm tot}}}

In physical matrix elements, the equation of motion for the photon
field gives
\begin{align}
\label{A3.11}
j^\mu 
  ={}&
    [1 - \Pi(0,\mub)] \ j^\mu_{\MSbar}
\qquad\mbox{(in physical matrix elements)}.
\end{align}
%
%
\begin{figure*}
\centering
\begin{tabular}[t]{c@{\hspace*{15mm}}c@{\hspace*{15mm}}c}
   \includegraphics[scale=0.25]{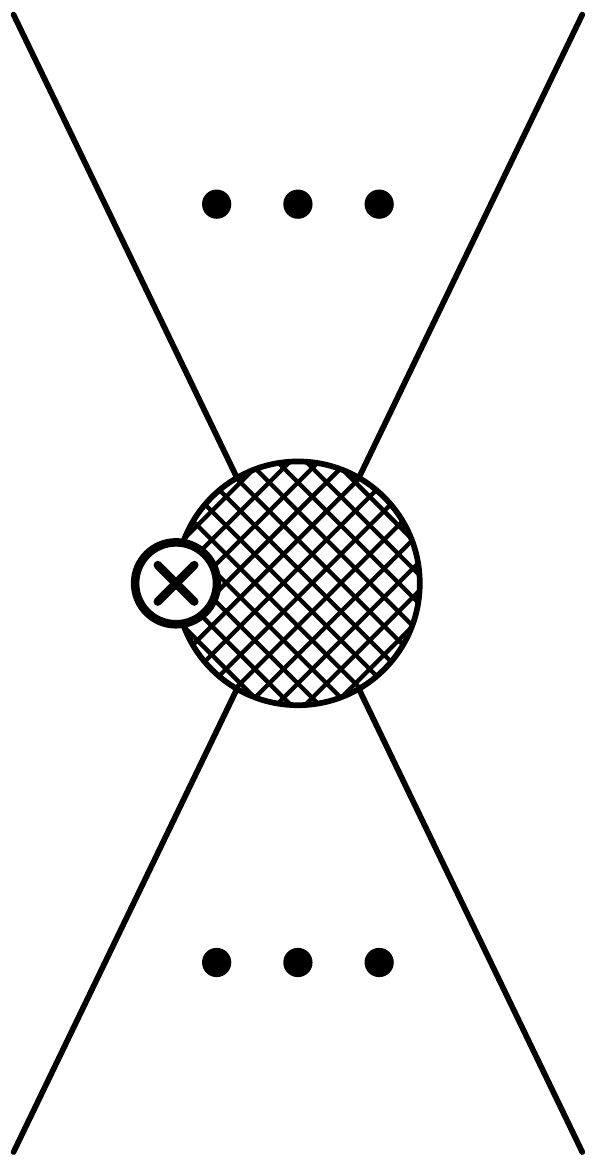}
   &\includegraphics[scale=0.25]{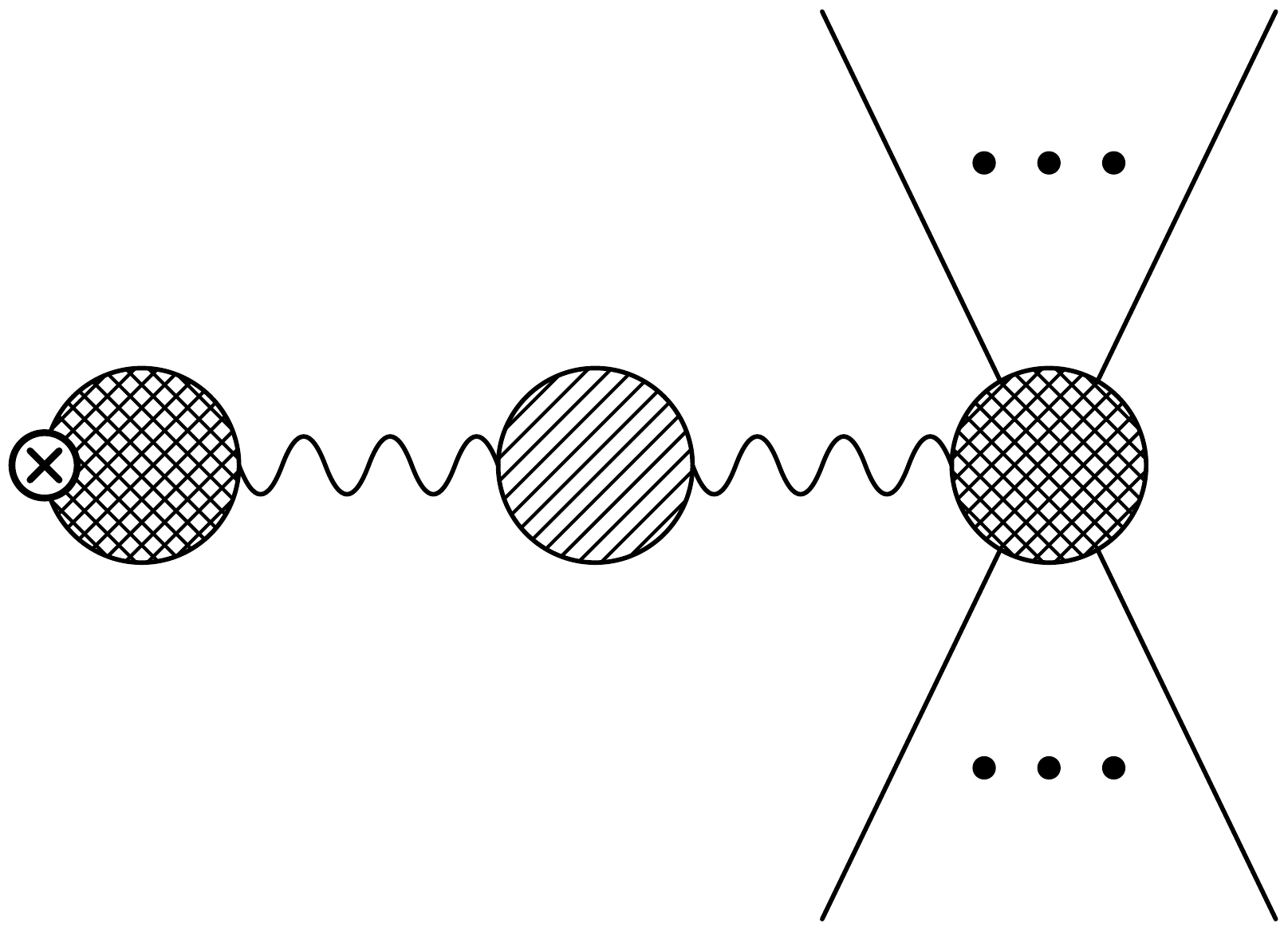}
   & \includegraphics[scale=0.25]{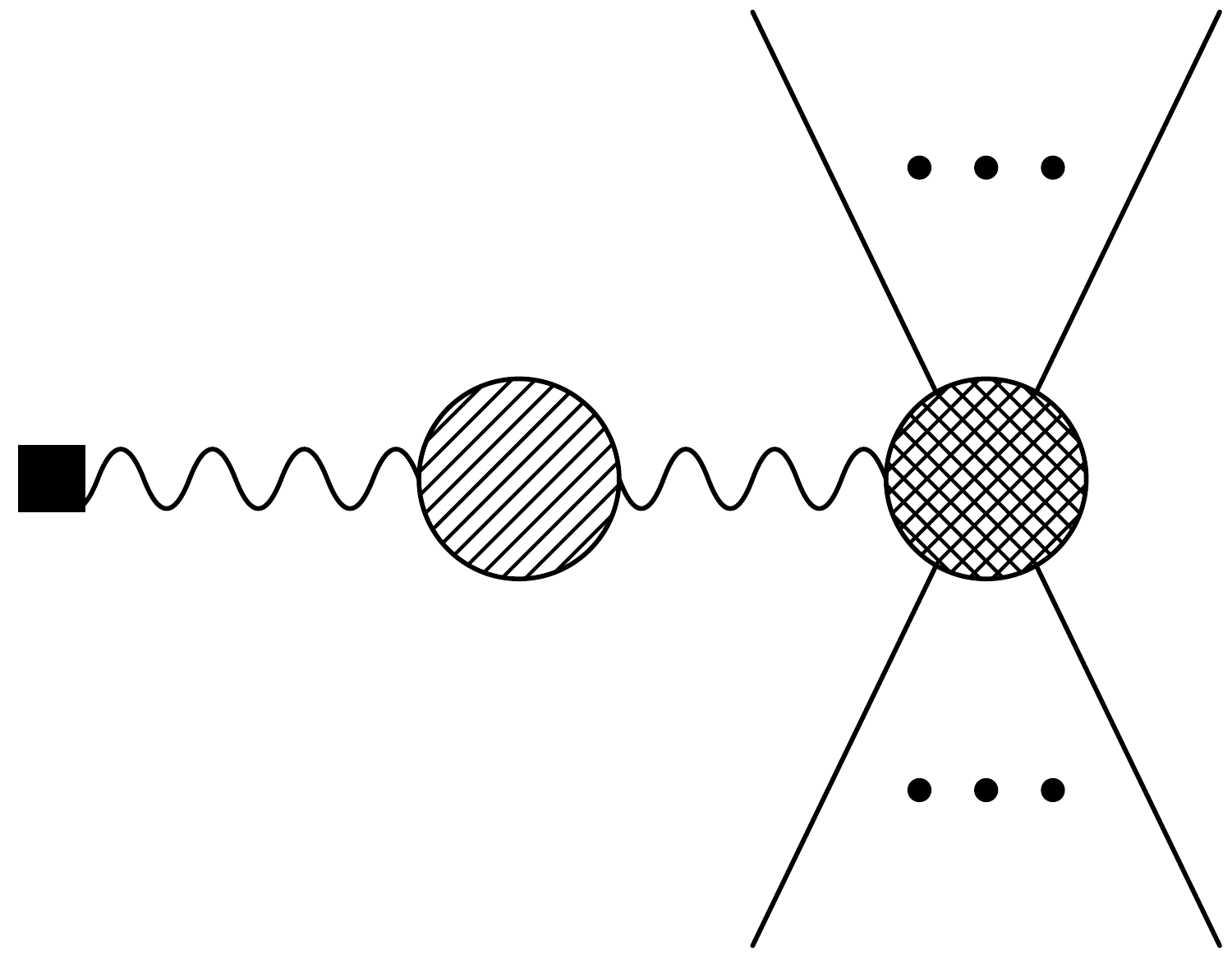}
\\[10pt]
   (a) & (b) & (c)
\end{tabular}
\caption{(a) Graphs that are one-particle-irreducible in the current
  channel for insertion of a current vertex in a Green function or
  matrix element; the standard non-renormalisation argument applies
  only to these. (b) These graphs, one-particle-reducible in the
  current channel, also contribute to matrix elements of the current
  and to its renormalisation. The two subgraphs that are cross hatched
  are irreducible in the photon line, while the other subgraph gives
  the full propagator corrections to the photon propagator. (c)
  Counterterm to (b).  The filled square corresponds to an operator
  proportional to $\partial_\nu F^{\nu\mu}$.
 \label{fig:03:2}
} 
\end{figure*}

$\eph(q^2)$ is defined in Eq.~(\ref{3.15}). The renormalisation group
equation for the inverse photon propagator
\begin{align}
i\Gamma^{(2)}_{\mu \nu}(q^2,\mub)
&= P_{\mu \nu}\,q^2 \left[1 - \Pi(q^2,\mub)\right]-\frac{1}{\xi} q_\mu q_\nu \,, &
P_{\mu\nu} &= g_{\mu \nu} - \frac{q_\mu q_\nu}{q^2}  \,,
\label{A3.16}
\end{align}
\begin{align}
\left[\mub \frac{\partial}{\partial \mub} + \mub \frac{\mathd e}{\mathd \mub}
\frac{\partial}{\partial e} -2 \gamma_A \right] \Gamma^{(2)}_{\mu \nu}(q^2,\mub) =0\,,
\label{A3.17}
\end{align}
and Eq.~(\ref{A3.3},\ref{A3.17}) imply that $e_{\text{ph}}(q^2)$ is
independent of $\mu$. $\eph(q^2=0)$ is the usual electron charge used
in classical physics. $\eph(q^2)$ depends on the exact vacuum
polarisation function, and is non-perturbatively related to $e(\mub)$
in the presence of strong interactions, unless $q^2$ is at large
Euclidean values, where perturbation theory holds.

We can express an arbitrary Green function or matrix element of
$j^\mu$ in terms of $G_{1\gamma I}^\mu$, which is the
1-photon-irreducible part in the current channel.  The total Green
function is
\begin{align}
G_j^\mu &= G^\mu_{1\gamma I} + \frac{\Pi(q^2,\mub)}{1 - \Pi(q^2,\mub)} P_{ \mu\nu} G^\nu_{1 P I} - \frac{\Pi(0,\mub)}{1 - \Pi(q^2,\mub)} P_{\mu\nu} G^\nu_{1 P I}\,,  
\label{A3.12}
\end{align}
from Graphs Fig.~\ref{fig:03:2}(a), (b) and (c), respectively, giving
\begin{equation}
\label{A3.12a}
G^\mu_{j}
=
G_{1\gamma I}^\mu  \frac{ 1 - \Pi(0,\mub) }{ 1 - \Pi(q^2,\mub) }
  - G_{1\gamma I}^\nu \frac{q_\nu q^\mu}{q^2}
    \frac{ \Pi(q^2,\mub) - \Pi(0,\mub) }{ 1 - \Pi(q^2,\mub) }.
\end{equation}
In physical matrix elements, the last term vanishes by current conservation, so
\begin{equation}
\label{A3.13}
G^\mu_{j} = G_{1\gamma I}^\mu \frac{ 1 - \Pi(0,\mub) }{ 1 -
  \Pi(q^2,\mub) } \quad\mbox{(in physical matrix elements)}.
\end{equation}
Since $j^\mu$ is $\mu$-independent, so is $G_j^\mu$ and hence also
$G_{1\gamma I}^\mu$. The $1\gamma I$ graph is the same whether one
uses the current $j^\mu_N$ or $j^\mu_{\MSbar}$ or $j^\mu$, since the
three operators only differ in their $1\gamma I$ part.

The one-photon-irreducible part is given by matrix elements of the
non-local current
\begin{align}
  j^\mu(q) ={}& j^\mu_{\MSbar} (q)
          + \frac{ \Pi(q^2,\mub) \,  \partial_\nu F^{\nu\mu}(q) }
                 { e\smu^\epsilon }\,,
\label{A3.14}
\end{align}
since it is Eq.~(\ref{A3.12}) with $\Pi(0,\mub) \to \Pi(q^2,\mub)$ in the last term.

We can now relate physical matrix elements of the current $j^\mu$
defined in Eq.~(\ref{A3.8}) and the $\MSbar$ current to
one-photon-irreducible matrix elements,
\begin{align}
\braket{X | j^\mu | Y}  &= \left[ \frac{\aph(q^2)}{\aph(0)} \right] \braket{X | j^\mu | Y}_{1\gamma I} \,, \nn
\braket{X | j^\mu_{\MSbar} | Y}  &= \left[ \frac{\aph(q^2)}{ \alpha(\mub) \smu^{2 \epsilon} } \right] \braket{X | j^\mu | Y}_{1\gamma I}\,,
\label{A3.18}
\end{align}
using Eq.~(\ref{3.15}).
$\braket{X | j^\mu | Y}$ is independent of $\mu$, as is $\braket{X |
  j^\mu | Y}_{1\gamma I}$, but \emph{not} $\braket{X | j^\mu_{\MSbar} |
  Y}$. 
 Similar expressions hold for matrix elements with multiple insertions
 of the electromagnetic current.

In scattering processes with external (i.e.\ on-shell) photons, the
LSZ reduction formula says that the $S$-matrix is given by
\begin{align}
\braket{X \gamma|S|Y} &=\mathfrak{R}^{1/2} e \smu^\epsilon \braket{X | j^\mu | Y}_{1\gamma I} \,,
\label{A3.29a}
\end{align}
where the wavefunction factor is
\begin{align}
\mathfrak{R}^{1/2} &= \left[ 1 - \Pi(0,\mub^2) \right]^{-1/2}= \frac{e_{\text{ph}}(0)}{e(\mub) \smu^{\epsilon}}\,,
\label{A3.30}
\end{align}
so that
\begin{align}
\braket{X \gamma|S|Y} &=e_{\text{ph}}(0) \braket{X | j^\mu | Y}_{1\gamma I} \,,
\label{A3.29}
\end{align}
and external photons couple with strength $e_{\text{ph}}(0)$. Internal
photon lines still have the coupling $e(\mub) \smu^\epsilon$.

It is convenient in our discussion of the photon PDF to use
one-photon-irreducible matrix elements, even though it is the matrix
element of a non-local current Eq.~(\ref{A3.14}). One can convert
expressions to those of the local current Eq.~(\ref{A3.8}) using
Eq.~(\ref{A3.18}).

The proton hadronic tensor is defined in Eq.~(\ref{eq:hadtens}).  We
will define three versions of this equation, $W_{\mu \nu}$ which uses
the current $j^\mu$ in Eq.~(\ref{A3.8}), $W^{\mu\nu}_{\MSbar}$, which
uses the $\MSbar$ current Eq.~(\ref{A3.6}), and $W^{\mu\nu}_{1 \gamma
  I}$, which uses the non-local current Eq.~(\ref{A3.14}), or
equivalently, is given diagrammatically in terms of one-photon
irreducible graphs. The relation between these is
\begin{align}
W^{\mu \nu} &= \left[ \frac{\aph(q^2)}{\aph(0)} \right]^2 W^{\mu \nu}_{1\gamma I} \,, \nn
W^{\mu \nu}_{\MSbar} &= \left[ \frac{\aph(q^2)}{ \alpha(\mub) \smu^{2 \epsilon} } \right]^2 W^{\mu \nu}_{1\gamma I}\,.
\label{A3.20}
\end{align}
from Eq.~(\ref{A3.18}). All three hadronic tensors are renormalised
using the $\MSbar$ scheme. The difference is which current is used in
the hadronic matrix element in Eq.~(\ref{eq:hadtens}).

One has to be careful about the definition of $W_{\mu\nu}$ and the
structure functions in the presence of electromagnetic
corrections. $W_{\mu \nu}$ defined using the $\mu$-independent current
Eq.~(\ref{A3.8}), and \emph{including all electromagnetic corrections
  to the matrix element} is $\mu$-independent, so the structure
functions in the decomposition Eq.~(\ref{3.21}) are also
$\mu$-independent, and depend only on $\xbj$ and $Q^2$. Note that if
one had instead used the $\MSbar$ current, the structure functions
would depend on $\mu$ as well as $\xbj$ and $Q^2$.

The one-photon-irreducible $F_2$ structure function is denoted $F_2^{1
  \gamma I}$, and is related to $F_2$ by
\begin{align}
F_2(\xbj,Q^2) &= \left[ \frac{\aph(q^2)}{\aph(0)} \right]^2 F_2^{1 \gamma I}(\xbj,Q^2)\,,
\label{A3.23}
\end{align}
Similarly, the $\MSbar$ structure function is
\begin{align}
F_2^{\MSbar}(\xbj,Q^2,\mub) &= \left[ \frac{\aph(q^2)}{ \alpha(\mub) \smu^{2 \epsilon} } \right]^2 F_2^{1 \gamma I}(\xbj,Q^2)\,,
\label{A3.24}
\end{align}
and depends on $\mu$. Similar results hold for the other structure
functions.

The structure function $F_2$ (and similarly for the others) can be
computed using the operator product expansion (OPE) as the convolution
of short distance coefficient functions $C_{2,i}(x,Q^2,\mub)$ and PDFs
$f_{i}(x,\mub)$. Graphically, it is easiest to compute $C^{1 \gamma
  I}_{2,i}(x,Q^2,\mub)$, and then obtain $C_{2,i}(x,Q^2,\mub)$ from
Eq.~(\ref{A3.23}).

\section{Kinematics}\label{sec:kin}

In this section, we summarise the kinematics for the scattering
process $k + p \to k^\prime + X$, where $p$ is the incoming proton
momentum with $p^2=\mpr^2$, $k$ is an incoming massless particle
momentum with $k^2=0$, and $k^\prime$ is an outgoing particle momentum
with $(k^\prime)^2=M^2$. In deep inelastic scattering (DIS), $M=0$,
but we will need to consider cases where $M \not =0$ as well. The
momentum transfer $q$ is defined as $q=k-k^\prime$, so that the final
hadronic state has invariant mass $m_X^2=(p+q)^2 \ge \mpr^2$.

We adopt light-cone coordinates, and introduce two null vectors $n$
and $\overline n$ which define the collision axis, with $n$ along the
$k$ direction, such that
\begin{align}
  k^\mu &= \frac12 k^- {n}^\mu, &
  p^\mu &= \frac12 p^+ {\overline n}^\mu + \frac12 p^- {n}^\mu, &
  q^\mu &=\frac12 q^+ {\overline n}^\mu + \frac12 q^- {n}^\mu  +\qperp^\mu,
  \label{2.1}
\end{align}
and $n^2=\bar{n}^2=0$ and $n \cdot \bar{n}=2$. Then $p^+=n \cdot p$,
$p^- =\overline n \cdot p$, etc. We adopt the convention
\begin{align}
  \qperp^2=-\qperp^\mu \qperp^\nu g_{\mu\nu} > 0\,,
  \label{2.2}
\end{align}
and use the standard definitions
\begin{align}
  Q^2 =& -q^2 = -q^+q^- + \qperp^2, &
  \xbj &= \frac{Q^2}{2 p\cdot q} = \frac{Q^2}{p^+ q^-+p^- q^+}, &
  y &= \frac{p \cdot q}{p \cdot k}
  \label{eq:xbjdef}\,.
\end{align}
We also introduce the variable
\begin{align}
  \ki &= -\frac{q\cdot k}{p\cdot k}=-\frac{q^+}{p^+}\,.
   \label{eq:kidef}
\end{align}

The $\mathd^4 q$ integration measure in light-cone coordinates is
\begin{align}
  \mathd^4 q=\frac{\mathd q^+ \mathd q^-}{2} \mathd^2\qperp=\frac{\pi}{2}
  \mathd q^+ \mathd q^- \mathd \qperp^2 \frac{\mathd \phi}{2\pi}\,.
  \label{2.3}
\end{align}
Eq.~(\ref{2.3}) uses $q^+$, $q^-$ and $q_\perp^2$ as the independent
variables. In a fixed Lorentz frame, with $p^\pm$ given, $q^\pm$ and
$q_\perp^2$ can be used to determine $Q^2$, $\xbj$ and $\ki$ using
Eqs.~(\ref{eq:xbjdef}) and Eq.~(\ref{eq:kidef}). We can express
$q^\pm$, $\qperp^2$ as a function of $\ki$, $\xbj$ and $Q^2$ by
inverting these equations,
\begin{align}
  q^+ &= -p^+\ki, &
  q^- &= \frac{Q^2}{\xbj p^+}+\ki p^-, &
  \qperp^2 &= Q^2\left(1-\frac{\ki}{\xbj}\right)-\ki^2 \mpr^2.
  \label{2.4}
\end{align}
We thus have a one-to-one mapping of the variables $\qperp^2$, $q^+$,
$q^-$ into $Q^2$, $\xbj$ and $\ki$. However, the inverse mapping
yields a valid kinematic point if and only if $\qperp^2$ is
non-negative. Therefore we must add the constraint
\begin{equation}
 Q^2\left(1-\frac{\ki}{\xbj}\right)-\ki^2 \mpr^2 >0.
   \label{2.5}
\end{equation}
The relevant Jacobian is easily computed to be $Q^2/\xbj^2$, so that
the phase space in terms of the new variables becomes
\begin{equation}
  \mathd^4 q= \frac{\pi}{2}\; \mathd \ki \, Q^2 \mathd Q^2\,
  \frac{\mathd \xbj}{\xbj^2}\frac{\mathd \phi}{2\pi}\,.
  \label{2.6}
\end{equation}

In computing the total scattering cross section, the integral over $q$
is restricted by $\theta$ and $\delta$ functions (see Eq.~(\ref{4.4}))
which ensure that the final state has positive energy and the correct
invariant mass. The energy theta functions are easily evaluated if we
choose $n$ and $\bar{n}$ such that $p^+=p^-=\mpr$, i.e.\ we work in
the proton rest frame.

Let $k=(E,0,0,E)$ and $k^\prime=(E^\prime,0,p^\prime \sin \theta,
p^\prime \cos \theta)$ with $(k^\prime)^2=M^2$. Since $p_X^0=p^0+q^0 =
\mpr + E - E^\prime \ge \mpr$, we have $E \ge E^\prime$.
\begin{align}
Q^2 &= 2 E(E^\prime-p^\prime \cos \theta) - M^2 \ge 2 E(E^\prime-p^\prime ) - M^2 = \left[ \frac{2E}{E^\prime + p^\prime} - 1 \right] M^2\,.
\label{2.25}
\end{align}
Since $2 E \ge 2 E^\prime \ge E^\prime + p^\prime \ge 0$, the factor
in square brackets is positive, and
\begin{align}
Q^2 \ge 0\,.
\label{2.26}
\end{align}
We have
\begin{align}
k^0 &= \frac{p \cdot k}{\mpr} > 0 \implies p \cdot k > 0\,.
\label{2.7}
\end{align}
From
\begin{equation}
  q^0=\frac{q^++q^-}{2}=\frac{Q^2}{2\xbj \mpr}\,,
  \label{2.8}
\end{equation}
we get
\begin{align}
2(k^\prime)^0=2(k^0-q^0) =\frac{2 p\cdot k}{m_p} -\frac{Q^2}{\xbj m_p} > 0\,, 
\label{2.9}
\end{align}
and
\begin{align}
2(p^0 + q^0) \ge 2 \mpr \implies \frac{Q^2}{\xbj} \ge 0 \implies \xbj \ge 0\,.
\label{2.9a}
\end{align}
The invariant mass inequality $m_X^2 =(p+q)^2 \ge \mpr^2$ gives
\begin{eqnarray}
Q^2\frac{1-\xbj}{\xbj} \ge 0 \implies \xbj \le 1\,,
\label{2.10}
\end{eqnarray}
and the invariant mass equality $(k-q)^2=M^2$ gives
\begin{eqnarray}
2 (p \cdot k)\chi = M^2+Q^2 \,.
\label{2.11}
\end{eqnarray}
If instead $k^\prime$ is integrated over some mass range, then we only have $M^2 \ge 0$, and Eq.~(\ref{2.11}) becomes the inequality
\begin{eqnarray}
2 (p \cdot k)\chi \ge Q^2 \,.
\label{2.11a}
\end{eqnarray}

From Eq.~(\ref{eq:xbjdef}),
\begin{eqnarray}
y = \frac{Q^2}{2 \xbj (p \cdot k)}
\label{2.12}
\end{eqnarray}
and so Eqs.~(\ref{2.26},\ref{2.7},\ref{2.9},\ref{2.9a}) give
\begin{eqnarray}
0 \le y \le 1.
\label{2.13}
\end{eqnarray}
The $q_\perp^2 \ge 0$ condition Eq.~(\ref{2.5}) gives the inequality
\begin{align}
\frac{Q^2}{\xbj} \left(\xbj-\chi \right) \ge 0,
\label{2.14}
\end{align}
and so, from Eq.~(\ref{2.9a}),
\begin{align}
\xbj \ge \chi\,.
\label{2.15}
\end{align}

In summary the phase space, including the constraints but assuming
that $k'$ is integrated over some mass range, is as follows: the
integration measure is given by Eq.~(\ref{2.6}) and the integration
region obeys the following constraints: $0 \le \xbj \le 1$,
Eq.~(\ref{2.5}) and Eq.~(\ref{2.11a}).

Here we will also assume that the final state $k^\prime$ is a particle
with mass $M$. In this case, the phase space will also include a
factor
\begin{eqnarray} \label{eq:deltaM}
  2\pi \delta((k-q)^2-M^2)= \frac{2\pi}{2 p\cdot k}
  \delta\left(\ki - \frac{Q^2+M^2}{2 p \cdot k}\right)\,,
\end{eqnarray}
that will lead to the elimination of the $\mathd \ki$ integration.  In this case,
Eq.~(\ref{2.5}) becomes
\begin{align}
\xbj \ge \frac{2 p \cdot k (M^2+Q^2)Q^2}{4(p \cdot k)^2 Q^2 - \mpr^2(M^2+Q^2)^2}\,.
\label{2.20}
\end{align}
The minimum allowed value of $\xbj$ is
\begin{equation}
\xbj = \frac{M^2}{2(p\cdot k - M m_p)}\,.  
\label{eq:xbjmin1}
\end{equation}

Furthermore, we will introduce a variable that plays the role of the
photon momentum fraction in the partonic interpretation of the process
\begin{equation} \label{eq:phxdef}
\xph = \frac{M^2}{2 p\cdot k}.
\end{equation}

The limits on the $Q^2$ integration at fixed $\xbj$ are easily worked
out from the last theta function in Eq.~(\ref{2.20}) to yield
\begin{equation} \label{eq:q2maxmin}
  \Qtwomaxmin = \frac{\xbj -\xph - 
    \frac{2\xbj\xph^2\mpr^2}{M^2} \pm \sqrt{( \xbj - \xph)^2
      - \frac{4\xbj^2 \xph^2 \mpr^2}{M^2}}}{\frac{2\xph}{M^2}\left(1 +
      \frac{\xbj \xph\, \mpr^2}{M^2}\right)}. 
\end{equation}
The limits obey the relation 
\begin{equation}
  \Qtwomin = \frac{\xbj \xph \,  \mpr^2 M^2}{\left( 1 + \frac{\xbj \xph \,
  \mpr^2}{M^2} \right) \Qtwomax}\,.  
  \label{2.21}
\end{equation}
Eq.~(\ref{eq:q2maxmin}) requires $\xbj - \xph$ to be positive,
otherwise $\Qtwomax < 0$. The positivity of the discriminant requires
\begin{equation} \label{eq:xbjmin}
  \xbj - \xph > \frac{2 \xbj  \xph \, \mpr}{M} \quad \Rightarrow \quad \xbj > 
  \frac{\xph}{1-\frac{2 \xph\, \mpr}{M}},
\end{equation}
which is equivalent to Eq.~\eqref{eq:xbjmin1}, and the form of
Eq.~(\ref{eq:q2maxmin}) guarantees that $\Qtwomax>\Qtwomin$.  Thus,
the only constraint on $\xbj$ is Eq.~(\ref{eq:xbjmin}).
For future convenience, we also introduce the variable
\begin{equation}
  \zed=\frac{\xph}{\xbj}
  \label{eq:zeddef}
\end{equation}
and trade $\xbj$ for $\zed$ in the phase space expression. 

Our final result, including the factor in Eq.~(\ref{eq:deltaM}),
becomes
\begin{multline}
  \int \frac{\mathd^4 q}{(2\pi)^4} 2\pi
  \delta((k-q)^2-M^2)\theta(k^0-q^0) \theta((p+q)^2-\mpr^2)
  \theta(p^0+q^0) =  \\
  \frac{1}{16\pi^2 M^2} \int_\xph^{1-\frac{2 \xph\, \mpr}{M}} \mathd \zed \int_{\Qtwomin}^{\Qtwomax} Q^2 \mathd Q^2 \int_{-\pi}^{\pi}
  \frac{\mathd \phi}{2\pi}\,,
  \label{2.22}
 \end{multline}
with
\begin{eqnarray}
  \Qtwomax &=&
  M^2\left(\frac{1-\zed}{\zed}\right)\frac{1-\frac{2\xph^2
      \mpr^2}{(1-\zed)M^2} +\sqrt{1-\frac{4 \xph^2 \mpr^2}{(1-\zed)^2
        M^2}}}{2\left(1+\frac{\xph^2 \mpr^2}{\zed \,M^2}\right)}\,,
  \nn \Qtwomin &=& \frac{\xph^2
    \mpr^2}{1-\zed}\frac{2}{1-\frac{2\xph^2 \mpr^2}{(1-\zed)M^2}
    +\sqrt{1-\frac{4 \xph^2 \mpr^2}{(1-\zed)^2 M^2}}}\,.
    \label{2.23}
\end{eqnarray}
Expanding the limits in $\mpr^2/M^2$ gives
\begin{align}
  \Qtwomax &\to \Qtwomaxexp = \frac{ M^2(1-\zed) }{ \zed } \,, &
  \Qtwomin &\to \Qtwominexp = \frac{ \mpr^2 \xph^2 }{1-\zed }\,. 
\label{2.24}
\end{align}
We will use $\Qtwomaxexp,\Qtwominexp$ as upper and lower limits when
we compute the DIS cross section neglecting power corrections.

\section{Photon PDF from $\gamma \gamma \to S$}\label{app:gamgam}

In this Appendix we illustrate the universality of our photon PDF
result. Consider another BSM probe, a heavy scalar $S$ with mass
$M$ that couples to $\gamma \gamma$ via the interaction Lagrangian
\begin{align}
\mathcal{L} &=  \frac{c e^2 (\mub) }{\Lambda} S\, F_{\mu \nu} F^{\mu \nu}\,.
\label{4.23}
\end{align}
The Lagrangian coefficient $c$ is not renormalised at one loop order,
but now there are higher order corrections since $\gamma \gamma$ can
interact via fermion loops at two-loop order. The interaction
Eq.~(\ref{4.23}) leads to the
$\gamma(k_1,\epsilon_1)+\gamma(k_2,\epsilon_2) \to S$ vertex
(neglecting irrelevant phase factors)
\begin{align}
  -\frac{2c e^2 (\mub) }{\Lambda} \left[ (k_1 \cdot k_2 )(\epsilon_1
    \cdot \epsilon_2) - (k_1 \cdot \epsilon_2) (k_2 \cdot
    \epsilon_1)\right]\,.
\label{4.24b}
\end{align}
The spin-averaged $\gamma\gamma \to S$ cross section to lowest order is
\begin{align}
  \sigma_{\gamma \gamma} &= \widehat \sigma(\gamma \gamma \to S)
  =\sigma_0 M^2 \delta(s-M^2) \,, & 
  \sigma_0 &= \frac{\pi c^2 e^4
    (\mub) }{2\Lambda^2}\,.
\label{4.25}
\end{align}
The cross section coefficient has again been called $\sigma_0$, so the
formul\ae\ can be easily compared with the $l \to L$ case.

\begin{figure}
\begin{center}
\includegraphics[width=6cm]{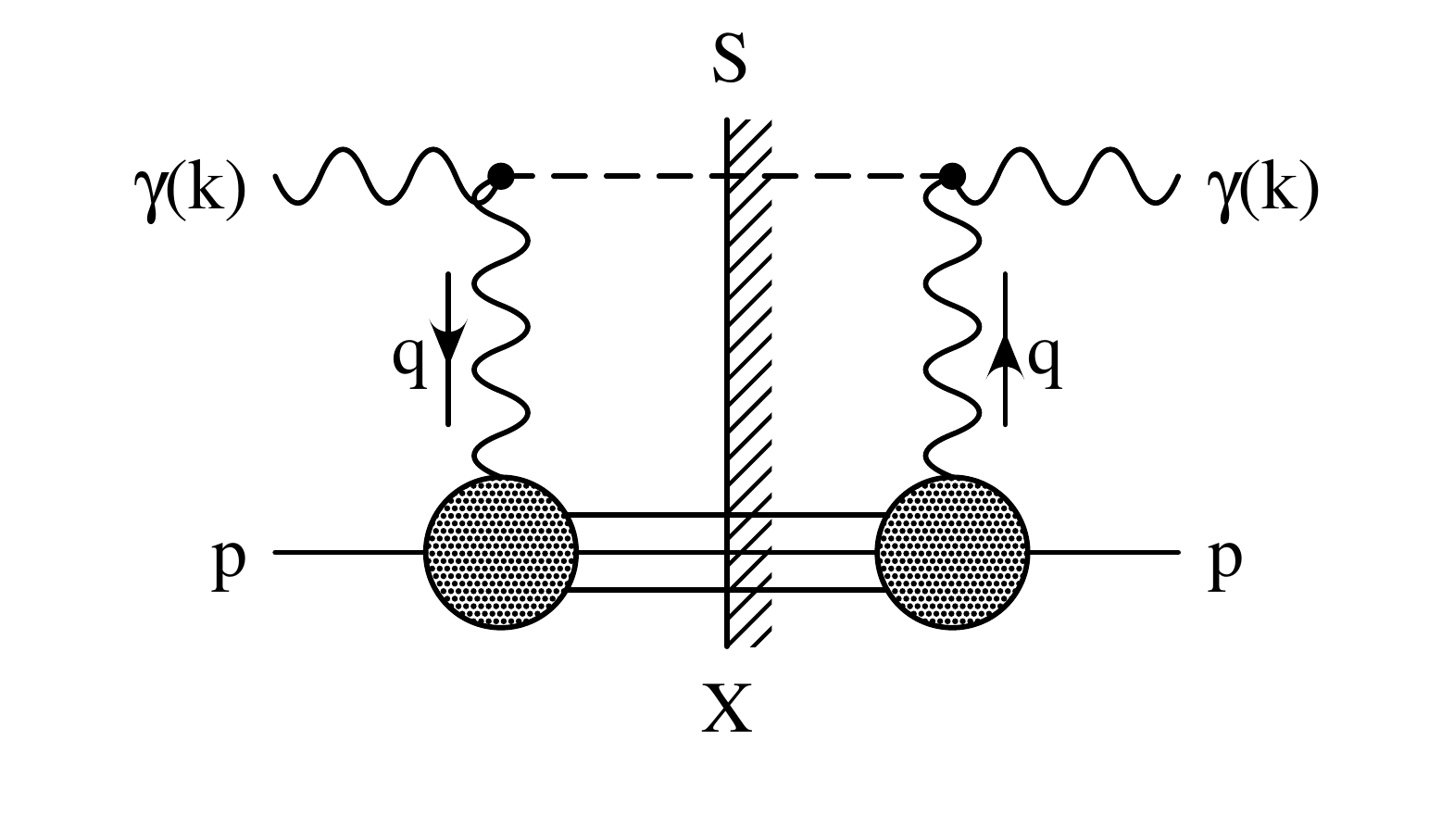}
\end{center}
\caption{\label{fig:04:4} Graph for the process $\gamma + p \to S + X$
  at lowest order in $\alpha$. }
\end{figure}

The $\gamma + p \to S + X$ cross section, from Fig.~\ref{fig:04:4}
%
%
including the QED vacuum polarisation bubbles and corrections to the
hadronic tensor is
\begin{multline}
  \sigma_{\gamma p} = \frac{\sigma_0}{2\pi \alpha(\mub)} 
  \int_{\xph}^{1-\frac{2 \xph\, \mpr}{M}} 
  \frac{d\zed}{\zed}
  \int^{\Qtwomax}_{\Qtwomin} 
  \frac{dQ^2}{Q^2} \aph^2(q^2)
  \Bigg[
  \biggl(
    -\zed^2
    -\frac{2 \zed ^2 Q^2}{M^2}
    -\frac{\zed ^2 Q^4}{M^4}
  \biggr)F_L(\xph/\zed,Q^2)
  +
  \\
  + \biggl(
    2
    -2 \zed 
    + \zed ^2
    +\frac{2 \xph ^2 \mpr^2}{Q^2}
    +\frac{2 \zed ^2 Q^2}{M^2}
    -\frac{2 \zed  Q^2}{M^2}
    +\frac{4 \xph ^2 \mpr^2}{M^2} 
    +\frac{\zed ^2 Q^4}{M^4} 
    +\frac{2 \xph ^2 Q^2 \mpr^2}{M^4}
  \biggr) F_2(\xph/\zed,Q^2)
  \Bigg]\,.
\label{4.26}
\end{multline}
We now follow the same procedure as before --- (a) define a ``physical
factorisation PDF'' from the terms not suppressed by $M^2$ (b) break
the integral up into three pieces as in Eq.~(\ref{4.18}) (c) evaluate
the two remaining integrals. Even though Eq.~(\ref{4.26}) is different
from Eq.~(\ref{4.18}), the $M^2$ unsuppressed terms are the same as
before, and lead to the same $f_{\gamma}^{\text{PF}}$ as in
Eq.~(\ref{4.16}). This gives the analogue of Eq.~(\ref{4.18x})
\begin{multline}
  \sigma_{\gamma p } = \sigma_0   \xph f^{\text{PF}}_{\gamma}(\xph,\mub) 
  +\frac{\alpha(\mub) }{2\pi}\sigma_0   \int_{\xph}^1
  \frac{d\zed}{\zed}
\Bigg[ 
\zed p_{\gamma q}(\zed) \left( \log \frac{M^2(1-\zed)^2 }{\zed \mub^2} \right)-\frac32 (1-\zed)^2 \Bigg]
F_2(\xph/\zed,\mub^2)\,.
  \label{4.27}
\end{multline}
We also need the partonic scattering cross section
%
%
\begin{figure}
\begin{center}
\includegraphics[width=3cm]{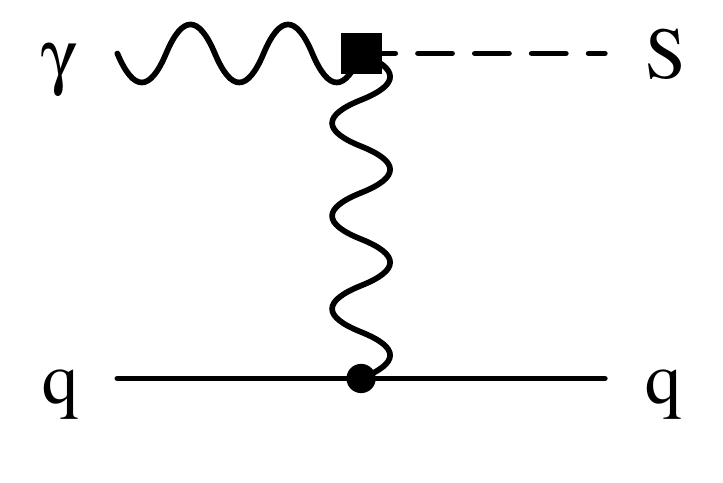}
\end{center}
\caption{\label{fig:04:3a} Lowest order graph for $\gamma + q \to S + q$. }
\end{figure}
\begin{multline}
\sigma_{ \gamma  q}
= \sigma_0 e_q^2 \frac{\alpha(\mub)}{2 \pi} \biggl[ 
\xph p_{\gamma q}(\xph) \left( -\frac{1}{\epsilon_{\text{IR}}}+\log \frac{M^2(1-\xph)^2 }{\xph \mub^2} \right)-\frac12 \xph^2+3 \xph -\frac32\biggr]\,.
\label{4.20a}
\end{multline}
From Eq.~(\ref{4.27}) and Eq.~(\ref{4.20a}), we obtain the same photon
PDF as before, Eq.~(\ref{eq:master}). Even though the non-logarithmic
terms in Eqs.~(\ref{4.27}) and Eq.~(\ref{4.20a}) are different from
those in Eqs.~(\ref{eq:sigmaprobeNLO}) and (\ref{4.18x}), the
difference of the two
\begin{align}
  \left[-\frac32 (1-\xph)^2\right] - \left[-\frac12 \xph^2+3 \xph
    -\frac32 \right] &= -\xph^2\,, & 
    \left[-\xph^2+3\xph-2\right] -
  \left[3 \xph-2 \right] &=-\xph^2\,,
\label{4.21b}
\end{align}
remains the same, and leads to the same $\MSbar$-conversion term. This
shows explicitly that, as expected, our derivation leads to a
process-independent result for the $\MSbar$ photon PDF. The structure
of radiative corrections for $\gamma + p \to S + X$ is more
complicated than for $l + p \to L +X$, so it is more difficult to
extend the $\gamma + p \to S + X$ result to higher
orders. Nevertheless, it must continue to give the same result for the
photon PDF, even at higher orders. An alternate derivation using PDF
operators that does not rely on any BSM process is given in
Sec.~\ref{sec:alt-deriv}. We have also checked the derivation of the
polarised photon PDF using the $\gamma \gamma \to S$ probe.

\section{Collinear-factorisation for $l q \to  L q$}
\label{sec:coll-q-e-to-L}
In this Appendix we give the derivation of the partonic
cross section given in Eq.~\eqref{eq:sigmahatprobe01}.
We start with the lowest order cross section for $l q \to L q $ in
$D$-dimensions ($D = 4-2\epsilon$), given by
\begin{equation}
  \label{eq:sigma-q-e-start}
  \hat \sigma_{l q }^{(0,1,\text{bare})}(\shat) 
  = 
  \frac{1}{2\shat}\cdot                  
  \frac14                                
    \left( \mu {\cal S}\right)^{2\epsilon} 
  \int_0^\pi d\theta \sin^{D-3}\!\theta\,
  \frac{2^{3-D} \pi^{1-D/2}}{\Gamma(D/2-1)}   
  \,\frac{|p'|^{D-3}}{4\sqrt{\shat}} 
  |\cM^2|\,,
\end{equation}
where the matrix element has been summed over spins, but without
averaging over initial spins.
The squared matrix-element is given by
\begin{equation}
  \label{eq:ME2-qe-gammaq}
    |\cM^2| = 256 \pi^2 \alpha^2 c^2 e_q^2 \frac{\left((D-2) M^2 \left(M^2+Q^2\right)-4 s
   \left(M^2+Q^2\right)+4 s^2\right)}{\Lambda ^2 Q^2}\,.
\end{equation}
The integral over $\theta$ gives
\begin{equation}
  \label{eq:L-epsilon-expansion}
  \hat \sigma_{lq}^{(0,1,\text{bare})}(\shat) 
  = \frac{\alpha \sigma_0 e_q^2}{2\pi}\left[ 
    -\frac{zp_{\gamma q}(z)}{\epsilon}
    -2 + 3z + 
    z p_{\gamma q}(z) \left(\ln\frac{M^2}{\mu^2} 
                             + \ln\frac{(1-z)^2}{z}\right)
    + \order{\epsilon}
  \right]\,,
\end{equation}
where we define $z \equiv M^2/\shat$, $\sigma_0$ is defined in
Eq.~\eqref{eq:sigma0} and $p_{\gamma q}(z)$ is given in
Eq.~\eqref{4.13}.

Since the leading-order cross section does not depend upon $\epsilon$,
the $\MSbar$ factorisation is simply achieved by dropping the
$1/\epsilon$ terms.  We obtain
\begin{align}
  \label{eq:5}
  \sigma_{lp}^{(\text{coll},q,\epsilon^{-1})}(\shat) 
  &= 
  - \frac{\alpha \sigma_0 e_q^2}{2\pi \epsilon}
  \int dx' \frac{M^2}{x' s} 
  p_{\gamma q}\left(\frac{M^2}{x' s}\right) f_{q}(x', \mub^2)
  \\
  &= 
  - \frac{\alpha \sigma_0 e_q^2}{2\pi \epsilon}
  \int dx'\, dz\, M^2 p_{\gamma q}\left(z\right) 
    \delta(x' z s - M^2) f_{q}(x',\mub^2)\,.
\end{align}
Now we introduce $x = z x'$ and we obtain
\begin{equation}
  \sigma_{l p}^{(\text{coll},q,\epsilon^{-1})}(\shat)  
  =
  - \frac{\alpha \sigma_0 e_q^2}{2\pi \epsilon}
  \int \frac{dx}{x}\, \frac{dz}{z}\, M^2 z p_{\gamma q}\left(z\right) 
    \delta(x s - M^2) \frac{x}{z} f_{q}(x/z,\mub^2)\,.
\end{equation}
Recall that $\sigma_{\gamma e \to L + X}^\text{(0)}(\shat) =  M^2 \sigma_0 
\delta(\shat - M^2)$ so that we rewrite our answer as
\begin{align}
  \sigma_{l p}^{(\text{coll},q,\epsilon^{-1})}(\shat)  
  &=
  - \frac{\alpha }{2\pi \epsilon} e_q^2
  \int \frac{dx}{x}\, 
  \frac{dz}{z}\, 
    z p_{\gamma q}\left(z\right) 
    \frac{x}{z} f_{q}(x/z,\mub^2)
  \sigma_{\gamma e \to L + X}^\text{0}(x s) \,,
  \\
  &= 
  \int dx\,  f_{\gamma}(x,\mub^2)\,  \sigma_{\gamma e \to L + X}^\text{0}(x s) \,,
\end{align}
where we have defined  
\begin{equation}
  xf_{\gamma}(x,\mub^2) \equiv - \frac{\alpha }{2\pi \epsilon} 
  \int \frac{dz}{z} zp_{\gamma q}(z)
  \frac{x}{z} f_{q}(x/z, \mub^2)\,.
\end{equation}
Thus if we remove just the $1/\epsilon$ part, the collinear-subtracted
part of $\hat \sigma_{lq}^{(1,0,\text{bare})}(\shat)$ is
given by
\begin{equation}
  \label{eq:coll-sub-L-qe}
  \hat \sigma_{lq}^{(0,1)}(\shat) 
  = \frac{\alpha \sigma_0 e_q^2}{2\pi}\left[ 
    -2 + 3z + 
    z p_{\gamma q}(z) \left(\ln\frac{M^2}{\mu^2} 
                             + \ln\frac{(1-z)^2}{z}\right)
  \right]\,.
\end{equation}
This formula holds both for quarks and leptons, by replacing the index
$q$ with $i\in \qlset$, and agrees with the result reported in
Eq.~\eqref{eq:sigmahatprobe01}.

\section{Low $Q^2$ behaviour of $F_2$ and $F_L$}
\label{sec:low-Q2-F2-FL}

$W_{\mu\nu}$ is the discontinuity of a forward amplitude in
$W^2=(p+q)^2$, and should be analytic in $Q^2$ and $W^2$ for $W^2$
away from the thresholds at $(\mpr + n m_\pi)^2$, and for $Q^2<(2
m_\pi)^2$.  In particular, it should be analytic as $Q^2\to 0$ at
fixed $W^2$ away from thresholds.  This implies that the coefficients
of its independent tensor structures should be analytic.
Looking at the tensor structure $q^\mu p^\nu$ in Eq.~(\ref{3.21}) we
immediately see that $F_2$ must vanish as $Q^2$. Considering instead
the tensor structure $q^\mu q^\nu$ one finds that $F_1 - F_2/(2 x)$
should vanish at least as $Q^2$ for small $Q^2$.
At small $Q^2$ and fixed $W^2$, $x$ behaves as $Q^2$, so that
\begin{eqnarray}
  F_L & =&  \left( 1 + \frac{4 \mpr^2 x^2}{Q^2} \right) F_2 - 2 x F_1 
  = - 2x\left(F_1 - \frac{F_2}{2 x}\right) + \frac{4 \mpr^2 x^2}{Q^2} F_2
\end{eqnarray}
vanishes at least as $Q^4$ as $Q^2\to 0$.

\section{Low $Q^2$ behaviour of $R=\sigma_L/\sigma_T$}\label{sec:R}

The cross section for $\gamma p$ scattering for transverse and
longitudinally polarised photons is
\begin{align}
\sigma_{T,L} &=  \left( \epsilon_{L,T}^{* \mu} \epsilon_{L,T}^\nu W_{\mu\nu}\right)\Phi  \,,
\label{R:1}
\end{align}
where the proportionality constant $\Phi$ depends on the convention
used for the incident photon flux for an off-shell (unphysical) photon
with momentum $Q^2$. In terms of structure functions,
\begin{align}
\sigma_{T} &= \Phi\,  F_1\,, & \sigma_L &= \Phi\, \frac{F_L}{2x}\,, & \frac{\sigma_L}{\sigma_T} &= \frac{F_L}{2x F_1}\,.
\label{R:6}
\end{align}

Pick a frame where $p=(m_p,0,0,0)$ and the photon has momentum
$q=(q^0,0,0,q^3)$ with longitudinal polarisation
\begin{align}
\epsilon_L=\frac{1}{Q}(q^3,0,0,q^0)\,.
\label{R:2}
\end{align}
We can rewrite this as
\begin{align}
\epsilon_L & = \frac{1}{Q} q^\mu + \frac{Q}{q^3+q^0} (1,0,0,-1)\,  =\frac{1}{Q} q^\mu + \frac{Q}{n \cdot q}  n^\mu \,,
\label{R:3}
\end{align}
where $n=(1,0,0,-1)$ is a null vector. Since $q^\mu W_{\mu \nu}=q^\nu W_{\mu \nu}=0$,
\begin{align}
\sigma_T &= \Phi\, W_{\perp \perp}\,, &
\sigma_L &=  \Phi\, \frac{Q^2}{(n \cdot q)^2} W_{nn}\,,
\label{R:4}
\end{align}
which shows that $\sigma_L/\sigma_T \propto Q^2$ as $Q^2 \to 0$ at
fixed $q^0$, or fixed $q^3$, or fixed $n \cdot q$.

\section{Summary of $\mathcal{O}(\alpha_s)$ and  $\mathcal{O}(\alpha)$
  coefficient and splitting functions}\label{subsec:coeff}

This section summarises known results for the $\mathcal{O}(\alpha_s)$
and $\mathcal{O}(\alpha)$ coefficient and splitting functions we need.

We start summarising the splitting functions for QCD processes that we
need
\begin{eqnarray}
  P_{ii}^{(1,0)}(x) &=& p_{qq}(x) = C_F \left(\frac{1+x^2}{1-x}\right)_+\,, \nonumber \\
  P_{ig}^{(1,0)}(x) &=& p_{qg}(x)= T_F \left(x^2+(1-x)^2\right)\,,
\end{eqnarray}
for $i\in\qset$,
and for QED
\begin{eqnarray}
  P_{ii}^{(0,1)}(x) &=\, e_i^2 p_{qq}^{\rm qed}(x)\,,\qquad p_{qq}^{\rm
                        qed}(x) &=  \left(\frac{1+x^2}{1-x}\right)_+\,,\nonumber \\ 
  P_{i\gamma}^{(0,1)}(x) &=\, n_i e_i^2 p_{q\gamma}(x)\,,\qquad
                             p_{q\gamma}(x) &= x^2+(1-x)^2\,,
\end{eqnarray}
for $i\in\qlset$.
We have also defined
\begin{equation}
    \Ptwoqa(x)=\left\{\begin{array}{ll} p_{qq}(x)\, e_a^2\; & \mbox{for}\; a\in\qset \\
                                 p_{qg}(x) \sum_{i\in\qset} e_i^2\; & \mbox{for}\; a=g
                                 \end{array}\right.\;,
  \label{G:9.51}
\end{equation}
and
\begin{equation}
  \Ptwoqael(x)=\left\{\begin{array}{ll} p^{\rm qed}_{qq}(x) e_a^4\; & \mbox{for}\; a\in\qlset \\
                                 p_{q\gamma}(x) \sum_{i\in\qlset} n_i e_i^4\; & \mbox{for}\; a=\gamma
                                 \end{array}\right.\;.
  \label{G:9.52}
\end{equation}

From Ref.~\cite{Bardeen:1978yd}, and converting to $x$ space:
\begin{align}
C_{L,i\in\qset}^{(1,0)} 
&= e_i^2\, C_F \, 2x\nn
C_{L,i\in\qlset}^{(0,1)}
  &= e_i^4 \, 2 x \nn
C_{2,i\in\qset}^{(1,0)} &=\frac12 e_i^2 C_F \biggl\{-3 \frac{1}{(1-x)_+}   + 4 \left[ \frac{\ln(1-x)}{1-x}\right]_+ -4 \frac{\ln x}{1-x} + 4x \nn
& - 2 (1+x) \ln (1-x) + 2 (1+x) \ln x   + 6  -(9+4 \zeta_2) \delta(1-x)\biggr\} \,, \nn
C_{2,i\in\qlset}^{(0,1)} &=\frac12 e_i^4 \biggl\{-3 \frac{1}{(1-x)_+}   + 4 \left[ \frac{\ln(1-x)}{1-x}\right]_+ -4 \frac{\ln x}{1-x} + 4x \nn
&- 2 (1+x) \ln (1-x) + 2 (1+x) \ln x  + 6  -(9+4 \zeta_2) \delta(1-x)\biggr\} \,, \nn
C_{L,g}^{(1,0)}  &= 4 T_R \left(\frac{1}{2}\sum_{i\in\qset} e_i^2\right)  x(1-x)   \,,\nn
C_{L,\gamma}^{(0,1)}  &= 4 \left(\frac{1}{2}\sum_{i\in\qlset} n_i e_i^4\right)  x(1-x)   \,,\nn
C_{2,g}^{(1,0)} &= T_R \left(\frac{1}{2}\sum_{i\in\qset} e_i^2\right) \biggl\{ \left[x^2+(1-x)^2\right]\ln \frac{1-x}{x} +8x(1-x)-1\biggr\} \,,\nn
C_{2,\gamma}^{(0,1)} &=  \left(\frac{1}{2}\sum_{i\in\qlset} n_i e_i^4\right) \biggl\{ \left[x^2+(1-x)^2\right]\ln \frac{1-x}{x} +8x(1-x)-1\biggr\}\,.
\label{583}
\end{align}
The $x$-space coefficient functions agree with
Ref.~\cite{Moch:1999eb}.  The $(0,1)$ coefficients can be obtained
from the $(1,0)$ ones using the replacement rules
\begin{eqnarray}
(1,0) & \longrightarrow & (0,1) \nonumber \\
g &\longrightarrow& \gamma \nonumber \\
T_R\left(\frac{1}{2}\sum_{i\in\qset} e_i^2\right)  &\longrightarrow& \left(\frac{1}{2}\sum_{i\in\qlset} n_i e_i^4\right) \nonumber \\
e_{i\in\qset}^2 C_f &\longrightarrow& e_{i\in\qlset}^4\;. \label{eq:replrules}
\end{eqnarray}
Since there was no quark vacuum polarisation contribution to the QCD
corrections at order $(1,0)$, the QED corrections are one-photon
irreducible.
This is consistent with our scheme for the hadronic tensor.

The order $\epsilon$ coefficient functions are~\cite{Zijlstra:1992qd}
\begin{subequations}\label{12}
  \begin{eqnarray}
    a_{2,i\in\qset}^{(1,0)} &=& e_i^2 C_F \Biggl[ \left( \frac{\ln^2(1-z)}{1-z}\right)_+ -
                                \frac32 \left( \frac{\ln(1-z)}{1-z}\right)_+ +\left(\frac72-3\zeta(2)\right)
                                \left( \frac{1}{1-z}\right)_+  \nn
    &-& \frac12(1+z) \log^2(1-z) 
        - \frac{1+z^2}{1-z} \log z \log (1-z) + \frac12  \frac{1+z^2}{1-z} \log^2 z +\frac32 \frac{1}{1-z} \log z \nn
    &+&(3+2z) \left(\log\frac{1-z}{z}-2\right)
        +\frac32(1+z) \zeta(2) +\left(9+\frac34 \zeta(2)\right)\delta(1-z)
        \Biggr]\,, 
    \\
    a_{2,g}^{(1,0)} &=& \frac12 T_R  \left(\frac{1}{2}\sum_{i\in\qset} e_i^2\right) \Biggl[ (1-2z+2z^2)\log^2\frac{1-z}{z} -2(1-8z+8z^2) \log\frac{1-z}{z} 
                            \nn
    &-&3(1-2z+2z^2)\zeta(2) 
        +6 -44z + 44 z^2 \Biggr]\,,
    \\
    a_{L,i\in\qset}^{(1,0)} &=& e_i^2 C_F \left[ 2 z
                                    \log\frac{1-z}{z} - 2z \right]\,, 
    \\
    a_{L,g}^{(1,0)} &=& T_R  \left(\frac{1}{2}\sum_{i\in\qset}
                        e_i^2\right)  4z(1-z)\left[ \log\frac{1-z}{z}-3
                        \right]\,. 
  \end{eqnarray}
\end{subequations}
We have recomputed Eq.~(\ref{12}) and found that the $\delta(1-z)$
term in $a_{2,q}^{(1,0)}$ has a coefficient $9+3 \zeta(2)/4$, rather
than $9+3 \zeta(2)/2$ given in Ref.~\cite{Zijlstra:1992qd}. With this
modification, the Adler sum rule
\begin{align}
\int_0^1 \rd z\ a_{2,q}^{(1,0)}(z) &=0
\end{align}
is now satisfied.

For the polarised case, from Ref.~\cite{Kodaira:1978sh}:
\begin{align}
C_{\Delta q}^{(1,0)} &= C_{2,q} - e_q^2  C_F\left( 1+x \right) \,.
\label{627.a}
\end{align}
$C_{\Delta q}$ is the same as the coefficient function for $F_3$ in $\nu$-scattering.
From Ref.~\cite{Manohar:1990jx}
\begin{align}
C_{\Delta g}^{(1,0)} 
 &= T_R  \left(\frac12 \sum_{i\in\qset} e_i^2\right) \left[ (2x-1) \ln \frac{1-x}{x}+3-4x\right] \,.
\label{627}
\end{align}
The order $\epsilon$ coefficient functions are~\cite{Zijlstra:1993sh}
\begin{subequations}  \label{12a}
  \begin{eqnarray}
    a_{\Delta i\in\qset}^{(1,0)} &=& a_{2,i}^{(1,0)}  + e_i^2 C_F
                                     \Biggl[
                                     -(1+z)\left(\log\frac{1-z}{z}-2\right)
                                     \Biggr] 
    \\
    a_{\Delta g}^{(1,0)} &=& \frac12 T_R  \left(\frac{1}{2}\sum_{i\in\qset} e_i^2\right)
                             \Biggl[ (2z-1)\log^2\frac{1-z}{z} +(6-8z) \log\frac{1-z}{z} -3(2z-1)\zeta(2)\nn
                         & -&12+6z \Biggr]
  \end{eqnarray}
\end{subequations}
The corresponding results for the $a^{(0,1)}_{\Delta i \in \qgset}$
coefficients are easily obtained from the above using the replacement
rules Eqs.~(\ref{eq:replrules}).

\section{Issues at low and moderate $\mu^2$ in PDF4LHC15}
\label{sec:PDF4LHC15-issues}

Our base PDF set, PDF4LHC15\_nnlo\_100~\cite{Butterworth:2015oua}, is
a combination of three underlying sets, CT14nnlo \cite{Dulat:2015mca},
MMHT2014nnlo68cl~\cite{Harland-Lang:2014zoa} and
NNPDF30\_nnlo\_as\_0118~\cite{Ball:2014uwa}.
This combination is intended for LHC applications, which mostly involve
high $\mu^2$ values (e.g. $(10\GeV)^2$ upwards).

In our work here, we need access to the PDF at low $\mu^2$ values in
order to then evolve it upwards with supplemental QED
corrections.\footnote{With many standard methods for numerical DGLAP
  evolution, upwards evolution is stable, while downwards evolution is
  often less so.
  In particular parton distributions from LHAPDF tend to have small
  irregularities (typically below a part in $10^3$) that make this
  especially problematic.
  One could imagine developing methods to make the
  downwards evolution more stable, however we have not investigated
  them.}
In doing so, we have encountered some issues, which we document here.

\begin{figure}
  \centering
  \includegraphics[width=0.49\textwidth,page=2]{Figs/flav-thresholds.pdf}%
  \hfill
  \includegraphics[width=0.49\textwidth,page=1]{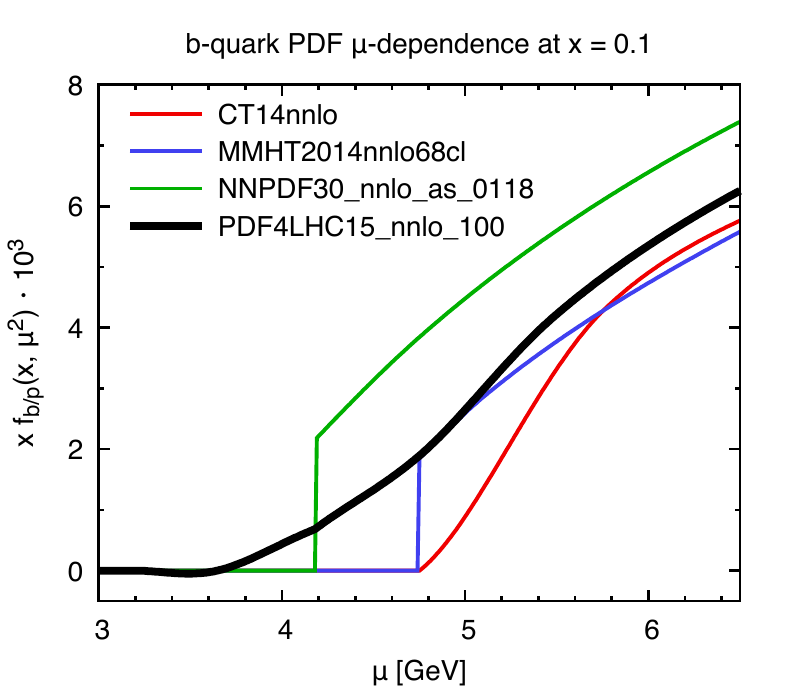}%
  \caption{%
    Left: the up-quark distribution at $x=0.1$ as a function
    of the factorisation scale $\mu$ in the PDF4LHC15\_nnlo\_100 set
    and in its three underlying input PDF sets, CT14nnlo,
    MMHT2014nnlo68cl and NNPDF30\_nnlo\_as\_0118.
    One sees anomalous behaviour for
    $\mu < 1.295\GeV$ in PDF4LHC15\_nnlo\_100 and CT14nnlo, associated
    with LHAPDF's extrapolation of the CT14nnlo below its range of
    validity.
    Right: similarly, but for the $b$-quark distribution (multiplied
    by $10^3$).
    It illustrates the different locations of the $b$-quark thresholds in
    the various sets.
    It also shows that only two of the sets display the expected
    discontinuous threshold at the $b$-quark mass.
}
  \label{fig:pdf4lhc-issues}
\end{figure}

A first point is that the set is quoted, within LHAPDF, as being valid
from $\mu = 1\GeV$.
However if one uses it at this scale, one encounters unexpected
behaviour such as a momentum sum of $0.94$ rather than $1$ and
inconsistency with other $\mu$ values if one evolves up with an
independent DGLAP code.
The origin of the problem turns out to be trivial, namely that one of the
input PDFs, CT14nnlo, is valid only from $\mu = \mu_0 = 1.295\GeV$ (while the
other two are valid from $ 1\GeV$).
When CT14nnlo is used below its starting scale, LHAPDF appears to
extrapolate it such that $f_{i}(x,\mu^2) \sim \mu^2/\mu_0^2 \,
f_{i}(x,\mu_0^2)$ as $\mu^2 \to 0$.
This is illustrated in Fig.~\ref{fig:pdf4lhc-issues} (left), which
shows the up quark PDF versus $\mu$.
One sees the CT14 curve dropping rapidly for $\mu^2 < \mu_0^2$,
whereas the other input PDFs vary much more slowly down to
$\mu^2 = 1\GeV^2$.
Since the PDF4LHC15 combination effectively averages the central
values of the three sets, it inherits part of the extrapolated CT14
behaviour.
This problem is common to all light flavours.
It would be trivial to fix, by increasing the lower $\mu^2$ limit of
the PDF set to the largest of the lower limits of the underlying sets.

A second issue concerns flavour thresholds.
The three underlying sets do not share the same thresholds.
For example CT14nnlo and MMHT2014nnlo68cl have their $b$-quark
threshold at $4.75\GeV$, while NNPDF30\_nnlo\_as\_0118 has its
threshold at $4.18\GeV$, as can be seen in
Fig.~\ref{fig:pdf4lhc-issues} (right).
Consequently, no single evolution that starts below the (highest) $b$
threshold can reproduce the results of the combined set across all
$\mu^2$ values.\footnote{One could imagine crafting an initial
  condition for the $b$ distribution such that the high-$\mu^2$
  results would coincide. The issues involved here are akin to those
  that arise in implementing downwards evolution.}

This suggests that the lowest scale from which one may start the evolution
is just above the highest of the different $b$ thresholds.
However yet another issue arises.
In Fig.~\ref{fig:pdf4lhc-issues} one sees that the MMHT and NNPDF
$b$-distributions have a discontinuity at the respective $b$ masses,
associated with a second order threshold term in the
evolution~\cite{Buza:1995ie,Buza:1996wv}.
The CT14 curve is instead continuous there.
For $\mu > 6\GeV$ it clearly approaches the MMHT2014 result, which
suggests that the underlying evolution contains the correct mass
threshold.
It seems likely, therefore, that the issue is related to the way that
the CT14nnlo set is included in LHAPDF.\footnote{One potential cause
  is that while the MMHT and NNPDF sets have separate $x,\mu^2$ grids
  for each number of light flavours, CT14 appears to use a single grid and so
  LHAPDF interpolates across the threshold.}
This issue seems to be reflected also in the PDF4LHC15 set.

Overall, therefore, we can only use PDF4LHC15\_nnlo\_100 as a starting
point for our evolution from a scale of about $\mu = 6\GeV$ upwards.

\bibliographystyle{JHEP}
\bibliography{photon}

\end{document}